\documentclass[english,3p, review]{elsarticle}
\usepackage[T1]{fontenc}
\usepackage[latin9]{inputenc}
\usepackage{array}
\usepackage{amsthm}
\usepackage{amsmath}
\usepackage{graphicx}
\usepackage{amssymb}

\makeatletter

\providecommand{\tabularnewline}{\\}

\theoremstyle{plain}
\theoremstyle{plain}
\newtheorem{thm}{Theorem}
  \theoremstyle{definition}
  \newtheorem{defn}[thm]{Definition}
  \theoremstyle{plain}
  \newtheorem{prop}[thm]{Proposition}

\newlength\myboxwidth
\newlength\myinnerboxwidth

\newcommand{\algobox}[2]{
\setlength\myboxwidth{\linewidth}%
\addtolength\myboxwidth{-2\parindent}%
\setlength\myinnerboxwidth{\myboxwidth}%
\addtolength\myinnerboxwidth{-1em}%
\begin{center}%
\noindent\framebox[\myboxwidth][c]{%
\noindent%
\begin{minipage}[]{\myinnerboxwidth}\textit{#1}\par\vspace{-2ex}#2\end{minipage}
}%
\end{center}%
}

\newcommand{\simplifiedmathchoice}[2]{\mathchoice{#1}{#2}{#1}{#1}}

\makeatother

\usepackage{babel}

\begin{document}
\global\long\def\rk{\mathrm{rank}}
\global\long\def\diag{\mathrm{diag}}
\global\long\def\adj{\mathrm{adj}}

\global\long\def\bigpar#1{\simplifiedmathchoice{\bigl(#1\bigr)}{(#1)}}

\global\long\def\bigbrk#1{\simplifiedmathchoice{\bigl[#1\bigr]}{[#1]}}

\global\long\def\bigcrl#1{\simplifiedmathchoice{\bigl\{#1\bigr\}}{\{#1\}}}

\global\long\def\bigabs#1{\simplifiedmathchoice{\bigl|#1\bigr|}{|#1|}}

\global\long\def\hvecd#1#2#3{\simplifiedmathchoice{\bigl[\!\begin{array}{cccc}
 \!#1  &  #2  &  \cdots &  #3\!\end{array}\!\!\bigr]}{[#1\ #2\ \cdots\ #3]}}

\global\long\def\fftinterpdatavector#1#2#3#4#5#6{[#1\ #2\ \cdots\ #3\ 0\ \cdots\ 0\ #4\ #5\ \cdots\ #6]}

\global\long\def\matcouple#1#2{\simplifiedmathchoice{\bigl[\!\!\begin{array}{cc}
 #1  &  #2\end{array}\!\!\bigr]}{[\!\!\begin{array}{cc}
 #1  &  #2\end{array}\!\!]}}

\global\long\def\matthreesome#1#2#3{\bigl[\!\!\begin{array}{ccc}
 #1  &  #2  &  #3\end{array}\!\!\bigr]}

\global\long\def\ran{\mathrm{ran}}

\global\long\def\vecd#1#2#3{\left[\!\begin{array}{c}
 #1\\
#2\\
\vdots\\
#3\end{array}\!\right]}

\global\long\def\subcol#1#2#3{#1_{#2,#3}}

\global\long\def\subrow#1#2#3{#1^{#2,#3}}
\global\long\def\subcolrow#1#2#3#4#5{#1_{#2,#3}^{#4,#5}}

\global\long\def\allzero#1#2{#1=0,1,\ldots,#2}

\global\long\def\allone#1#2{#1=1,2,\ldots,#2}

\global\long\def\alltwo#1#2{#1=2,3,\ldots,#2}

\global\long\def\pol#1#2#3{#1\sim\left(#2,#3\right)}

\global\long\def\cof#1#2#3{\Gamma_{#2,#3}^{(#1)}}

\global\long\def\ntx{M_{T}}
\global\long\def\ntxx{M_{T}^{2}}
\global\long\def\ntxxx{M_{T}^{3}}

\global\long\def\nrx{M_{R}}
\global\long\def\nrxx{M_{R}^{2}}
\global\long\def\nrxxx{M_{R}^{3}}

\global\long\def\ncar{N}

\global\long\def\setunit{\mathcal{U}}

\global\long\def\lcp{L_{\textrm{CP}}}

\global\long\def\ndat{D}
\global\long\def\setdat{\mathcal{D}}

\global\long\def\mattap{\mathbf{H}}

\global\long\def\mattf{\mathbf{H}}
\global\long\def\stkmattf{\bar{\mattf}}
\global\long\def\stkmatq{\bar{\mathbf{Q}}}

\global\long\def\symtx{c}
\global\long\def\vectx{\mathbf{c}}

\global\long\def\setconst{\mathcal{O}}

\global\long\def\symrx{d}
\global\long\def\vecrx{\mathbf{d}}

\global\long\def\vecnoise{\mathbf{w}}

\global\long\def\stddevnoise{\sigma_{w}}
\global\long\def\varnoise{\stddevnoise^{2}}

\global\long\def\setpil{\mathcal{E}}

\global\long\def\nbase{B}
\global\long\def\matbase{\mathbf{B}}
\global\long\def\setbase{\mathcal{B}}

\global\long\def\ntgt{T}
\global\long\def\mattgt{\mathbf{T}}
\global\long\def\settgt{\mathcal{T}}

\global\long\def\veccoeff{\mathbf{a}}

\global\long\def\matdft#1{\mathbf{W}_{\!#1}}

\global\long\def\nentries{J}

\global\long\def\dg{V}

\global\long\def\map{\mathcal{M}}

\global\long\def\qrnotilde#1{(\mathbf{Q}#1,\mathbf{R}#1)}

\global\long\def\qrtilde#1{(\tilde{\mathbf{Q}}#1,\tilde{\mathbf{R}}#1)}

\global\long\def\qrpartnotilde#1#2#3{(\subcol{\mathbf{Q}}{#1}{#2}#3,\subrow{\mathbf{R}}{#1}{#2}#3)}

\global\long\def\qrparttilde#1#2#3{(\subcol{\tilde{\mathbf{Q}}}{#1}{#2}#3,\subrow{\tilde{\mathbf{R}}}{#1}{#2}#3)}

\global\long\def\ofsn{\bigpar{s_{n}}}

\global\long\def\expval{\mathbb{E}}

\global\long\def\cgauss#1#2{\mathcal{CN}(#1,#2)}

\global\long\def\calg#1{C_{\textrm{#1}}}

\global\long\def\cqrtot{c_{\textrm{QR}}}
\global\long\def\cqralg#1{c_{\textrm{QR,#1}}}

\global\long\def\cqrk#1{c_{\textrm{QR}}^{\nrx\times#1}}
\global\long\def\cmmseqr{c_{\textrm{MMSE-QR}}^{\nrx\times\ntx}}

\global\long\def\cqrpm{c_{\textrm{QR}}^{P\times M}}
\global\long\def\cmmseqrpm{c_{\textrm{MMSE-QR}}^{P\times M}}
\global\long\def\cthreemmseqrpm{c_{\textrm{QR,III-MMSE}}^{P\times M}}

\global\long\def\credk#1{c_{\textrm{red}}^{(#1)}}
\global\long\def\credalg#1{c_{\textrm{red,#1}}}

\global\long\def\cip{c_{\textrm{IP}}}
\global\long\def\cipmax#1{c_{\textrm{IP,max,#1}}}
\global\long\def\cipfft{c_{\textrm{IP,FFT}}}

\global\long\def\ciphk#1{c_{\textrm{IP,}\mattf}^{#1,\ntx}}

\global\long\def\cipqr{c_{\textrm{IP,}\tilde{\mathbf{Q}}\tilde{\mathbf{R}}}}
\global\long\def\cipqrk#1{c_{\textrm{IP,}\tilde{\mathbf{q}}\tilde{\mathbf{r}}}^{(#1)}}
\global\long\def\cipqralg#1{c_{\textrm{IP,}\tilde{\mathbf{Q}}\tilde{\mathbf{R}},\textrm{#1}}}

\global\long\def\cipH{c_{\textrm{IP,}\mattf}}
\global\long\def\ciphalg#1{c_{\textrm{IP,}\mattf,\textrm{#1}}}

\global\long\def\ciptot{c_{\textrm{IP,}\mattf\tilde{\mathbf{Q}}\tilde{\mathbf{R}}}}

\global\long\def\cmapk#1{c_{\map}^{#1,\ntx}}
\global\long\def\cmapalg#1{c_{\map,\textrm{#1}}}

\global\long\def\cdemapk#1{c_{\map^{-1}}^{1,#1}}
\global\long\def\cdemapalg#1{c_{\map^{-1},\textrm{#1}}}

\global\long\def\cmaptot{c_{\map,\map^{-1}}}

\global\long\def\upratio{R}

\global\long\def\modulo{\,\mathrm{mod}\,}

\global\long\def\argmin#1{\underset{#1}{\mathrm{arg\, min}}}

\global\long\def\setidxbase{\mathcal{I}}

\global\long\def\setidx#1{\mathcal{S}(#1)}

\global\long\def\matperm{\mathbf{P}}

\global\long\def\stkmata{\bar{\mathbf{A}}}

\global\long\def\stkqrnotilde#1{(\stkmatq#1,\mathbf{R}#1)}

\global\long\def\stkqrtilde#1{(\tilde{\stkmatq}#1,\tilde{\mathbf{R}}#1)}

\global\long\def\stkqrpartnotilde#1#2#3{(\subcol{\stkmatq}{#1}{#2}#3,\subrow{\mathbf{R}}{#1}{#2}#3)}

\global\long\def\stkqrparttilde#1#2#3{(\subcol{\tilde{\stkmatq}}{#1}{#2}#3,\subrow{\tilde{\mathbf{R}}}{#1}{#2}#3)}

\global\long\def\stkvecq{\bar{\mathbf{q}}}

\global\long\def\cmultr{\chi_{\mathbb{R}}}
\global\long\def\cmultc{\chi_{\mathbb{C}}}

\global\long\def\matcirc#1#2{\mathbf{C}_{#1}^{#2}}

\begin{frontmatter}
\title{Interpolation-Based QR~Decomposition\\in MIMO-OFDM Systems\tnoteref{t1,t2}}
\author[ctg]{Davide Cescato}
\ead{dcescato@nari.ee.ethz.ch}
\author[ctg]{Helmut B\"olcskei\corref{cor1}}
\ead{boelcskei@nari.ee.ethz.ch}
\cortext[cor1]{Corresponding author. Tel.: +41 44 632 3433, fax: +41 44 632 1209.}
\address[ctg]{Communication Technology Laboratory, ETH Zurich, 8092 Zurich, Switzerland}
\tnotetext[t1]{This work was supported in part by the Swiss National Science Foundation under grant No.~200021-100025/1.}
\tnotetext[t2]{Parts of this paper were presented at the Sixth IEEE Workshop on Signal Processing Advances in Wireless Communications (SPAWC), New York, NY, June 2005.}
\begin{abstract}

Detection algorithms for multiple-input multiple-output (MIMO) wireless
systems based on orthogonal fre\-quency-division multiplexing (OFDM)
typically require the computation of a QR~decomposition for each
of the data-carrying OFDM tones. The resulting computational complexity
will, in general, be significant, as the number of data-carrying tones
ranges from 48 (as in the IEEE 802.11a/g standards) to 1728 (as in
the IEEE 802.16e standard). Motivated by the fact that the channel
matrices arising in MIMO-OFDM systems are highly oversampled polynomial
matrices, we formulate interpolation-based QR~decomposition algorithms.
An in-depth complexity analysis, based on a metric relevant for very
large scale integration~(VLSI) implementations, shows that the proposed
algorithms, for sufficiently high number of data-carrying tones and
sufficiently small channel order, provably exhibit significantly smaller
complexity than brute-force per-tone QR~decomposition.

\end{abstract}
\end{frontmatter}

\section{Introduction and Outline}

The use of orthogonal frequency-division multiplexing (OFDM) drastically
reduces data detection complexity in wideband multiple-input multiple-output
(MIMO) wireless systems by decoupling a frequency-selective fading
MIMO channel into a set of flat-fading MIMO channels. Nevertheless,
MIMO-OFDM detectors still pose significant challenges in terms of
computational complexity, as processing has to be performed on a per-tone
basis with the number of data-carrying tones ranging from 48 (as in
the IEEE 802.11a/g wireless local area network standards) to 1728
(as in the IEEE 802.16 wireless metropolitan area network standard).

Specifically, in the setting of coherent MIMO-OFDM detection, for
which the receiver is assumed to have perfect channel knowledge, linear
MIMO-OFDM detectors\citet{paulraj03} require matrix inversion, whereas
successive cancelation receivers\citet{wolniansky98} and sphere decoders\citet{fincke85,viterbo93}
require QR~decomposition, in all cases on each of the data-carrying
OFDM tones. The corresponding computations, termed as\emph{ preprocessing}
in the following, have to be performed at the rate of change of the
channel which, depending on the propagation environment, is typically
much lower than the rate at which the transmission of actual data
symbols takes place. Nevertheless, as payload data received during
the preprocessing phase must be stored in a dedicated buffer, preprocessing
represents a major bottleneck in terms of the size of this buffer
and the resulting detection latency\citet{perels05}.

In a very large scale integration (VLSI) implementation, the straightforward
approach to reducing the preprocessing latency is to employ parallel
processing over multiple matrix inversion or QR~decomposition units,
which, however, comes at the cost of increased silicon area. In\citet{borgmann04-11a},
the problem of reducing preprocessing complexity in linear MIMO-OFDM
receivers is addressed on an algorithmic level by formulating efficient
interpolation-based algorithms for matrix inversion that take the
polynomial nature of the MIMO-OFDM channel matrix explicitly into
account. Specifically, the algorithms proposed in\citet{borgmann04-11a}
exploit the fact that the channel matrices arising in MIMO-OFDM systems
are polynomial matrices that are highly oversampled on the unit circle.
The goal of the present paper is to devise computationally efficient
interpolation-based algorithms for QR~decomposition in MIMO-OFDM
systems. Although throughout the paper we focus on QR~decomposition
in the context of coherent MIMO-OFDM detectors, our results also apply
to transmit precoding schemes for MIMO-OFDM (under the assumption
of perfect channel knowledge at the transmitter) requiring per-tone
QR~decomposition\citet{windpassinger04-07}.

\paragraph*{Contributions}

Our contributions can be summarized as follows:
\begin{itemize}
\item We present a new result on the QR~decomposition of Laurent polynomial~(LP)
matrices, based on which interpolation-based algorithms for QR~decomposition
in MIMO-OFDM systems are formulated.
\item Using a computational complexity metric relevant for VLSI implementations,
we demonstrate that, for a wide range of system parameters, the proposed
interpolation-based algorithms exhibit significantly smaller complexity
than brute-force per-tone QR~decomposition.
\item We present different strategies for efficient LP~interpolation that
take the specific structure of the problem at hand into account and
thereby enable (often significant) computational complexity savings
of interpolation-based QR~decomposition.
\item We provide a numerical analysis of the trade-off between the computational
complexity of the inter\-polation-based QR~decomposition algorithms
presented and the performance of corresponding MIMO-OFDM detectors.
\end{itemize}

\paragraph*{Outline of the paper}

In Section~\ref{sec:preliminaries}, we present the mathematical
preliminaries needed in the rest of the paper. In Section~\ref{sec:mimoofdm},
we briefly review the use of QR~decomposition in MIMO-OFDM receivers,
and we formulate the problem statement. In Section~\ref{sec:interpqr},
we present our main technical result on the QR~decomposition of LP~matrices.
This result is then used in Section~\ref{sec:algorithms} to formulate
interpolation-based algorithms for QR~decomposition of MIMO-OFDM
channel matrices. Section~\ref{sec:complexity} contains an in-depth
computational complexity analysis of the proposed algorithms. In Section~\ref{sec:mmsecase},
we describe the application of the new approach to the QR~decomposition
of the augmented MIMO-OFDM channel matrices arising in the context
of minimum mean-square error~(MMSE) receivers. In Section~\ref{sec:efficientinterpolation},
we discuss methods for LP~interpolation that exploit the specific
structure of the problem at hand and exhibit low VLSI implementation
complexity. Section~\ref{sec:numresults} contains numerical results
on the computational complexity of the proposed interpolation-based
QR~decomposition algorithms along with a discussion of the trade-off
between algorithm complexity and MIMO-OFDM receiver performance. We
conclude in Section~\ref{sec:conclusions}.

\section{Mathematical Preliminaries}

\label{sec:preliminaries}

\subsection{Notation}

$\mathbb{C}^{P\times M}$ denotes the set of complex-valued $P\times M$
matrices. $\setunit\triangleq\{s\in\mathbb{C}:|s|=1\}$ indicates
the unit circle. $\emptyset$ is the empty set. $\left|\mathcal{A}\right|$
stands for the cardinality of the set~$\mathcal{A}$. $\!\modulo\!$
is the modulo operator. All logarithms are to the base~2. $\expval[\cdot]$
denotes the expectation operator. $\cgauss{\mathbf{0}}{\mathbf{K}}$
stands for the multivariate, circularly-symmetric complex Gaussian
distribution with covariance matrix~$\mathbf{K}$. Throughout the
paper, we use the following conventions. First, if $k_{2}<k_{1}$,
$\sum_{k=k_{1}}^{k_{2}}\alpha_{k}=0$, regardless of~$\alpha_{k}$.
Second, sequences of integers of the form $k_{1},k_{1}+\Delta,\ldots,k_{2}$,
with~$\Delta>0$, simplify to the sequence $k_{1},k_{2}$ if $k_{2}=k_{1}+\Delta$,
to the single value~$k_{1}$ if $k_{2}=k_{1}$, and to the empty
sequence if $k_{2}<k_{1}$.

\textbf{$\mathbf{A}^{*}\!$}, $\mathbf{A}^{T}\!$, $\mathbf{A}^{H}\!$,
$\mathbf{A}^{\dagger}\!$, $\rk(\mathbf{A})$, and $\ran(\mathbf{A})$
denote the entrywise conjugate, the transpose, the conjugate transpose,
the pseudoinverse, the rank, and the range space, respectively, of
the matrix~$\mathbf{A}$. $[\mathbf{A}]_{p,m}$ indicates the entry
in the $p$th row and $m$th column of~$\mathbf{A}$. $\subrow{\mathbf{A}}{p_{1}}{p_{2}}$
and~$\subcol{\mathbf{A}}{m_{1}}{m_{2}}$ stand for the submatrix
given by the rows $p_{1},p_{1}+1,\ldots,p_{2}$ of\textbf{~$\mathbf{A}$}
and the submatrix given by the columns $m_{1},m_{1}+1,\ldots,m_{2}$
of\textbf{~$\mathbf{A}$}, respectively. Furthermore, we set $\subcolrow{\mathbf{A}}{m_{1}}{m_{2}}{p_{1}}{p_{2}}\triangleq\subrow{\bigpar{\subcol{\mathbf{A}}{m_{1}}{m_{2}}}}{p_{1}}{p_{2}}\!$
and $\subcol{\mathbf{A}}{m_{1}}{m_{2}}^{H}\triangleq\bigpar{\subcol{\mathbf{A}}{m_{1}}{m_{2}}}^{H}\!$.
A $P\times M$ matrix~$\mathbf{A}$ is said to be upper triangular
if all entries below its main diagonal $\{[\mathbf{A}]_{k,k}:\allone k{\min(P,M)}\}$
are equal to zero. $\det(\mathbf{A})$ and~$\adj(\mathbf{A})$ denote
the determinant and the adjoint of a square matrix~$\mathbf{A}$,
respectively. $\diag\bigpar{a_{1},a_{2},\ldots,a_{M}}$ indicates
the $M\times M$ diagonal matrix with the scalar~$a_{m}$ as its
$m$th main diagonal element. $\mathbf{I}_{M}$ stands for the $M\times M$
identity matrix, $\mathbf{0}$ denotes the all-zeros matrix of appropriate
size, and~$\matdft M$ is the $M\times M$ discrete Fourier transform
matrix, given by $\bigbrk{\matdft M}_{p+1,q+1}=e^{-j2\pi pq/M}$ ($\allzero{p,q}{M-1}$).
Finally, orthogonality and norm of complex-valued vectors $\mathbf{a}_{1},\mathbf{a}_{2}$
are induced by the inner product $\mathbf{a}_{1}^{H}\mathbf{a}_{2}$.

\subsection{QR Decomposition}

\label{sub:QRDecomposition}

Throughout this section, we consider a matrix $\mathbf{A}=\hvecd{\mathbf{a}_{1}}{\mathbf{a}_{2}}{\mathbf{a}_{M}}\in\mathbb{C}^{P\times M}$
with $P\geq M$, where $\mathbf{a}_{k}$ denotes the $k$th column
of~$\mathbf{A}$ ($\allone kM).$ In the remainder of the paper,
the term QR~decomposition refers to the following:
\begin{defn}
\label{def:QR} We call any factorization $\mathbf{A}=\mathbf{QR}$,
for which the matrices $\mathbf{Q}\in\mathbb{C}^{P\times M}$ and
$\mathbf{R}\in\mathbb{C}^{M\times M}$ satisfy the following conditions,
a \emph{QR~decomposition} of~$\mathbf{A}$ with \emph{QR~factors}
$\mathbf{Q}$ and~$\mathbf{R}$:
\begin{enumerate}
\item \label{enu:QR_cond_Q}the nonzero columns of~$\mathbf{Q}$ are orthonormal
\item \label{enu:QR_cond_R}$\mathbf{R}$ is upper triangular with real-valued
nonnegative entries on its main diagonal
\item \label{enu:QR_cond_R_QH}$\mathbf{R}=\mathbf{Q}^{H}\mathbf{A}$
\end{enumerate}
\end{defn}
Practical algorithms for QR~decomposition are either based on Gram-Schmidt
(GS) orthonormalization or on unitary transformations~(UT). We next
briefly review both classes of algorithms. GS-based QR~decomposition
is summarized as follows. For $\allone kM$, the $k$th column of~$\mathbf{Q}$,
denoted by~$\mathbf{q}_{k}$, is determined by\begin{equation}
\mathbf{y}_{k}\triangleq\mathbf{a}_{k}-\sum_{i=1}^{k-1}\mathbf{q}_{i}^{H}\mathbf{a}_{k}\mathbf{q}_{i}\label{eq:y_k}\end{equation}
with\begin{equation}
\mathbf{q}_{k}=\begin{cases}
\frac{\mathbf{y}_{k}}{\sqrt{\mathbf{y}_{k}^{H}\mathbf{y}_{k}}}, & \quad\mathbf{y}_{k}\neq\mathbf{0}\\
\mathbf{0}, & \quad\mathbf{y}_{k}=\mathbf{0}\end{cases}\label{eq:q_k}\end{equation}
whereas the $k$th row of~$\mathbf{R}$, denoted by~$\mathbf{r}_{k}^{T}$,
is given by\begin{equation}
\mathbf{r}_{k}^{T}=\mathbf{q}_{k}^{H}\mathbf{A}.\label{eq:r_k}\end{equation}
UT-based QR~decomposition of~$\mathbf{A}$ is performed by left-multiplying~$\mathbf{A}$
by the product $\mathbf{\Theta}_{U}\cdots\mathbf{\Theta}_{2}\mathbf{\Theta}_{1}$
of $P\times P$ unitary matrices~$\mathbf{\Theta}_{u}$, where the
sequence of matrices $\mathbf{\Theta}_{1},\mathbf{\Theta}_{2},\ldots,\mathbf{\Theta}_{U}$
and the parameter~$U$ are not unique and are chosen such that the
$P\times M$ matrix $\mathbf{\Theta}_{U}\cdots\mathbf{\Theta}_{2}\mathbf{\Theta}_{1}\mathbf{A}$
is upper triangular with nonnegative real-valued entries on its main
diagonal. The matrices~$\mathbf{\Theta}_{u}$ are typically either
Givens rotation matrices\citet{golub96} or Householder reflection
matrices\citet{golub96}. With $\mathbf{R}\triangleq\subrow{(\mathbf{\Theta}_{U}\cdots\mathbf{\Theta}_{2}\mathbf{\Theta}_{1}\mathbf{A})}1M$
and $\mathbf{Q}\triangleq\subcol{((\mathbf{\Theta}_{U}\cdots\mathbf{\Theta}_{2}\mathbf{\Theta}_{1})^{H})}1M$,
we obtain that $\mathbf{Q}^{H}\mathbf{A}=\mathbf{R}$ and, since $\mathbf{\Theta}_{U}\cdots\mathbf{\Theta}_{2}\mathbf{\Theta}_{1}$
is unitary, that $\mathbf{Q}^{H}\mathbf{Q}=\mathbf{I}_{M}$. Therefore,
$\mathbf{Q}$ and~$\mathbf{R}$ are QR~factors of~$\mathbf{A}$.
For $P>M$, we note that the $P\times(P-M)$ matrix $\mathbf{Q}^{\perp}\triangleq\subcol{((\mathbf{\Theta}_{U}\cdots\mathbf{\Theta}_{2}\mathbf{\Theta}_{1})^{H})}{M+1}P$
satisfies $(\mathbf{Q}^{\perp})^{H}\mathbf{Q}^{\perp}=\mathbf{I}_{P-M}$
and $\mathbf{Q}^{H}\mathbf{Q}^{\perp}=\mathbf{0}$. In practice, UT-based
QR~decomposition of~$\mathbf{A}$ can be performed as follows\citet{golub96,burg05}.
A $P\times M$ matrix~$\mathbf{X}$ and a $P\times P$ matrix~$\mathbf{Y}$
are initialized as $\mathbf{X}\leftarrow\mathbf{A}$ and $\mathbf{Y}\leftarrow\mathbf{I}_{P}$,
respectively, and the counter~$u$ is set to zero. Then, $u$ is
incremented by one, and $\mathbf{X}$ and $\mathbf{Y}$ are updated
according to $\mathbf{X}\leftarrow\mathbf{\Theta}_{u}\mathbf{X}$
and $\mathbf{Y}\leftarrow\mathbf{\Theta}_{u}\mathbf{Y}$, for an appropriately
chosen matrix~$\mathbf{\Theta}_{u}$. This update step is repeated
until~$\mathbf{X}$ becomes upper-triangular with nonnegative real-valued
entries on its main diagonal. The parameter~$U$ is obtained as the
final value of the counter~$u$, and the final values of $\mathbf{X}$
and~$\mathbf{Y}$ are\[
\mathbf{X}=\left[\begin{array}{c}
\mathbf{R}\\
\mathbf{0}\end{array}\right],\mbox{ }\mathbf{Y}=\left[\begin{array}{c}
\mathbf{Q}^{H}\\
(\mathbf{Q}^{\perp})^{H}\end{array}\right]\!.\]
Since the $u$th update step can be represented as $\matcouple{\mathbf{X}}{\mathbf{Y}}\leftarrow\mathbf{\Theta}_{u}\matcouple{\mathbf{X}}{\mathbf{Y}}$,
we can describe UT-based QR~decomposition of~$\mathbf{A}$ by means
of the formal relation\begin{equation}
\mathbf{\Theta}_{U}\cdots\mathbf{\Theta}_{2}\mathbf{\Theta}_{1}\matcouple{\mathbf{A}}{\mathbf{I}_{P}}=\left[\begin{array}{cc}
\mathbf{R} & \mathbf{Q}^{H}\\
\mathbf{0} & (\mathbf{Q}^{\perp})^{H}\end{array}\right]\label{eq:utbasedqr}\end{equation}
which, from now on, will be called \emph{standard form }of UT-based
QR~decomposition, and will be needed in Section~\ref{sub:regularizedqr}
in the context of regularized QR~decomposition. The standard form~(\ref{eq:utbasedqr})
shows that for $P>M$, UT-based QR~decomposition yields the $(P-M)\times P$
matrix $(\mathbf{Q}^{\perp})^{H}$ as a by-product. For $P=M$, the
right-hand side~(RHS) of~(\ref{eq:utbasedqr}) reduces to $\matcouple{\mathbf{R}}{\mathbf{Q}^{H}}$.

We note that since $\mathbf{y}_{1}=\mathbf{0}$ is equivalent to $\mathbf{a}_{1}=\mathbf{0}$
and $\mathbf{y}_{k}=\mathbf{0}$ is equivalent to $\rk(\subcol{\mathbf{A}}1{k-1})=\rk(\subcol{\mathbf{A}}1k)$
($\alltwo kM$)\citet{Horn85}, GS-based QR~decomposition sets $M-\rk(\mathbf{A})$
columns of~$\mathbf{Q}$ and the corresponding $M-\rk(\mathbf{A})$
rows of~$\mathbf{R}$ to zero. In contrast, UT-based QR~decomposition
yields a matrix $\mathbf{Q}$ such that $\mathbf{Q}^{H}\mathbf{Q}=\mathbf{I}_{M}$,
regardless of the value of $\rk(\mathbf{A})$, and sets $M-\rk(\mathbf{A})$
entries on the main diagonal of~$\mathbf{R}$ to zero\citet{golub96}.
Hence, for $\rk(\mathbf{A})<M$, different QR~decomposition algorithms
will in general produce different QR~factors.
\begin{prop}
\label{pro:qrfullrank}If $\rk(\mathbf{A})=M$, Conditions~\ref{enu:QR_cond_Q}
and \ref{enu:QR_cond_R} of Definition~\ref{def:QR} simplify, respectively,
to
\begin{enumerate}
\item $\mathbf{Q}^{H}\mathbf{Q}=\mathbf{I}_{M}$
\item $\mathbf{R}$ is upper triangular with $[\mathbf{R}]_{k,k}>0,\allone kM$
\end{enumerate}
whereas Condition~\ref{enu:QR_cond_R_QH} is redundant. Moreover,
$\mathbf{A}$ has unique QR~factors.\end{prop}
\begin{proof}
Since $\mathbf{A}=\mathbf{QR}$ implies $\rk(\mathbf{A})\leq\min\{\rk(\mathbf{Q}),\rk(\mathbf{R})\}$,
it follows from $\rk(\mathbf{A})=M$ that $\rk(\mathbf{Q})=\rk(\mathbf{R})=M$.
Now, $\rk(\mathbf{Q})=M$ implies that the $P\times M$ matrix~$\mathbf{Q}$
can not contain all-zero columns, and hence Condition~\ref{enu:QR_cond_Q}
is equivalent to $\mathbf{Q}^{H}\mathbf{Q}=\mathbf{I}_{M}$. Moreover,
$\rk(\mathbf{R})=M$ implies $\det(\mathbf{R})\neq0$ and, since $\mathbf{R}$
is upper triangular, we have $\det(\mathbf{R})=\prod_{k=1}^{M}[\mathbf{R}]_{k,k}$.
Hence, Condition~\ref{enu:QR_cond_R} becomes $[\mathbf{R}]_{k,k}>0,\allone kM$.
Condition~\ref{enu:QR_cond_R_QH} is redundant since $\mathbf{A}=\mathbf{Q}\mathbf{R}$,
together with $\mathbf{Q}^{H}\mathbf{Q}=\mathbf{I}_{M}$, implies
$\mathbf{Q}^{H}\mathbf{A}=\mathbf{R}$. The uniqueness of $\mathbf{Q}$
and~$\mathbf{R}$ is proven in\citet[Sec. 2.6]{Horn85}.
\end{proof}
We conclude by noting that for full-rank~$\mathbf{A}$, the uniqueness
of $\mathbf{Q}$ and~$\mathbf{R}$ implies that $\mathbf{A}=\mathbf{Q}\mathbf{R}$
can be called \emph{the} QR~decomposition of~$\mathbf{A}$ with
\emph{the} QR~factors $\mathbf{Q}$ and~$\mathbf{R}$.

\subsection{Laurent Polynomials and Interpolation}

\label{sub:lpandinterp}

In the remainder of the paper, the term \emph{interpolation} indicates
LP~interpolation, as presented in this section. Interpolation is
a central component of the algorithms for efficient QR~decomposition
of polynomial matrices presented in Sections \ref{sec:algorithms}
and~\ref{sec:mmsecase}. In the following, we review basic results
on interpolation and establish the corresponding notation. In Section~\ref{sec:efficientinterpolation},
we will present various strategies for computationally efficient interpolation
tailored to the problem at hand.
\begin{defn}
\label{def:laurentpolynomial}Given a matrix-valued function $\mathbf{A}:\;\setunit\rightarrow\mathbb{C}^{P\times M}$
and integers $\dg_{1},\dg_{2}\geq0$, the notation $\pol{\mathbf{A}(s)}{\dg_{1}}{\dg_{2}}$
indicates that there exist coefficient matrices $\mathbf{A}_{v}\in\mathbb{C}^{P\times M},v=-\dg_{1},-\dg_{1}+1,\ldots,\dg_{2}$,
such that\begin{equation}
\mathbf{A}(s)=\sum_{v=-\dg_{1}}^{\dg_{2}}\mathbf{A}_{v}s^{-v},\qquad s\in\setunit.\label{eq:AsLP}\end{equation}
If $\pol{\mathbf{A}(s)}{\dg_{1}}{\dg_{2}}$, then $\mathbf{A}(s)$
is a \emph{Laurent polynomial}~(LP) matrix with \emph{maximum degree}
$\dg_{1}+\dg_{2}$.
\end{defn}
Before discussing interpolation, we briefly list the following statements
which follow directly from Definition~\ref{def:laurentpolynomial}.
First, $\pol{\mathbf{A}(s)}{\dg_{1}}{\dg_{2}}$ implies $\pol{\mathbf{A}(s)}{\dg_{1}'}{\dg_{2}'}$
for any $\dg_{1}'\geq\dg_{1},\dg_{2}'\ge\dg_{2}$. Moreover, since
for $s\in\setunit$ we have $s^{*}=s^{-1}\!$, $\pol{\mathbf{A}(s)}{\dg_{1}}{\dg_{2}}$
implies $\pol{\mathbf{A}^{H}(s)}{\dg_{2}}{\dg_{1}}$. Finally, given
LP matrices $\pol{\mathbf{A}_{1}(s)}{\dg_{11}}{\dg_{12}}$ and $\pol{\mathbf{A}_{2}(s)}{\dg_{21}}{\dg_{22}}$,
if $\mathbf{A}_{1}(s)$ and~$\mathbf{A}_{2}(s)$ have the same dimensions,
then $\pol{(\mathbf{A}_{1}(s)+\mathbf{A}_{2}(s))}{\max\bigpar{\dg_{11},\dg_{21}}}{\max\bigpar{\dg_{12},\dg_{22}}}$,
whereas if the dimensions of $\mathbf{A}_{1}(s)$ and~$\mathbf{A}_{2}(s)$
are such that the matrix product $\mathbf{A}_{1}(s)\mathbf{A}_{2}(s)$
is defined, then $\pol{\mathbf{A}_{1}(s)\mathbf{A}_{2}(s)}{\dg_{11}+\dg_{21}}{\dg_{12}+\dg_{22}}$.

In the remainder of this section, we review basic results on interpolation
by considering the LP $\pol{a(s)}{\dg_{1}}{\dg_{2}}$ with maximum
degree $\dg\triangleq\dg_{1}+\dg_{2}$. The following results can
be directly extended to the interpolation of LP matrices through entrywise
application. Borrowing terminology from signal analysis, we call the
value of~$a(s)$ at a given point $s_{0}\in\setunit$ the\emph{ sample}
$a(s_{0})$.
\begin{defn}
\emph{Interpolation} of the LP~$\pol{a(s)}{\dg_{1}}{\dg_{2}}$ from
the set $\setbase=\bigcrl{b_{0},b_{1},\ldots,b_{\nbase-1}}\subset\setunit$,
containing $\nbase$ distinct \emph{base points}, to the set $\settgt=\bigcrl{t_{0},t_{1},\ldots,t_{\ntgt-1}}\subset\setunit$,
containing $\ntgt$ distinct \emph{target points,} is the process
of obtaining the samples $a(t_{0}),a(t_{1}),\ldots,a(t_{\ntgt-1})$
from the samples $a(b_{0}),a(b_{1}),\ldots,a(b_{\nbase-1})$, with
knowledge of $\dg_{1}$ and~$\dg_{2}$, but without explicit knowledge
of the coefficients $a_{-\dg_{1}},a_{-\dg_{1}+1},\ldots,a_{\dg_{2}}$
that determine~$a(s)$ according to~(\ref{eq:AsLP}).
\end{defn}
In the following, we assume that $\nbase\geq\dg+1$. By defining the
vectors $\veccoeff\triangleq\hvecd{a_{-\dg_{1}}}{a_{-\dg_{1}+1}}{a_{\dg_{2}}}^{T}\!$,
$\mathbf{a}_{\setbase}\triangleq\hvecd{a\bigpar{b_{0}}}{a\bigpar{b_{1}}}{a\bigpar{b_{\nbase-1}}}^{T}\!$,
and $\mathbf{a}_{\settgt}\triangleq\hvecd{a\bigpar{t_{0}}}{a\bigpar{t_{1}}}{a\bigpar{t_{\ntgt-1}}}^{T}\!$,
we note that $\mathbf{a}_{\setbase}=\matbase\veccoeff$, with the
$\nbase\times(\dg+1)$ \emph{base point matrix} \begin{align}
\matbase & \triangleq\left[\begin{array}{cccc}
b_{0}^{\dg_{1}} & b_{0}^{\dg_{1}-1} & \cdots & b_{0}^{-\dg_{2}}\\
b_{1}^{\dg_{1}} & b_{1}^{\dg_{1}-1} & \cdots & b_{1}^{-\dg_{2}}\\
\vdots & \vdots & \ddots & \vdots\\
b_{\nbase-1}^{\dg_{1}} & b_{\nbase-1}^{\dg_{1}-1} & \cdots & b_{\nbase-1}^{-\dg_{2}}\end{array}\right]\label{eq:matbase}\end{align}
 and $\mathbf{a}_{\settgt}=\mattgt\veccoeff$, with the $\ntgt\times(\dg+1)$
\emph{target point matrix} \begin{align}
\mattgt & \triangleq\left[\begin{array}{cccc}
t_{0}^{\dg_{1}} & t_{0}^{\dg_{1}-1} & \cdots & t_{0}^{-\dg_{2}}\\
t_{1}^{\dg_{1}} & t_{1}^{\dg_{1}-1} & \cdots & t_{1}^{-\dg_{2}}\\
\vdots & \vdots & \ddots & \vdots\\
t_{\ntgt-1}^{\dg_{1}} & t_{\ntgt-1}^{\dg_{1}-1} & \cdots & t_{\ntgt-1}^{-\dg_{2}}\end{array}\right]\!.\label{eq:mattgt}\end{align}
Now, $\matbase$ can be written as $\matbase=\mathbf{D}_{\setbase}\mathbf{V}_{\setbase}$,
where $\mathbf{D}_{\setbase}\triangleq\diag(b_{0}^{\dg_{1}},b_{1}^{\dg_{1}},\ldots,b_{\nbase-1}^{\dg_{1}})$
and $\mathbf{V}_{\setbase}$ is the $\nbase\times(V+1)$ Vandermonde
matrix\[
\mathbf{V}_{\setbase}\triangleq\left[\begin{array}{cccc}
1 & b_{0}^{-1} & \cdots & b_{0}^{-(\dg_{1}+\dg_{2})}\\
1 & b_{1}^{-1} & \cdots & b_{1}^{-(\dg_{1}+\dg_{2})}\\
\vdots & \vdots & \ddots & \vdots\\
1 & b_{\nbase-1}^{-1} & \cdots & b_{\nbase-1}^{-(\dg_{1}+\dg_{2})}\end{array}\right]\!.\]
Since the base points $b_{0},b_{1},\ldots,b_{\nbase-1}$ are distinct,
$\mathbf{V}_{\setbase}$ has full rank\citet{Horn85}. Hence, $\rk(\mathbf{V}_{\setbase})=\dg+1$,
which, together with the fact that~$\mathbf{D}_{\setbase}$ is nonsingular,
implies that $\rk(\matbase)=\dg+1$. Therefore, the coefficient vector~$\veccoeff$
is uniquely determined by the~$\nbase$ samples of~$a(s)$ at the
base points $b_{0},b_{1},\ldots,b_{\nbase-1}$ according to $\veccoeff=\matbase^{\dagger}\mathbf{a}_{\setbase}$,
and interpolation of~$a(s)$ from $\setbase$ to~$\settgt$ can
be performed by computing\begin{equation}
\mathbf{a}_{\settgt}=\mattgt\matbase^{\dagger}\mathbf{a}_{\setbase}.\label{eq:lpinterp}\end{equation}
In the remainder of the paper, we call the $\ntgt\times\nbase$ matrix
$\mattgt\matbase^{\dagger}$ the \emph{interpolation matrix}.

We conclude this section by noting that in the special case $\dg_{1}=\dg_{2}$,
we have $\matbase=\matbase^{*}\mathbf{E}$ and $\mattgt=\mattgt^{*}\mathbf{E}$,
where~the $(\dg+1)\times(\dg+1)$ matrix~$\mathbf{E}$ is obtained
by flipping~$\mathbf{I}_{\dg+1}$ upside down. Since the operation
of taking the pseudoinverse commutes with entrywise conjugation, it
follows that $\matbase^{\dagger}=\mathbf{E}\bigpar{\matbase^{\dagger}}^{*}$
and, as a consequence of $\mathbf{E}^{2}=\mathbf{I}_{\dg+1}$, we
obtain $\mattgt\matbase^{\dagger}=\bigpar{\mattgt\matbase^{\dagger}}^{*}\!$,
i.e., the interpolation matrix is real-valued.

\section{Problem Statement}

\label{sec:mimoofdm}

\subsection{MIMO-OFDM System Model}

\label{sub:systemmodel}

We consider a MIMO system\citet{paulraj03} with~$\ntx$ transmit
and~$\nrx$ receive antennas. Throughout the paper, we focus on the
case $\nrx\geq\ntx$. The matrix-valued impulse response of the frequency-selective
MIMO channel is given by the taps $\mattap_{l}\in\mathbb{C}^{\nrx\times\ntx}$
($l=0,1,\ldots,L$) with the corresponding matrix-valued transfer
function\[
\mattf\bigpar{e^{j2\pi\theta}}=\sum_{l=0}^{L}\mattap_{l}e^{-j2\pi l\theta},\qquad0\leq\theta<1\]
which satisfies $\pol{\mattf(s)}0L$. In a MIMO-OFDM system with~$\ncar$
OFDM~tones and a cyclic prefix of length $\lcp\geq L$ samples, the
equivalent input-output relation for the~$n$th tone is given by\[
\vecrx_{n}=\mattf\ofsn\vectx_{n}+\vecnoise_{n},\qquad n=0,1,\ldots,\ncar-1\]
with the transmit signal vector $\vectx_{n}\!\triangleq\!\hvecd{\symtx_{n,1}}{\symtx_{n,2}}{\symtx_{n,\ntx}}^{T}\!$,
the receive signal vector $\vecrx_{n}\!\triangleq\!\hvecd{\symrx_{n,1}}{\symrx_{n,2}}{\symrx_{n,\nrx}}^{T}\!\!$,
the additive noise vector~$\vecnoise_{n}$, and $s_{n}\triangleq e^{j2\pi n/\ncar}\!$.
Here, $\symtx_{n,m}$ stands for the complex-valued data symbol, taken
from a finite constellation~$\setconst$, transmitted by the~$m$th
antenna on the~$n$th tone and~$\symrx_{n,m}$ is the signal observed
at the~$m$th receive antenna on the~$n$th tone. For $\allzero n{\ncar-1}$,
we assume that~$\vectx_{n}$ contains statistically independent entries
and satisfies $\expval\bigbrk{\vectx_{n}}=\mathbf{0}$ and $\expval\bigbrk{\vectx_{n}^{H}\vectx_{n}}=1$.
Again for $\allzero n{\ncar-1}$, we assume that~$\vecnoise_{n}$
is statistically independent of~$\vectx_{n}$ and contains entries
that are independent and identically distributed (i.i.d.) as $\cgauss 0{\varnoise}$,
where~$\varnoise$ denotes the noise variance and is assumed to be
known at the receiver.

In practice, $\ncar$ is typically chosen to be a power of two in
order to allow for efficient OFDM processing based on the Fast Fourier
Transform (FFT). Moreover, a small subset of the~$\ncar$ tones is
typically set aside for pilot symbols and virtual tones at the frequency
band edges, which help to reduce out-of-band interference and relax
the pulse-shaping filter requirements. We collect the indices corresponding
to the~$\ndat$ tones carrying payload data into the set $\setdat\subseteq\{0,1,\ldots,\ncar-1\}$.
Typical OFDM systems have $\ndat\geq3\lcp$.

\subsection{QR~Decomposition in MIMO-OFDM Detectors}

\label{sub:receivers}

Widely used algorithms for coherent detection in MIMO-OFDM systems
include successive cancelation~(SC) detectors\citet{paulraj03},
both zero-forcing (ZF) and MMSE\citet{wolniansky98,hassibi00-06},
and sphere decoders, both in the original formulation\citet{fincke85,viterbo93}
requiring ZF-based preprocessing, as well as in the MMSE-based form
proposed in\citet{studer08-02}. These detection algorithms require
QR~decomposition in the preprocessing step, or, more specifically,
computation of matrices $\mathbf{Q}\ofsn$ and~$\mathbf{R}\ofsn$,
for all $n\in\setdat$, defined as follows. In the ZF case, $\mathbf{Q}\ofsn$
and~$\mathbf{R}\ofsn$ are QR~factors of~$\mattf\ofsn$, whereas
in the MMSE case, $\mathbf{Q}\ofsn$ and~$\mathbf{R}\ofsn$ are obtained
as follows: $\stkmatq\ofsn\mathbf{R}\ofsn$ is the unique QR~decomposition
of the full-rank, $\bigpar{\nrx+\ntx}\times\ntx$ \emph{MMSE-augmented
channel matrix}\begin{equation}
\stkmattf\ofsn\triangleq\left[\begin{array}{c}
\mattf\ofsn\\
\sqrt{\ntx}\stddevnoise\mathbf{I}_{\ntx}\end{array}\right]\label{eq:Hmmseaugmented}\end{equation}
and~$\mathbf{Q}\ofsn$ is given by~$\subrow{\stkmatq}1{\nrx}\ofsn$.
Taking the first~$\nrx$ rows on both sides of the equation $\stkmattf\ofsn=\stkmatq\ofsn\mathbf{R}\ofsn$
yields the factorization $\mattf\ofsn=\mathbf{Q}\ofsn\mathbf{R}\ofsn$,
which is unique because of the uniqueness of $\stkmatq\ofsn$ and~$\mathbf{R}\ofsn$,
and which we call the \emph{MMSE-QR decomposition} of~$\mattf\ofsn$
with the \emph{MMSE-QR~factors} $\mathbf{Q}\ofsn$ and~$\mathbf{R}\ofsn$.

In the following, we briefly describe how $\mathbf{Q}\ofsn$ and~$\mathbf{R}\ofsn$,
either derived as QR~decomposition or as MMSE-QR~decomposition of~$\mattf\ofsn$,
are used in the detection algorithms listed above. SC~detectors essentially
solve the linear system of equations $\mathbf{Q}^{H}\ofsn\vecrx_{n}=\mathbf{R}\ofsn\hat{\vectx}_{n}$
by back-substitution (with rounding of the intermediate results to
elements of~$\setconst$\citet{paulraj03}) to obtain $\hat{\vectx}_{n}\in\setconst^{\ntx}\!$.
Sphere decoders exploit the upper triangularity of~$\mathbf{R}\ofsn$
to find the symbol vector $\hat{\vectx}_{n}\in\setconst^{\ntx}$ that
minimizes $\Vert\mathbf{Q}^{H}\ofsn\vecrx_{n}-\mathbf{R}\ofsn\hat{\vectx}_{n}\Vert^{2}$
through an efficient tree search\citet{viterbo93}.

\subsection{Problem Statement}

\label{sub:problemstatement}We assume that the MIMO-OFDM receiver
has perfect knowledge of the samples~$\mattf\ofsn$ for $n\in\setpil\subseteq\{0,1,\ldots,\ncar-1\}$,
with $\left|\setpil\right|\geq L+1$, from which~$\mattf\ofsn$ can
be obtained at any data-carrying tone $n\in\setdat$ through interpolation
of $\pol{\mattf(s)}0L$. We note that interpolation of~$\mattf(s)$
is not necessary if $\setdat\subseteq\setpil$. We next formulate
the problem statement by focusing on ZF-based detectors, which require
QR~decomposition of the MIMO-OFDM channel matrices $\mattf\ofsn$.
The problem statement for the MMSE case is analogous with QR~decomposition
replaced by MMSE-QR decomposition.

The MIMO-OFDM receiver needs to compute QR~factors $\mathbf{Q}\ofsn$
and~$\mathbf{R}\ofsn$ of~$\mattf\ofsn$ for all data-carrying tones
$n\in\setdat$. A straightforward approach to solving this problem
consists of first interpolating~$\mattf(s)$ to obtain~$\mattf\ofsn$
at the tones $n\in\setdat$ and then performing QR~decomposition
on a per-tone basis. This method will henceforth be called \emph{brute-force
per-tone QR~decomposition}. The interpolation-based QR~decomposition
algorithms presented in this paper are motivated by the following
observations. First, performing QR~decomposition on an $M\times M$
matrix requires~$O(M^{3})$ arithmetic operations\citet{golub96},
whereas the number of arithmetic operations involved in computing
one sample of an $M\times M$ LP~matrix by interpolation is proportional
to the number of matrix entries~$M^{2}$, as interpolation of an
LP~matrix is performed entrywise. This comparison suggests that we
may obtain fundamental savings in computational complexity by replacing
QR~decomposition by interpolation. Second, consider a flat-fading
channel, so that $L=0$ and hence $\mattf\ofsn=\mattap_{0}$ for all
$\allzero n{\ncar-1}$. In this case, a single QR~decomposition $\mattap_{0}=\mathbf{QR}$
yields QR~factors of~$\mattf\ofsn$ for all data-carrying tones
$n\in\setdat$. A question that now arises naturally is whether for
$L>0$ QR~factors $\mathbf{Q}\ofsn$ and~$\mathbf{R}\ofsn$, $n\in\setdat$,
can be obtained from a smaller set of QR~factors through interpolation.
We will see that the answer is in the affirmative and will, moreover,
demonstrate that interpolation-based QR~decomposition algorithms
can yield significant computational complexity savings over brute-force
per-tone QR~decomposition for a wide range of values of the parameters
$\ntx$, $\nrx$, $L$, $\ncar$, and~$\ndat$, which will be referred
to as \emph{the system parameters} throughout the paper. The key to
formulating interpolation-based algorithms and realizing these complexity
savings is a result on QR~decomposition of LP~matrices formalized
in Theorem~\ref{thm:qrinterp} in the next section.

\section{QR Decomposition through Interpolation}

\label{sec:interpqr}

\subsection{Additional Properties of QR~Decomposition}

\label{sub:qrfurtherproperties}

We next set the stage for the formulation of our main technical result
by presenting additional properties of QR~decomposition of a matrix
$\mathbf{A}\in\mathbb{C}^{P\times M}\!$, with $P\geq M$, that are
directly implied by Definition~\ref{def:QR}.
\begin{prop}
\label{pro:qrpartial}Let $\mathbf{A}=\mathbf{QR}$ be a QR~decomposition
of~$\mathbf{A}$. Then, for a given $k\in\{1,2,\ldots,M\}$, $\subcol{\mathbf{A}}1k=\subcol{\mathbf{Q}}1k\subcolrow{\mathbf{R}}1k1k$
is a QR~decomposition of~$\subcol{\mathbf{A}}1k$.\end{prop}
\begin{proof}
From $\mathbf{A}=\mathbf{Q}\mathbf{R}$ it follows that $\subcol{\mathbf{A}}1k=\subcol{(\mathbf{Q}\mathbf{R})}1k=\subcol{\mathbf{Q}}1k\subcolrow{\mathbf{R}}1k1k+\subcol{\mathbf{Q}}{k+1}M\subcolrow{\mathbf{R}}1k{k+1}M$,
which simplifies to $\subcol{\mathbf{A}}1k=\subcol{\mathbf{Q}}1k\subcolrow{\mathbf{R}}1k1k$,
since the upper triangularity of~$\mathbf{R}$ implies $\subcolrow{\mathbf{R}}1k{k+1}M=\mathbf{0}$.
$\subcol{\mathbf{Q}}1k$ and~$\subcolrow{\mathbf{R}}1k1k$ satisfy
Conditions \ref{enu:QR_cond_Q} and~\ref{enu:QR_cond_R} of Definition~\ref{def:QR}
since all columns of~$\subcol{\mathbf{Q}}1k$ are also columns of~$\mathbf{Q}$
and since $\subcolrow{\mathbf{R}}1k1k$ is a principal submatrix of~$\mathbf{R}$,
respectively. Finally, $\mathbf{R}=\mathbf{Q}^{H}\mathbf{A}$ implies
$\subcolrow{\mathbf{R}}1k1k=\subcolrow{(\mathbf{Q}^{H}\mathbf{A})}1k1k=\subcol{\mathbf{Q}}1k^{H}\subcol{\mathbf{A}}1k$
and hence Condition~\ref{enu:QR_cond_R_QH} of Definition~\ref{def:QR}
is satisfied.\end{proof}
\begin{prop}
\label{pro:qrreduction}Let $\mathbf{A}=\mathbf{QR}$ be a QR~decomposition
of~$\mathbf{A}$. Then, for $M>1$ and for a given $k\in\{2,3,\ldots,M\}$,
$\subcol{\mathbf{A}}kM-\subcol{\mathbf{Q}}1{k-1}\subcolrow{\mathbf{R}}kM1{k-1}=\subcol{\mathbf{Q}}kM\subcolrow{\mathbf{R}}kMkM$
is a QR~decomposition of $\subcol{\mathbf{A}}kM-\subcol{\mathbf{Q}}1{k-1}\subcolrow{\mathbf{R}}kM1{k-1}$.\end{prop}
\begin{proof}
$\mathbf{A}=\subcol{\mathbf{Q}}1{k-1}\subrow{\mathbf{R}}1{k-1}+\subcol{\mathbf{Q}}kM\subrow{\mathbf{R}}kM$
implies $\subcol{\mathbf{A}}kM=\subcol{\mathbf{Q}}1{k-1}\subcolrow{\mathbf{R}}kM1{k-1}+\subcol{\mathbf{Q}}kM\subcolrow{\mathbf{R}}kMkM$
and hence $\subcol{\mathbf{A}}kM-\subcol{\mathbf{Q}}1{k-1}\subcolrow{\mathbf{R}}kM1{k-1}=\subcol{\mathbf{Q}}kM\subcolrow{\mathbf{R}}kMkM$.
$\subcol{\mathbf{Q}}kM$ and~$\subcolrow{\mathbf{R}}kMkM$ satisfy
Conditions \ref{enu:QR_cond_Q} and~\ref{enu:QR_cond_R} of Definition~\ref{def:QR}
since all columns of~$\subcol{\mathbf{Q}}kM$ are also columns of~$\mathbf{Q}$
and since $\subcolrow{\mathbf{R}}kMkM$ is a principal submatrix of~$\mathbf{R}$,
respectively. Moreover, $\mathbf{R}=\mathbf{Q}^{H}\mathbf{A}$ implies
$\subcolrow{\mathbf{R}}kMkM=(\mathbf{Q}^{H}\mathbf{A}\subcolrow )kMkM=\subcol{\mathbf{Q}}kM^{H}\subcol{\mathbf{A}}kM$.
Using $\subcol{\mathbf{Q}}kM^{H}\subcol{\mathbf{Q}}1{k-1}=\mathbf{0}$,
which follows from the fact that the nonzero columns of~$\mathbf{Q}$
are orthonormal, we can write $\subcolrow{\mathbf{R}}kMkM=\subcol{\mathbf{Q}}kM^{H}\subcol{\mathbf{A}}kM-\subcol{\mathbf{Q}}kM^{H}\subcol{\mathbf{Q}}1{k-1}\subcolrow{\mathbf{R}}kM1{k-1}=\subcol{\mathbf{Q}}kM^{H}(\subcol{\mathbf{A}}kM-\subcol{\mathbf{Q}}1{k-1}\subcolrow{\mathbf{R}}kM1{k-1})$.
Hence, Condition~\ref{enu:QR_cond_R_QH} of Definition~\ref{def:QR}
is satisfied.
\end{proof}
In order to characterize QR~decomposition of~$\mathbf{A}$ in the
general case $\rk(\mathbf{A})\leq M$, we introduce the following
concept.
\begin{defn}
\label{def:orderedrank}The \emph{ordered column rank} of~$\mathbf{A}$
is the number \[
K\triangleq\begin{cases}
0, & \quad\rk(\subcol{\mathbf{A}}11)=0\\
\max\bigcrl{k\in\{1,2,\ldots,M\}:\rk(\subcol{\mathbf{A}}1k)=k}, & \quad\mbox{else.}\end{cases}\]

\end{defn}
For later use, we note that $K=0$ is equivalent to $\mathbf{a}_{1}=\mathbf{0}$,
and that $K<M$ is equivalent to~$\mathbf{A}$ being rank-deficient.
\begin{prop}
\label{pro:qrproperties}QR~factors $\mathbf{Q}$ and~$\mathbf{R}$
of a matrix~$\mathbf{A}$ of ordered column rank~$K>0$ satisfy
the following properties:
\begin{enumerate}
\item $\subcol{\mathbf{Q}}1K^{H}\subcol{\mathbf{Q}}1K=\mathbf{I}_{K}$\label{enu:Q_Kunitary}
\item $[\mathbf{R}]_{k,k}>0$ for $\allone kK$\label{enu:R_kk_positive}
\item $\subcol{\mathbf{Q}}1K$ and~$\subrow{\mathbf{R}}1K$ are unique\label{enu:QR_K_unique}
\item $\ran(\subcol{\mathbf{Q}}1k)=\ran(\subcol{\mathbf{A}}1k)$ for $\allone kK$\label{enu:cspan_q_K_cspan_a_K}
\item if $K<M$, $[\mathbf{R}]_{K+1,K+1}=0$\label{enu:R_rankdef_zero}
\end{enumerate}
\end{prop}
\begin{proof}
Since $\subcol{\mathbf{Q}}1K$ and~$\subcolrow{\mathbf{R}}1K1K$
are QR~factors of~$\subcol{\mathbf{A}}1K$, as stated in Proposition~\ref{pro:qrpartial},
and since $\rk(\subcol{\mathbf{A}}1K)=K$, Properties~\ref{enu:Q_Kunitary}
and~\ref{enu:R_kk_positive}, as well as the uniqueness of~$\subcol{\mathbf{Q}}1K$
stated in Property~\ref{enu:QR_K_unique}, are obtained directly
by applying Proposition~\ref{pro:qrfullrank} to the full-rank matrix~$\subcol{\mathbf{A}}1K$.
The uniqueness of~$\subrow{\mathbf{R}}1K$ stated in Property~\ref{enu:QR_K_unique}
is implied by the uniqueness of~$\subcol{\mathbf{Q}}1K$ and by $\subrow{\mathbf{R}}1K=\subcol{\mathbf{Q}}1K^{H}\mathbf{A}$,
which follows from Condition~\ref{enu:QR_cond_R_QH} of Definition~\ref{def:QR}.
For $\allone kK$, $\ran(\subcol{\mathbf{Q}}1k)=\ran(\subcol{\mathbf{A}}1k)$
is a trivial consequence of $\subcol{\mathbf{A}}1k=\subcol{\mathbf{Q}}1k\subcolrow{\mathbf{R}}1k1k$
and of $\rk(\subcolrow{\mathbf{R}}1k1k)=k$, which follows from the
fact that~$\subcolrow{\mathbf{R}}1k1k$ is upper triangular with
nonzero entries on its main diagonal. This proves Property~\ref{enu:cspan_q_K_cspan_a_K}.
If $K<M$,  Condition~\ref{enu:QR_cond_R_QH} of Definition~\ref{def:QR}
implies $[\mathbf{R}]_{K+1,K+1}=\mathbf{q}_{K+1}^{H}\mathbf{a}_{K+1}$.
If $\mathbf{q}_{K+1}=\mathbf{0}$, $[\mathbf{R}]_{K+1,K+1}=0$ follows
trivially. If $\mathbf{q}_{K+1}\neq\mathbf{0}$, Condition~\ref{enu:QR_cond_Q}
of Definition~\ref{def:QR} implies that $\mathbf{q}_{K+1}$ is orthogonal
to $\ran(\subcol{\mathbf{Q}}1K)$, whereas the definition of~$K$
implies that $\mathbf{a}_{K+1}\in\ran(\subcol{\mathbf{A}}1K)$. Since
$\ran(\subcol{\mathbf{Q}}1K)=\ran(\subcol{\mathbf{A}}1K)$, we obtain
$\mathbf{q}_{K+1}^{H}\mathbf{a}_{K+1}=[\mathbf{R}]_{K+1,K+1}=0$,
which proves Property~\ref{enu:R_rankdef_zero}.
\end{proof}
We emphasize that for $K>0$, the uniqueness of $\subcol{\mathbf{Q}}1K$
and~$\subrow{\mathbf{R}}1K$ has two significant consequences. First,
the GS~orthonormalization procedure (\ref{eq:y_k})--(\ref{eq:r_k}),
evaluated for $\allone kK$, determines the submatrices $\subcol{\mathbf{Q}}1K$
and~$\subrow{\mathbf{R}}1K$ of the matrices $\mathbf{Q}$ and~$\mathbf{R}$
produced by \emph{any} QR~decomposition algorithm. Second, the nonuniqueness
of~$\mathbf{Q}$ and~$\mathbf{R}$ in the case of rank-deficient~$\mathbf{A}$,
demonstrated in Section~\ref{sub:QRDecomposition}, is restricted
to the submatrices $\subcol{\mathbf{Q}}{K+1}M$ and~$\subrow{\mathbf{R}}{K+1}M\!$.

Finally, we note that Property~\ref{enu:R_rankdef_zero} of Proposition~\ref{pro:qrproperties}
is valid for the case $K=0$ as well. In fact, Condition~\ref{enu:QR_cond_R_QH}
of Definition~\ref{def:QR} implies $[\mathbf{R}]_{1,1}=\mathbf{q}_{1}^{H}\mathbf{a}_{1}$.
Since $K=0$ implies $\mathbf{a}_{1}=\mathbf{0}$, we immediately
obtain $[\mathbf{R}]_{1,1}=0$.

\subsection{QR~Decomposition of an LP Matrix}

\label{sub:qrdecpolynomialmatrix}

In the remainder of Section~\ref{sec:interpqr}, we consider a $P\times M$
LP matrix $\pol{\mathbf{A}(s)}{\dg_{1}}{\dg_{2}}$, $s\in\setunit$,
with $P\geq M$, and QR~factors $\mathbf{Q}(s)$ and~$\mathbf{R}(s)$
of~$\mathbf{A}(s)$. Despite~$\mathbf{A}(s)$ being an LP~matrix,
$\mathbf{Q}(s)$ and $\mathbf{R}(s)$ will, in general, not be LP~matrices.
To see this, consider the case where $\rk(\mathbf{A}(s))=M$ for all
$s\in\setunit$. It follows from the results in Sections~\ref{sub:QRDecomposition}
and~\ref{sub:qrfurtherproperties} that, in this case, $\mathbf{Q}(s)$
and~$\mathbf{R}(s)$ are unique and determined through (\ref{eq:y_k})--(\ref{eq:r_k}).
The division and the square root operation in~(\ref{eq:q_k}), in
general, prevent~$\mathbf{Q}(s)$, and hence also $\mathbf{R}(s)=\mathbf{Q}^{H}(s)\mathbf{A}(s)$,
from being~LP matrices. Nevertheless, in this section we will show
that there exists a mapping~$\map$ that transforms $\mathbf{Q}(s)$
and~$\mathbf{R}(s)$ into corresponding LP matrices $\tilde{\mathbf{Q}}(s)$
and~$\tilde{\mathbf{R}}(s)$. The mapping~$\map$ constitutes the
basis for the formulation of interpolation-based QR~decomposition
algorithms for MIMO-OFDM systems.

In the following, we consider QR~factors of~$\mathbf{A}(s_{0})$
for a given $s_{0}\in\setunit$. In order to keep the notation compact,
we omit the dependence of all involved quantities on~$s_{0}$. We
start by defining the auxiliary variables~$\Delta_{k}$ as \begin{equation}
\Delta_{k}\triangleq\Delta_{k-1}[\mathbf{R}]_{k,k}^{2},\qquad\allone kM\label{eq:Delta_k_map}\end{equation}
 with $\Delta_{0}\triangleq1$. Next, we introduce the vectors\begin{align}
\tilde{\mathbf{q}}_{k} & \triangleq\Delta_{k-1}\left[\mathbf{R}\right]_{k,k}\mathbf{q}_{k},\qquad\allone kM\label{eq:q_k_map}\\
\tilde{\mathbf{r}}_{k}^{T} & \triangleq\Delta_{k-1}\left[\mathbf{R}\right]_{k,k}\mathbf{r}_{k}^{T},\qquad\allone kM\label{eq:r_k_map}\end{align}
and define the mapping $\map:\qrnotilde{}\mapsto\qrtilde{}$ by $\tilde{\mathbf{Q}}\triangleq\hvecd{\tilde{\mathbf{q}}_{1}}{\tilde{\mathbf{q}}_{2}}{\tilde{\mathbf{q}}_{M}}$
and $\tilde{\mathbf{R}}\triangleq\hvecd{\tilde{\mathbf{r}}_{1}}{\tilde{\mathbf{r}}_{2}}{\tilde{\mathbf{r}}_{M}}^{T}\!$.

Now, we consider the ordered column rank~$K$ of~$\mathbf{A}$,
and note that Property~\ref{enu:R_kk_positive} in Proposition~\ref{pro:qrproperties}
implies that, if $K>0$, $\Delta_{k-1}\left[\mathbf{R}\right]_{k,k}>0$
for $\allone kK$, as seen by unfolding the recursion in~(\ref{eq:Delta_k_map}).
Hence, for $K>0$ and $\allone kK$, we can compute $\mathbf{q}_{k}$
and~$\mathbf{r}_{k}^{T}$ from $\tilde{\mathbf{q}}_{k}$ and~$\tilde{\mathbf{r}}_{k}^{T}$,
respectively, according to\begin{align}
\mathbf{q}_{k} & =(\Delta_{k-1}\left[\mathbf{R}\right]_{k,k})^{-1}\,\tilde{\mathbf{q}}_{k}\label{eq:q_kdemap}\\
\mathbf{r}_{k}^{T} & =(\Delta_{k-1}\left[\mathbf{R}\right]_{k,k})^{-1}\,\tilde{\mathbf{r}}_{k}^{T}\label{eq:r_kdemap}\end{align}
where $\Delta_{k-1}\left[\mathbf{R}\right]_{k,k}$ is obtained from
the entries on the main diagonal of~$\tilde{\mathbf{R}}$ as\begin{equation}
\Delta_{k-1}\left[\mathbf{R}\right]_{k,k}=\begin{cases}
\sqrt{[\tilde{\mathbf{R}}]_{k,k}}, & \quad k=1\\
\sqrt{[\tilde{\mathbf{R}}]_{k-1,k-1}[\tilde{\mathbf{R}}]_{k,k}}, & \quad\alltwo k{K.}\end{cases}\label{eq:Delta_scalingfactorforinvmap}\end{equation}
If $K=M$, i.e., for full-rank~$\mathbf{A}$, we have $\Delta_{k-1}\left[\mathbf{R}\right]_{k,k}\neq0$
for all $\allone kM$, and the mapping~$\map$ is invertible. In
the case $K<M$, Property~\ref{enu:R_rankdef_zero} in Proposition~\ref{pro:qrproperties}
states that $\left[\mathbf{R}\right]_{K+1,K+1}=0$, which combined
with (\ref{eq:Delta_k_map})--(\ref{eq:r_k_map}) implies that $\Delta_{k}=0$,
$\tilde{\mathbf{q}}_{k}=\mathbf{0}$, and $\tilde{\mathbf{r}}_{k}^{T}=\mathbf{0}$
for $k=K+1,K+2,\ldots,M$. Hence, the mapping~$\map$ is not invertible
for $K<M$, since the information contained in $\subcol{\mathbf{Q}}{K+1}M$
and~$\subrow{\mathbf{R}}{K+1}M$ can not be extracted from $\subcol{\tilde{\mathbf{Q}}}{K+1}M=\mathbf{0}$
and~$\subrow{\tilde{\mathbf{R}}}{K+1}M=\mathbf{0}$. Nevertheless,
we can recover $\subcol{\mathbf{Q}}{K+1}M$ and~$\subrow{\mathbf{R}}{K+1}M$
as follows. For $0<K<M$, setting $k=K+1$ in Proposition~\ref{pro:qrreduction}
shows that $\subcol{\mathbf{Q}}{K+1}M$ and~$\subcolrow{\mathbf{R}}{K+1}M{K+1}M$
can be obtained by QR~decomposition of $\subcol{\mathbf{A}}{K+1}M-\subcol{\mathbf{Q}}1K\subcolrow{\mathbf{R}}{K+1}M1K$.
Then, $\subrow{\mathbf{R}}{K+1}M$ is obtained as $\subrow{\mathbf{R}}{K+1}M=\matcouple{\subcolrow{\mathbf{R}}1K{K+1}M}{\subcolrow{\mathbf{R}}{K+1}M{K+1}M}$
with $\subcolrow{\mathbf{R}}1K{K+1}M=\mathbf{0}$ because of the upper
triangularity of~$\mathbf{R}$. For $K=0$, since $\tilde{\mathbf{Q}}$
and~$\tilde{\mathbf{R}}$ are all-zero matrices, $\subcol{\mathbf{Q}}{K+1}M=\mathbf{Q}$
and~$\subcolrow{\mathbf{R}}{K+1}M{K+1}M=\mathbf{R}$ must be obtained
by performing QR~decomposition on~$\mathbf{A}$. In the remainder
of the paper, we denote by \emph{inverse mapping~}$\map^{-1}:\qrtilde{}\mapsto\qrnotilde{}$
the procedure%
\footnote{Note that for $K<M$, the inverse mapping~$\map^{-1}$ requires explicit
knowledge of~$\subcol{\mathbf{A}}{K+1}M$.%
} formulated in the following steps:
\begin{enumerate}
\item If $K>0$, for $\allone kK$, compute the scaling factor $(\Delta_{k-1}\left[\mathbf{R}\right]_{k,k})^{-1}$
using~(\ref{eq:Delta_scalingfactorforinvmap}) and scale $\tilde{\mathbf{q}}_{k}$
and~$\tilde{\mathbf{r}}_{k}^{T}$ according to (\ref{eq:q_kdemap})
and~(\ref{eq:r_kdemap}), respectively.
\item If $0<K<M$, compute $\subcol{\mathbf{Q}}{K+1}M$ and~$\subcolrow{\mathbf{R}}{K+1}M{K+1}M$
by performing QR~decomposition on $\subcol{\mathbf{A}}{K+1}M-\subcol{\mathbf{Q}}1K\subcolrow{\mathbf{R}}{K+1}M1K$,
and construct $\subrow{\mathbf{R}}{K+1}M=\matcouple{\mathbf{0}}{\subcolrow{\mathbf{R}}{K+1}M{K+1}M}$.
\item If $K=0$, compute $\mathbf{Q}$ and~$\mathbf{R}$ by performing
QR~decomposition on~$\mathbf{A}$.
\end{enumerate}
We note that the nonuniqueness of QR~decomposition in the case $K<M$
has the following consequence. Given QR~factors $\mathbf{Q}_{1}$
and~$\mathbf{R}_{1}$ of~$\mathbf{A}$, the application of the mapping~$\map$
to~$(\mathbf{Q}_{1},\mathbf{R}_{1})$ followed by application of
the inverse mapping~$\map^{-1}$ yields matrices $\mathbf{Q}_{2}$
and~$\mathbf{R}_{2}$ that may not be equal to $\mathbf{Q}_{1}$
and~$\mathbf{R}_{1}$, respectively. However, $\mathbf{Q}_{2}$ and~$\mathbf{R}_{2}$
are QR~factors of~$\mathbf{A}$ in the sense of Definition~\ref{def:QR}.

We are now ready to present the main technical result of this paper.
This result paves the way for the formulation of interpolation-based
QR~decomposition algorithms.
\begin{thm}
\label{thm:qrinterp}Given $\mathbf{A}:\;\setunit\rightarrow\mathbb{C}^{P\times M}$
with $P\geq M$, such that $\pol{\mathbf{A}(s)}{\dg_{1}}{\dg_{2}}$
with maximum degree $\dg=\dg_{1}+\dg_{2}$. The functions $\Delta_{k}(s)$,
$\tilde{\mathbf{q}}_{k}(s)$, and~$\tilde{\mathbf{r}}_{k}^{T}(s)$,
obtained by applying the mapping~$\map$ as in (\ref{eq:Delta_k_map})--(\ref{eq:r_k_map})
to QR~factors $\mathbf{Q}(s)$ and~$\mathbf{R}(s)$ of~$\mathbf{A}(s)$
for all $s\in\setunit$, satisfy the following properties:
\begin{enumerate}
\item $\pol{\Delta_{k}(s)}{k\dg}{k\dg}$\label{enu:Delta_k_LP}
\item $\pol{\tilde{\mathbf{q}}_{k}(s)}{(k-1)\dg+\dg_{1}}{(k-1)\dg+\dg_{2}}$\label{enu:qtilde_k_LP}
\item $\pol{\tilde{\mathbf{r}}_{k}^{T}(s)}{k\dg}{k\dg}.$\label{enu:rtilde_k_LP}
\end{enumerate}
\end{thm}
We emphasize that Theorem~\ref{thm:qrinterp} applies to any QR~factors
satisfying Definition~\ref{def:QR} and is therefore not affected
by the nonuniqueness of QR~decomposition arising in the rank-deficient
case.

Before proceeding to the proof, we note that Theorem~\ref{thm:qrinterp}
implies that the maximum degrees of the LP~matrices $\tilde{\mathbf{Q}}(s)$
and~$\tilde{\mathbf{R}}(s)$ are $(2M-1)\dg$ and~$2M\dg$, respectively.
We can therefore conclude that $2M\dg+1$ base points are enough for
interpolation of both $\tilde{\mathbf{Q}}(s)$ and~$\tilde{\mathbf{R}}(s)$.
We mention that the results presented in\citet{Davis03}, in the context
of narrowband MIMO systems, involving a QR~decomposition algorithm
that avoids divisions and square root operations, can be applied to
the problem at hand as well. This leads to an alternative mapping
of $\mathbf{Q}(s)$ and~$\mathbf{R}(s)$ to LP~matrices with maximum
degrees significantly higher than~$2M\dg$.

\subsection{Proof of Theorem~\ref{thm:qrinterp}}

\label{sub:prooflpinterp}

The proof consists of three steps, summarized as follows. In Step~1,
we focus on a given $s_{0}\in\setunit$ and aim at writing $\Delta_{k}(s_{0})$,
$\tilde{\mathbf{q}}_{k}(s_{0})$, and~$\tilde{\mathbf{r}}_{k}^{T}(s_{0})$
as functions of~$\mathbf{A}(s_{0})$ for all $(K(s_{0}),k)\in\mathcal{K}\triangleq\{0,1,\ldots,M\}\times\{1,2,\ldots,M\}$,
where~$K(s_{0})$ denotes the ordered column rank of~$\mathbf{A}(s_{0})$.
Step~1 is split into Steps 1a and~1b, in which the two disjoint
subsets $\mathcal{K}_{1}\triangleq\{(K',k')\in\mathcal{K}:0<K'\leq M,1\leq k'\leq K'\}$
and $\mathcal{K}_{2}\triangleq\{(K',k')\in\mathcal{K}:0\leq K'<M,K'+1\leq k'\leq M\}$
(with $\mathcal{K}_{1}\cup\mathcal{K}_{2}=\mathcal{K}$) are considered,
respectively. In Step~1a, we note that for $(K(s_{0}),k)\in\mathcal{K}_{1}$,
$\subcol{\mathbf{Q}}1{K(s_{0})}(s_{0})$ and~$\subrow{\mathbf{R}}1{K(s_{0})}(s_{0})$
are unique and can be obtained by evaluating~(\ref{eq:y_k})--(\ref{eq:r_k})
for $\allone k{K(s_{0})}$. By unfolding the recursions in~(\ref{eq:y_k})--(\ref{eq:r_k})
and in~(\ref{eq:Delta_k_map})--(\ref{eq:r_k_map}), we write $\Delta_{k}(s_{0})$,
$\tilde{\mathbf{q}}_{k}(s_{0})$, and~$\tilde{\mathbf{r}}_{k}^{T}(s_{0})$
as functions of~$\mathbf{A}(s_{0})$ for $(K(s_{0}),k)\in\mathcal{K}_{1}$.
In Step~1b, we show that the expressions for $\Delta_{k}(s_{0})$,
$\tilde{\mathbf{q}}_{k}(s_{0})$, and~$\tilde{\mathbf{r}}_{k}^{T}(s_{0})$,
derived in Step~1a for $(K(s_{0}),k)\in\mathcal{K}_{1}$, are also
valid for $(K(s_{0}),k)\in\mathcal{K}_{2}$ and hence, as a consequence
of $\mathcal{K}_{1}\cup\mathcal{K}_{2}=\mathcal{K}$, for all $(K(s_{0}),k)\in\mathcal{K}$.
In Step~2, we note that the derivations in Step~1 carry over to
all $s_{0}\in\setunit$, and generalize the expressions obtained in
Step~1 to expressions for $\Delta_{k}(s)$, $\tilde{\mathbf{q}}_{k}(s)$,
and~$\tilde{\mathbf{r}}_{k}^{T}(s)$ that hold for $\allone kM$
and for all $s\in\setunit$. Making use of $\pol{\mathbf{A}(s)}{\dg_{1}}{\dg_{2}}$,
in Step~3 it is finally shown that $\Delta_{k}(s)$, $\tilde{\mathbf{q}}_{k}(s)$,
and~$\tilde{\mathbf{r}}_{k}^{T}(s)$ satisfy Properties \ref{enu:Delta_k_LP}--\ref{enu:rtilde_k_LP}
in the statement of Theorem~\ref{thm:qrinterp}.

\paragraph*{Step 1a}

Throughout Steps 1a and~1b, in order to simplify the notation, we
drop the dependence of all quantities on~$s_{0}$. In Step~1a, we
assume that $(K,k)\in\mathcal{K}_{1}$ and, unless stated otherwise,
all equations and statements involving~$k$ are valid for all $\allone kK$.\\
We start by listing preparatory results. We recall from Section~\ref{sub:qrfurtherproperties}
that the submatrices $\subcol{\mathbf{Q}}1K$ and~$\subrow{\mathbf{R}}1K$
are unique and that, consequently, $\mathbf{q}_{k}$ and~$\mathbf{r}_{k}^{T}$
are determined by (\ref{eq:y_k})--(\ref{eq:r_k}). From $\mathbf{q}_{k}\neq\mathbf{0}$,
implied by Property~\ref{enu:Q_Kunitary} in Proposition~\ref{pro:qrproperties},
and from~(\ref{eq:q_k}) we deduce that $\mathbf{y}_{k}\neq\mathbf{0}$.
Then, from~(\ref{eq:y_k}) and~(\ref{eq:q_k}) we obtain \begin{align}
\mathbf{y}_{k}^{H}\mathbf{y}_{k} & =\mathbf{y}_{k}^{H}\mathbf{a}_{k}-\sum_{i=1}^{k-1}\mathbf{q}_{i}^{H}\mathbf{a}_{k}\sqrt{\mathbf{y}_{k}^{H}\mathbf{y}_{k}}\mathbf{q}_{k}^{H}\mathbf{q}_{i}=\mathbf{y}_{k}^{H}\mathbf{a}_{k}\label{eq:y_k_y_k}\end{align}
as $\mathbf{q}_{k}^{H}\mathbf{q}_{i}=0$ for $\allone i{k-1}$. Consequently,
we can write~$[\mathbf{R}]_{k,k}$, using (\ref{eq:q_k}) and (\ref{eq:r_k}),
as\begin{equation}
[\mathbf{R}]_{k,k}=\mathbf{q}_{k}^{H}\mathbf{a}_{k}=\frac{\mathbf{y}_{k}^{H}\mathbf{a}_{k}}{\sqrt{\mathbf{y}_{k}^{H}\mathbf{y}_{k}}}=\sqrt{\mathbf{y}_{k}^{H}\mathbf{y}_{k}}\label{eq:R_kk_y_k}\end{equation}
thus implying $[\mathbf{R}]_{k,k}\mathbf{q}_{k}=\mathbf{y}_{k}$ and
hence, by (\ref{eq:q_k_map}),\begin{equation}
\tilde{\mathbf{q}}_{k}=\Delta_{k-1}\mathbf{y}_{k}.\label{eq:q_tilde_k}\end{equation}
Furthermore, using (\ref{eq:Delta_k_map}) and~(\ref{eq:R_kk_y_k}),
we can write $\Delta_{k}=\Delta_{k-1}\mathbf{y}_{k}^{H}\mathbf{y}_{k}$
or alternatively, in recursion-free form,\begin{align}
\Delta_{k} & =\prod_{i=1}^{k}\mathbf{y}_{i}^{H}\mathbf{y}_{i}.\label{eq:Delta_k_prod}\end{align}
Next, we note that~(\ref{eq:y_k}) implies\begin{equation}
\mathbf{y}_{k}=\mathbf{a}_{k}+\sum_{i=1}^{k-1}\alpha_{i}^{(k)}\mathbf{a}_{i}\label{eq:y_k_a_i}\end{equation}
with unique coefficients $\alpha_{i}^{(k)},\allone i{k-1}$, since
$\mathbf{y}_{1}=\mathbf{a}_{1}$ and since for $k>1$, we have $\rk(\subcol{\mathbf{A}}1{k-1})=k-1$
and, as stated in Property~\ref{enu:cspan_q_K_cspan_a_K} of Proposition~\ref{pro:qrproperties},
$\ran(\subcol{\mathbf{Q}}1{k-1})=\ran(\subcol{\mathbf{A}}1{k-1})$.
Next, we  consider the relation between $\{\mathbf{a}_{1},\mathbf{a}_{2},\ldots,\mathbf{a}_{k}\}$
and $\{\mathbf{y}_{1},\mathbf{y}_{2},\ldots,\mathbf{y}_{k}\}$. Inserting~(\ref{eq:q_k})
into~(\ref{eq:y_k}) yields\[
\mathbf{y}_{k}=\mathbf{a}_{k}-\sum_{i=1}^{k-1}\frac{\mathbf{y}_{i}^{H}\mathbf{a}_{k}}{\mathbf{y}_{i}^{H}\mathbf{y}_{i}}\mathbf{y}_{i}.\]
Hence, using~(\ref{eq:y_k_y_k}), we obtain\begin{align}
\mathbf{a}_{k'} & =\mathbf{y}_{k'}+\sum_{i=1}^{k'-1}\frac{\mathbf{y}_{i}^{H}\mathbf{a}_{k'}}{\mathbf{y}_{i}^{H}\mathbf{y}_{i}}\mathbf{y}_{i}\nonumber \\
 & =\sum_{i=1}^{k'}\frac{\mathbf{y}_{i}^{H}\mathbf{a}_{k'}}{\mathbf{y}_{i}^{H}\mathbf{y}_{i}}\mathbf{y}_{i},\qquad\allone{k'}k.\label{eq:akprimelincomb}\end{align}
We next note that~(\ref{eq:akprimelincomb}) can be rewritten, for
$\allone{k'}k$, in vector-matrix form as\begin{equation}
\hvecd{\mathbf{a}_{1}}{\mathbf{a}_{2}}{\mathbf{a}_{k}}=\hvecd{\mathbf{y}_{1}}{\mathbf{y}_{2}}{\mathbf{y}_{k}}\,\mathbf{V}_{k}\label{eq:a_1k_y_1k}\end{equation}
with the $k\times k$ matrix\[
\mathbf{V}_{k}\triangleq\left[\begin{array}{cccc}
\frac{\mathbf{y}_{1}^{H}\mathbf{a}_{1}}{\mathbf{y}_{1}^{H}\mathbf{y}_{1}} & \frac{\mathbf{y}_{1}^{H}\mathbf{a}_{2}}{\mathbf{y}_{1}^{H}\mathbf{y}_{1}} & \cdots & \frac{\mathbf{y}_{1}^{H}\mathbf{a}_{k}}{\mathbf{y}_{1}^{H}\mathbf{y}_{1}}\\
0 & \frac{\mathbf{y}_{2}^{H}\mathbf{a}_{2}}{\mathbf{y}_{2}^{H}\mathbf{y}_{2}} & \cdots & \frac{\mathbf{y}_{2}^{H}\mathbf{a}_{k}}{\mathbf{y}_{2}^{H}\mathbf{y}_{2}}\\
\vdots & \vdots & \ddots & \vdots\\
0 & 0 & \cdots & \frac{\mathbf{y}_{k}^{H}\mathbf{a}_{k}}{\mathbf{y}_{k}^{H}\mathbf{y}_{k}}\end{array}\right]\]
satisfying $\det(\mathbf{V}_{k})=1$ because of~$\mathbf{y}_{k}\neq\mathbf{0}$
and of~(\ref{eq:y_k_y_k}). Next, we can write~$\mathbf{V}_{k}$
as $\mathbf{V}_{k}=\mathbf{D}_{k}^{-1}\mathbf{U}_{k}$ with the $k\times k$
nonsingular matrices $\mathbf{D}_{k}\triangleq\diag\bigpar{\mathbf{y}_{1}^{H}\mathbf{y}_{1},\mathbf{y}_{2}^{H}\mathbf{y}_{2},\ldots,\mathbf{y}_{k}^{H}\mathbf{y}_{k}}$
and \begin{align}
\mathbf{U}_{k} & \triangleq\left[\begin{array}{cccc}
\mathbf{y}_{1}^{H}\mathbf{a}_{1} & \mathbf{y}_{1}^{H}\mathbf{a}_{2} & \cdots & \mathbf{y}_{1}^{H}\mathbf{a}_{k}\\
0 & \mathbf{y}_{2}^{H}\mathbf{a}_{2} & \cdots & \mathbf{y}_{2}^{H}\mathbf{a}_{k}\\
\vdots & \vdots & \ddots & \vdots\\
0 & 0 & \cdots & \mathbf{y}_{k}^{H}\mathbf{a}_{k}\end{array}\right]\!.\label{eq:U_k}\end{align}
We next express~$\Delta_{k}$ as a function of~$\subcol{\mathbf{A}}1k$.
From (\ref{eq:y_k_y_k}), (\ref{eq:Delta_k_prod}), and~(\ref{eq:U_k}),
we obtain\begin{align}
\Delta_{k} & =\prod_{i=1}^{k}\mathbf{y}_{i}^{H}\mathbf{a}_{i}=\det(\mathbf{U}_{k}).\label{eq:det_U_k}\end{align}
Furthermore, (\ref{eq:q_k}), (\ref{eq:r_k}), and~(\ref{eq:R_kk_y_k})
imply \begin{align*}
\mathbf{y}_{k'}^{H}\mathbf{a}_{i} & =\sqrt{\mathbf{y}_{k'}^{H}\mathbf{y}_{k'}}\mathbf{q}_{k'}^{H}\mathbf{a}_{i}=[\mathbf{R}]_{k',k'}[\mathbf{R}]_{k',i}\end{align*}
which evaluates to zero for $1\leq i<k'\leq k$ because of the upper
triangularity of~$\mathbf{R}$. Hence, $\mathbf{U}_{k}$ can be written
as \begin{align}
\mathbf{U}_{k} & =\left[\begin{array}{cccc}
\mathbf{y}_{1}^{H}\mathbf{a}_{1} & \mathbf{y}_{1}^{H}\mathbf{a}_{2} & \cdots & \mathbf{y}_{1}^{H}\mathbf{a}_{k}\\
\mathbf{y}_{2}^{H}\mathbf{a}_{1} & \mathbf{y}_{2}^{H}\mathbf{a}_{2} & \cdots & \mathbf{y}_{2}^{H}\mathbf{a}_{k}\\
\vdots & \vdots & \ddots & \vdots\\
\mathbf{y}_{k}^{H}\mathbf{a}_{1} & \mathbf{y}_{k}^{H}\mathbf{a}_{2} & \cdots & \mathbf{y}_{k}^{H}\mathbf{a}_{k}\end{array}\right]\!.\label{eq:U_k_full}\end{align}
By combining~(\ref{eq:det_U_k}) and~(\ref{eq:U_k_full}), we obtain\begin{align}
\Delta_{k} & =\det(\mathbf{U}_{k})=\det\!\vecd{\mathbf{y}_{1}^{H}\subcol{\mathbf{A}}1k}{\mathbf{y}_{2}^{H}\subcol{\mathbf{A}}1k}{\mathbf{y}_{k}^{H}\subcol{\mathbf{A}}1k}\!=\det\!\vecd{\mathbf{a}_{1}^{H}\subcol{\mathbf{A}}1k}{\mathbf{a}_{2}^{H}\subcol{\mathbf{A}}1k}{\mathbf{a}_{k}^{H}\subcol{\mathbf{A}}1k}\!\label{eq:y_ka_k_replace}\\
 & =\det\bigpar{\subcol{\mathbf{A}}1k^{H}\subcol{\mathbf{A}}1k}\label{eq:Delta_k_A}\end{align}
where the third equality in~(\ref{eq:y_ka_k_replace}) can be shown
by induction as follows. We start by noting that $\mathbf{y}_{1}=\mathbf{a}_{1}$,
which implies that in the first row of $\mathbf{U}_{k}$, $\mathbf{y}_{1}$
can be replaced by~$\mathbf{a}_{1}$. For $k'>1$, assuming that
we have already replaced $\mathbf{y}_{1},\mathbf{y}_{2},\ldots,\mathbf{y}_{k'-1}$
by $\mathbf{a}_{1},\mathbf{a}_{2},\ldots,\mathbf{a}_{k'-1}$, respectively,
we can replace $\mathbf{y}_{k'}$ by~$\mathbf{a}_{k'}$ since, as
a consequence of~(\ref{eq:y_k_a_i}), the $k'$th row of~$\mathbf{U}_{k}$
can be written as \[
\mathbf{y}_{k'}^{H}\subcol{\mathbf{A}}1k=\mathbf{a}_{k'}^{H}\subcol{\mathbf{A}}1k+\sum_{i=1}^{k'-1}\bigl(\alpha_{i}^{(k')}\bigr)^{*}\bigpar{\mathbf{a}_{i}^{H}\subcol{\mathbf{A}}1k}.\]
Hence, replacing $\mathbf{y}_{k'}^{H}\subcol{\mathbf{A}}1k$ by~$\mathbf{a}_{k'}^{H}\subcol{\mathbf{A}}1k$
amounts to subtracting a linear combination of the first $k'-1$ rows
of~$\mathbf{U}_{k}$ from the~$k'$th row of~$\mathbf{U}_{k}$.
This operation does not affect the value of~$\det(\mathbf{U}_{k})$\citet{Horn85}.

Similarly to what we have done for~$\Delta_{k}$, we will next show
that $\tilde{\mathbf{q}}_{k}$ can be expressed in terms of~$\subcol{\mathbf{A}}1k$
only. We start by noting that, since $\mathbf{V}_{k}$ is nonsingular,
we can rewrite~(\ref{eq:a_1k_y_1k}) as\begin{equation}
\hvecd{\mathbf{y}_{1}}{\mathbf{y}_{2}}{\mathbf{y}_{k}}=\hvecd{\mathbf{a}_{1}}{\mathbf{a}_{2}}{\mathbf{a}_{k}}\,\mathbf{V}_{k}^{-1}\!.\label{eq:y_1k_a_1k}\end{equation}
Next, from $\mathbf{V}_{k}=\mathbf{D}_{k}^{-1}\mathbf{U}_{k}$ we
obtain that\begin{align*}
\mathbf{V}_{k}^{-1} & =\mathbf{U}_{k}^{-1}\mathbf{D}_{k}=\frac{\adj(\mathbf{U}_{k})}{\det(\mathbf{U}_{k})}\mathbf{D}_{k}\end{align*}
and hence, by~(\ref{eq:det_U_k}), that\begin{align}
\mathbf{V}_{k}^{-1} & =\frac{1}{\Delta_{k}}\underbrace{\left[\begin{array}{cccc}
\cof k11 & \cof k21 & \cdots & \cof kk1\\
0 & \cof k22 & \cdots & \cof kk2\\
\vdots & \vdots & \ddots & \vdots\\
0 & 0 & \cdots & \cof kkk\end{array}\right]}_{\adj(\mathbf{U}_{k})}\mathbf{D}_{k}\label{eq:V_k_inv}\end{align}
where $\adj(\mathbf{U}_{k})$ is upper triangular since~$\mathbf{U}_{k}$
is upper triangular, and $\cof knm$ denotes the cofactor of~$\mathbf{U}_{k}$
relative to the matrix entry~$[\mathbf{U}_{k}]_{n,m}$ ($\allone nk$;
$m=n,n+1,\ldots,k$)\citet{Horn85}. Note that in order to handle
the case $k=1$ correctly, for which $\adj(\mathbf{U}_{1})=\cof 111$,
$\det(\mathbf{U}_{1})=\mathbf{U}_{1}=\Delta_{1}$, and $\mathbf{U}_{1}^{-1}=1/\Delta_{1}$,
we define $\cof 111\triangleq1$. From~(\ref{eq:y_1k_a_1k}) and~(\ref{eq:V_k_inv})
it follows that\begin{align*}
\mathbf{y}_{k} & =\frac{1}{\Delta_{k}}\mathbf{y}_{k}^{H}\mathbf{y}_{k}\sum_{i=1}^{k}\cof kki\mathbf{a}_{i}\\
 & =\frac{1}{\Delta_{k-1}}\sum_{i=1}^{k}\cof kki\mathbf{a}_{i}\end{align*}
 and therefore, by~(\ref{eq:q_tilde_k}), we get\begin{equation}
\tilde{\mathbf{q}}_{k}=\sum_{i=1}^{k}\cof kki\mathbf{a}_{i}\label{eq:q_tilde_k_cof}\end{equation}
which evaluates to $\tilde{\mathbf{q}}_{1}=\mathbf{a}_{1}$ for $k=1$.
Next, for $k>1$ we denote by~$\subcol{\mathbf{A}}1{k\backslash i}$
the matrix obtained by removing the $i$th column of~$\subcol{\mathbf{A}}1k$,
and we express~$\cof kki$ as a function of~$\mathbf{a}_{1},\mathbf{a}_{2},\ldots,\mathbf{a}_{k}$
according to \begin{align*}
\cof kki & =(-1)^{k+i}\det\!\vecd{\mathbf{y}_{1}^{H}\subcol{\mathbf{A}}1{k\backslash i}}{\mathbf{y}_{2}^{H}\subcol{\mathbf{A}}1{k\backslash i}}{\mathbf{y}_{k-1}^{H}\subcol{\mathbf{A}}1{k\backslash i}}\\
 & =(-1)^{k+i}\det\bigpar{\subcol{\mathbf{A}}1{k-1}^{H}\subcol{\mathbf{A}}1{k\backslash i}}\end{align*}
where the last equality is derived analogously to~(\ref{eq:y_ka_k_replace})
and (\ref{eq:Delta_k_A}). Thus, (\ref{eq:q_tilde_k_cof}) can be
written as\begin{equation}
\tilde{\mathbf{q}}_{k}=\begin{cases}
\mathbf{a}_{k}, & \quad k=1\\
\sum_{i=1}^{k}(-1)^{k+i}\det\bigpar{\subcol{\mathbf{A}}1{k-1}^{H}\subcol{\mathbf{A}}1{k\backslash i}}\mathbf{a}_{i}, & \quad k>1.\end{cases}\label{eq:q_tilde_k_A}\end{equation}
Finally, we obtain\begin{equation}
\tilde{\mathbf{r}}_{k}^{T}=\tilde{\mathbf{q}}_{k}^{H}\mathbf{A}\label{eq:r_tilde_k_A}\end{equation}
as implied by (\ref{eq:r_k}), (\ref{eq:q_k_map}), and (\ref{eq:r_k_map}).
The results of Step~1a are the relations~(\ref{eq:Delta_k_A}),
(\ref{eq:q_tilde_k_A}), and~(\ref{eq:r_tilde_k_A}), which are valid
for $(K,k)\in\mathcal{K}_{1}$.

\paragraph*{Step 1b}

We next show that (\ref{eq:Delta_k_A}), (\ref{eq:q_tilde_k_A}),
and~(\ref{eq:r_tilde_k_A}) hold for $(K,k)\in\mathcal{K}_{2}$ as
well. Throughout Step~1b we assume that $(K,k)\in\mathcal{K}_{2}$,
and, unless specified otherwise, all equations and statements involving~$k$
are valid for $k=K+1,K+2,\ldots,M$. We know from Section~\ref{sub:qrfurtherproperties}
that $[\mathbf{R}]_{K+1,K+1}=0$. According to the definition of~$\map$,
$[\mathbf{R}]_{K+1,K+1}=0$ implies $\Delta_{k}=0$, $\tilde{\mathbf{q}}_{k}=\mathbf{0}$,
and $\tilde{\mathbf{r}}_{k}^{T}=\mathbf{0}$. It is therefore to be
shown that the RHS of~(\ref{eq:Delta_k_A}) evaluates to zero, and
that the RHS expressions of (\ref{eq:q_tilde_k_A}) and~(\ref{eq:r_tilde_k_A})
evaluate to all-zero vectors. We start by noting that since $k>K$,
$\subcol{\mathbf{A}}1k$ is rank-deficient. Since $\rk(\subcol{\mathbf{A}}1k^{H}\subcol{\mathbf{A}}1k)=\rk(\subcol{\mathbf{A}}1k)<k$,
we obtain that $\det(\subcol{\mathbf{A}}1k^{H}\subcol{\mathbf{A}}1k)$
on the RHS of~(\ref{eq:Delta_k_A}) evaluates to zero. Next, for
$k>\max(K,1)$, the expression \begin{equation}
\sum_{i=1}^{k}(-1)^{k+i}\det\bigpar{\subcol{\mathbf{A}}1{k-1}^{H}\subcol{\mathbf{A}}1{k\backslash i}}\mathbf{a}_{i}\label{eq:qtildekassumofai}\end{equation}
on the RHS of~(\ref{eq:q_tilde_k_A}) is a vector whose $p$th component
can be written, by inverse Laplace expansion\citet{Horn85}, as\begin{align}
\sum_{i=1}^{k}(-1)^{k+i}\det\bigpar{\subcol{\mathbf{A}}1{k-1}^{H}\subcol{\mathbf{A}}1{k\backslash i}}[\mathbf{A}]_{p,i} & =\det\left[\begin{array}{cccc}
\subcol{\mathbf{A}}1{k-1}^{H}\mathbf{a}_{1} & \subcol{\mathbf{A}}1{k-1}^{H}\mathbf{a}_{2} & \cdots & \subcol{\mathbf{A}}1{k-1}^{H}\mathbf{a}_{k}\\
{}[\mathbf{A}]_{p,1} & [\mathbf{A}]_{p,2} & \cdots & [\mathbf{A}]_{p,k}\end{array}\right]\label{eq:invlaplexpansion}\end{align}
for all $\allone pP$. Now, again for $k>\max(K,1)$, since $\subcol{\mathbf{A}}1k$
is rank-deficient, $\mathbf{a}_{k}$ can be written as a linear combination\[
\mathbf{a}_{k}=\sum_{k'=1}^{k-1}\beta^{(k')}\mathbf{a}_{k'}\]
(for some coefficients~$\beta^{(k')}$, $\allone{k'}{k-1}$) which
implies that, for all $\allone pP$, the argument of the determinant
on the RHS of~(\ref{eq:invlaplexpansion}) has\[
\left[\begin{array}{c}
\subcol{\mathbf{A}}1{k-1}^{H}\mathbf{a}_{k}\\
{}[\mathbf{A}]_{p,k}\end{array}\right]=\sum_{k'=1}^{k-1}\beta^{(k')}\left[\begin{array}{c}
\subcol{\mathbf{A}}1{k-1}^{H}\mathbf{a}_{k'}\\
{}[\mathbf{A}]_{p,k'}\end{array}\right]\]
as its last column. Since this column is a linear combination of the
first $k-1$ columns, the determinant on the RHS of~(\ref{eq:invlaplexpansion})
is equal to zero for all $\allone pP$, and hence the expression in~(\ref{eq:qtildekassumofai})
is equal to an all-zero vector for $k>\max(K,1)$. Moreover, if $K=0$
and $k=1$, we have $\mathbf{a}_{1}=\mathbf{0}$ on the RHS of~(\ref{eq:q_tilde_k_A}).
Hence, the RHS of~(\ref{eq:q_tilde_k_A}) evaluates to an all-zero
vector for all $(K,k)\in\mathcal{K}_{2}$. Thus, (\ref{eq:q_tilde_k_A})
simplifies to $\tilde{\mathbf{q}}_{k}=\mathbf{0}$, which in turn
implies that the RHS of~(\ref{eq:r_tilde_k_A}) evaluates to an all-zero
vector as well. We have therefore shown that (\ref{eq:Delta_k_A}),
(\ref{eq:q_tilde_k_A}), and~(\ref{eq:r_tilde_k_A}) hold for $(K,k)\in\mathcal{K}_{2}$.
Finally, since $\mathcal{K}_{1}\cup\mathcal{K}_{2}=\mathcal{K}$,
the results of Steps~1a and~1b imply that (\ref{eq:Delta_k_A}),
(\ref{eq:q_tilde_k_A}), and~(\ref{eq:r_tilde_k_A}) are valid for
$(K,k)\in\mathcal{K}$.

\paragraph*{Step 2}

We note that the derivations presented in Steps 1a and~1b for a given~$s_{0}\in\setunit$
do not depend on~$s_{0}$ and can hence be carried over to all $s_{0}\in\setunit$.
Thus, we can rewrite (\ref{eq:Delta_k_A}), (\ref{eq:q_tilde_k_A}),
and~(\ref{eq:r_tilde_k_A}), respectively, as \begin{align}
\Delta_{k}(s) & =\det\bigpar{\subcol{\mathbf{A}}1k^{H}(s)\subcol{\mathbf{A}}1k(s)}\label{eq:Delta_k_LP}\\
\tilde{\mathbf{q}}_{k}(s) & =\begin{cases}
\mathbf{a}_{k}(s), & \quad k=1\\
\sum_{i=1}^{k}(-1)^{k+i}\det\bigpar{\subcol{\mathbf{A}}1{k-1}^{H}(s)\subcol{\mathbf{A}}1{k\backslash i}(s)}\mathbf{a}_{i}(s), & \quad k>1\end{cases}\label{eq:q_tilde_k_LP}\\
\tilde{\mathbf{r}}_{k}^{T}(s) & =\tilde{\mathbf{q}}_{k}^{H}(s)\mathbf{A}(s)\label{eq:r_tilde_k_LP}\end{align}
for $\allone kM$ and $s\in\setunit$.

\paragraph*{Step 3}

For $\allone kM$, we note that $\pol{\mathbf{A}(s)}{\dg_{1}}{\dg_{2}}$,
along with $\dg=\dg_{1}+\dg_{2}$, implies $\pol{\subcol{\mathbf{A}^{H}}1k(s)\subcol{\mathbf{A}}1k(s)}{\dg}{\dg}$.
Now, the determinant on the RHS of~(\ref{eq:Delta_k_LP}) can be
expressed through Laplace expansion as a sum of products of~$k$
entries of $\pol{\subcol{\mathbf{A}^{H}}1k(s)\subcol{\mathbf{A}}1k(s)}{\dg}{\dg}$.
Therefore, we get $\pol{\Delta_{k}(s)}{k\dg}{k\dg}$ for $\allone kM$.
Analogously, for $\alltwo kM$ we obtain $\pol{\det\bigpar{\subcol{\mathbf{A}}1{k-1}^{H}(s)\subcol{\mathbf{A}}1{k\backslash i}(s)}}{(k-1)\dg}{(k-1)\dg}$.
The latter result, combined with $\pol{\mathbf{A}(s)}{\dg_{1}}{\dg_{2}}$
in (\ref{eq:q_tilde_k_LP}) yields $\pol{\tilde{\mathbf{q}}_{k}(s)}{(k-1)\dg+\dg_{1}}{(k-1)\dg+\dg_{2}}$,
which holds for $k=1$ as well as a trivial consequence of~(\ref{eq:q_tilde_k_LP})
and~$\pol{\mathbf{A}(s)}{\dg_{1}}{\dg_{2}}$. Finally, from $\pol{\tilde{\mathbf{q}}_{k}(s)}{(k-1)\dg+\dg_{1}}{(k-1)\dg+\dg_{2}}$
and~(\ref{eq:r_tilde_k_LP}), using $\pol{\mathbf{A}(s)}{\dg_{1}}{\dg_{2}}$
and $\dg=\dg_{1}+\dg_{2}$, we obtain $\pol{\tilde{\mathbf{r}}_{k}^{T}(s)}{k\dg}{k\dg}$
for $\allone kM$.\hfill{}$\qed$

\section{Application to MIMO-OFDM}

\label{sec:algorithms}

We are now ready to show how the results derived in the previous section
lead to algorithms that exploit the polynomial nature of the MIMO~channel
transfer function $\pol{\mattf(s)}0L$ to perform efficient interpolation-based
computation of QR~factors of~$\mattf\ofsn$, for all $n\in\setdat$,
given knowledge of~$\mattf\ofsn$ for $n\in\setpil$. We note that
the algorithms described in the following apply to QR~decomposition
of generic polynomial matrices that are oversampled on the unit circle.

Within the algorithms to be presented, interpolation involves base
points and target points on~$\setunit$ that correspond to OFDM~tones
indexed by integers taken from the set $\{0,1,\ldots,\ncar-1\}$.
For a given set $\mathcal{X}\subseteq\{0,1,\ldots,\ncar-1\}$ of OFDM
tones, we define $\setidx{\mathcal{X}}\triangleq\{s_{n}:n\in\mathcal{X}\}$
to denote the set of corresponding points on~$\setunit.$ With this
definition in place, we start by summarizing the brute-force approach
described in Section~\ref{sub:problemstatement}.

\algobox{Algorithm~I: Brute-force per-tone  QR~decomposition}{ 
\begin{enumerate}
\item Interpolate $\mattf(s)$ from $\setidx{\setpil}$ to $\setidx{\setdat}$.
\item For each $n\in\setdat$, perform QR~decomposition on $\mattf\ofsn$
to obtain $\mathbf{Q}\ofsn$ and $\mathbf{R}\ofsn$.
\end{enumerate}
}

It is obvious that for large $\ndat$, performing QR~decomposition
on a per-tone basis will result in high computational complexity.
However, in the practically relevant case $L\ll\ndat$ the OFDM system
effectively highly oversamples the MIMO channel's transfer function,
so that $\mattf\ofsn$ changes slowly across~$n$. This observation,
combined with the results in Section~\ref{sec:interpqr}, constitutes
the basis for a new class of algorithms that perform QR~decomposition
at a small number of tones and obtain the remaining QR~factors through
interpolation. More specifically, the basic idea of interpolation-based
QR~decomposition is as follows. By applying Theorem~\ref{thm:qrinterp}
to the $\nrx\times\ntx$ LP~matrix $\pol{\mattf(s)}0L$, we obtain
$\pol{\tilde{\mathbf{q}}_{k}(s)}{(k-1)L}{kL}$ and $\pol{\tilde{\mathbf{r}}_{k}^{T}(s)}{kL}{kL}$
for $\allone k{\ntx}$. In order to simplify the exposition, in the
remainder of the paper we consider~$\tilde{\mathbf{q}}_{k}(s)$ as
satisfying $\pol{\tilde{\mathbf{q}}_{k}(s)}{kL}{kL}$. The resulting
statements \begin{equation}
\pol{\tilde{\mathbf{q}}_{k}(s),\tilde{\mathbf{r}}_{k}^{T}(s)}{kL}{kL},\qquad\allone k{\ntx}\label{eq:qkandrkLP}\end{equation}
imply that both $\tilde{\mathbf{q}}_{k}(s)$ and~$\tilde{\mathbf{r}}_{k}^{T}(s)$
can be interpolated from at least $2kL+1$ base points, and that,
as a consequence of $\dg_{1}=\dg_{2}=kL$, the corresponding interpolation
matrices are real-valued. For $\allone k{\ntx}$, the interpolation-based
algorithms to be presented compute $\tilde{\mathbf{q}}_{k}\ofsn$
and~$\tilde{\mathbf{r}}_{k}^{T}\ofsn$, through QR~decomposition
followed by application of the mapping~$\map$, at a subset of OFDM
tones of cardinality at least $2kL+1$, then interpolate $\tilde{\mathbf{q}}_{k}(s)$
and~$\tilde{\mathbf{r}}_{k}^{T}(s)$ to obtain $\tilde{\mathbf{q}}_{k}\ofsn$
and~$\tilde{\mathbf{r}}_{k}^{T}\ofsn$ at the remaining tones, and
finally apply the inverse mapping~$\map^{-1}$ at these tones. In
the following, the sets $\setidxbase_{k}\subseteq\{0,1,\ldots,\ncar-1\}$,
with $\setidxbase_{k-1}\subseteq\setidxbase_{k}$ and $\nbase_{k}\triangleq|\setidxbase_{k}|\geq2kL+1$
($k=1,2,\ldots,\ntx$), contain the indices corresponding to the OFDM
tones chosen as base points. For completeness, we define $\setidxbase_{0}\triangleq\emptyset$.
Specific choices of the sets~$\setidxbase_{k}$ will be discussed
in detail in Section~\ref{sec:efficientinterpolation}.

We start with a conceptually simple algorithm for interpolation-based
QR~decomposition, derived from the observation that the~$\ntx$
statements in~(\ref{eq:qkandrkLP}) can be unified into the single
statement $\pol{\tilde{\mathbf{Q}}(s),\tilde{\mathbf{R}}(s)}{\ntx L}{\ntx L}$.
This implies that we can interpolate $\tilde{\mathbf{Q}}(s)$ and~$\tilde{\mathbf{R}}(s)$
from a single set of base points of cardinality~$\nbase_{\ntx}$.
The corresponding algorithm can be formulated as follows:

\algobox{Algorithm~II: Single interpolation step}{
\begin{enumerate}
\item Interpolate $\mattf(s)$ from~$\setidx{\setpil}$ to~$\setidx{\setidxbase_{\ntx}}$.\label{enu:alg2interpH}
\item For each $n\in\setidxbase_{\ntx}$, perform QR~decomposition on $\mattf\ofsn$
to obtain $\mathbf{Q}\ofsn$ and $\mathbf{R}\ofsn$.\label{enu:alg2qr}
\item For each $n\in\setidxbase_{\ntx}$, apply $\map:\qrnotilde{\ofsn}\mapsto\qrtilde{\ofsn}$.\label{enu:alg2map}
\item Interpolate $\tilde{\mathbf{Q}}(s)$ and~$\tilde{\mathbf{R}}(s)$
from~$\setidx{\setidxbase_{\ntx}}$ to~$\setidx{\setdat\backslash\setidxbase_{\ntx}}$.\label{enu:alg2interpQR}
\item For each $n\in\setdat\backslash\setidxbase_{\ntx}$, apply $\map^{-1}:\qrtilde{\ofsn}\mapsto\qrnotilde{\ofsn}$.\label{enu:alg2demap}
\end{enumerate}
}

This formulation of Algorithm~II assumes that~$\mattf\ofsn$ has
full rank for all $n\in\setdat\backslash\setidxbase_{\ntx}$, which
allows to perform all inverse mappings~$\map^{-1}$ in Step~\ref{enu:alg2demap}
using (\ref{eq:q_kdemap})--(\ref{eq:Delta_scalingfactorforinvmap})
only. If, however, for a given $n\in\setdat\backslash\setidxbase_{\ntx}$,
$\mattf\ofsn$ is rank-deficient with ordered column rank $K<\ntx$,
we have $\subcol{\tilde{\mathbf{Q}}}{K+1}{\ntx}\ofsn=\mathbf{0}$
and $\subrow{\tilde{\mathbf{R}}}{K+1}{\ntx}\ofsn=\mathbf{0}$. Hence,
according to the results in Section~\ref{sub:qrdecpolynomialmatrix},
$\subcol{\mathbf{Q}}{K+1}{\ntx}\ofsn$ and $\subrow{\mathbf{R}}{K+1}{\ntx}\ofsn$
must be computed through QR~decomposition of $\subcol{\mattf}{K+1}{\ntx}\ofsn-\subcol{\mathbf{Q}}1K\ofsn\subcolrow{\mathbf{R}}{K+1}{\ntx}1K\ofsn$
for $K>0$ or of $\mattf\ofsn$ for $K=0$. This, in turn, requires~$\subcol{\mattf}{K+1}{\ntx}\ofsn$
to be obtained by interpolating~$\subcol{\mattf}{K+1}{\ntx}(s)$
from~$\setidx{\setpil}$ to the single target point~$s_{n}$ in
an additional step. For simplicity of exposition, in the remainder
of the paper we will assume that~$\mattf\ofsn$ is full-rank for
all $n\in\setdat$.

Departing from Algorithm~II, which interpolates $\tilde{\mathbf{q}}_{k}(s)$
and~$\tilde{\mathbf{r}}_{k}^{T}(s)$ from~$\nbase_{\ntx}$ base
points, we next present a more sophisticated algorithm that involves
interpolation of $\tilde{\mathbf{q}}_{k}(s)$ and~$\tilde{\mathbf{r}}_{k}^{T}(s)$
from $\nbase_{k}\leq\nbase_{\ntx}$ base points ($\allone k{\ntx}$),
in agreement with~(\ref{eq:qkandrkLP}). The resulting Algorithm~III
consists of~$\ntx$ iterations. In the first iteration, the tones
$n\in\setidxbase_{1}$ are considered. At each of these tones, QR~decomposition
is performed on~$\mattf\ofsn$, resulting in $\mathbf{Q}\ofsn$ and~$\mathbf{R}\ofsn$,
which are then mapped to $\qrtilde{\ofsn}$ by applying~$\map$.
Next, $\tilde{\mathbf{q}}_{1}(s)$ and~$\tilde{\mathbf{r}}_{1}^{T}(s)$
are interpolated from the tones $n\in\setidxbase_{1}$ to the remaining
tones $n\in\setdat\backslash\setidxbase_{1}$. In the~$k$th iteration
($\alltwo k{\ntx}$), the tones $n\in\setidxbase_{k}\backslash\setidxbase_{k-1}$
are considered. At each of these tones, $\subcol{\mathbf{Q}}1{k-1}\ofsn$
and~$\subrow{\mathbf{R}}1{k-1}\ofsn$ are obtained%
\footnote{The mapping $\map$ and its inverse $\map^{-1}$ are defined on submatrices
of $\mathbf{Q}\ofsn$ and~$\mathbf{R}\ofsn$ according to~(\ref{eq:Delta_k_map})--(\ref{eq:Delta_scalingfactorforinvmap}).%
} by applying~$\map^{-1}$ to $\qrparttilde 1{k-1}{\ofsn}$, already
known from the previous iterations, whereas the submatrices $\subcol{\mathbf{Q}}k{\ntx}\ofsn$
and~$\subcolrow{\mathbf{R}}k{\ntx}k{\ntx}\ofsn$ are obtained by
performing QR~decomposition on the matrix $\subcol{\mattf}k{\ntx}\ofsn-\subcol{\mathbf{Q}}1{k-1}\ofsn\subcolrow{\mathbf{R}}k{\ntx}1{k-1}\ofsn$,
in accordance with Proposition~\ref{pro:qrreduction}, and $\subrow{\mathbf{R}}k{\ntx}\ofsn$
is given, for $k>1$, by $\matcouple{\mathbf{0}}{\subcolrow{\mathbf{R}}k{\ntx}k{\ntx}\ofsn}$
. Next, the submatrices $\subcol{\tilde{\mathbf{Q}}}k{\ntx}\ofsn$
and~$\subrow{\tilde{\mathbf{R}}}k{\ntx}\ofsn$ are computed by applying~$\map$
to $\qrpartnotilde k{\ntx}{\ofsn}$. Since the samples $\tilde{\mathbf{q}}_{k}(s_{n})$
and~$\tilde{\mathbf{r}}_{k}^{T}(s_{n})$ are now known at all tones
$n\in\setidxbase_{k}$, $\tilde{\mathbf{q}}_{k}(s)$ and~$\tilde{\mathbf{r}}_{k}^{T}(s)$
can be interpolated from the tones $n\in\setidxbase_{k}$ to the remaining
tones $n\in\setdat\backslash\setidxbase_{k}$, thereby completing
the $k$th iteration. After~$\ntx$ iterations, we know $\tilde{\mathbf{Q}}\ofsn$
and~$\tilde{\mathbf{R}}\ofsn$ at all tones $n\in\setdat$, as well
as $\mathbf{Q}\ofsn$ and~$\mathbf{R}\ofsn$ at the tones $n\in\setidxbase_{\ntx}$.
The last step consists of applying~$\map^{-1}$ to $\qrtilde{\ofsn}$
to obtain $\mathbf{Q}\ofsn$ and~$\mathbf{R}\ofsn$ at the remaining
tones $n\in\setdat\backslash\setidxbase_{k}$. The algorithm is formulated
as follows:

\algobox{Algorithm~III: Multiple interpolation steps}{
\begin{enumerate}
\item Set $k\leftarrow1$.
\item Interpolate $\subcol{\mattf}k{\ntx}(s)$ from~$\setidx{\setpil}$
to~$\setidx{\setidxbase_{k}\backslash\setidxbase_{k-1}}$.\label{enu:alg3loopstart}
\item If $k=1$, go to Step~\ref{enu:alg3qr}. Otherwise, for each $n\in\setidxbase_{k}\backslash\setidxbase_{k-1}$,
apply $\map^{-1}:\qrparttilde 1{k-1}{\ofsn}\mapsto\qrpartnotilde 1{k-1}{\ofsn}$.\label{enu:alg3demap}
\item For each $n\in\setidxbase_{k}\backslash\setidxbase_{k-1}$, overwrite
$\subcol{\mattf}k{\ntx}\ofsn$ by $\subcol{\mattf}k{\ntx}\ofsn-\subcol{\mathbf{Q}}1{k-1}\ofsn\subcolrow{\mathbf{R}}k{\ntx}1{k-1}\ofsn$.\label{enu:alg3red}
\item For each $n\in\setidxbase_{k}\backslash\setidxbase_{k-1}$, perform
QR~decomposition on $\subcol{\mattf}k{\ntx}\ofsn$ to obtain $\subcol{\mathbf{Q}}k{\ntx}\ofsn$
and $\subcolrow{\mathbf{R}}k{\ntx}k{\ntx}\ofsn$, and, if $k>1$,
construct $\subrow{\mathbf{R}}k{\ntx}\ofsn=\matcouple{\mathbf{0}}{\subcolrow{\mathbf{R}}k{\ntx}k{\ntx}\ofsn}.$\label{enu:alg3qr}
\item For each $n\in\setidxbase_{k}\backslash\setidxbase_{k-1}$, apply
$\map:\qrpartnotilde k{\ntx}{\ofsn}\mapsto\qrparttilde k{\ntx}{\ofsn}$.
\item Interpolate $\tilde{\mathbf{q}}_{k}(s)$ and $\tilde{\mathbf{r}}_{k}^{T}(s)$
from~$\setidx{\setidxbase_{k}}$ to~$\setidx{\setdat\backslash\setidxbase_{k}}$.\label{enu:alg3ipqr}
\item If $k=\ntx$, proceed to the next step. Otherwise, set $k\leftarrow k+1$
and go back to Step~\ref{enu:alg3loopstart}.
\item For each $n\in\setdat\backslash\setidxbase_{\ntx}$, apply $\map^{-1}:\qrtilde{\ofsn}\mapsto\qrnotilde{\ofsn}$.
\end{enumerate}
}

In comparison with Algorithm~II, Algorithm~III performs QR~decompositions
on increasingly smaller matrices. The corresponding computational
complexity savings are, however, traded against an increase in interpolation
effort and the computational overhead associated with Step~\ref{enu:alg3red},
which will be referred to as the \emph{reduction step} in what follows.
Moreover, the complexity of applying $\map$ and~$\map^{-1}$ differs
for the two algorithms. A detailed complexity analysis provided in
the next section will show that, depending on the system parameters,
Algorithm~III can exhibit smaller complexity than Algorithm II.

We conclude this section with some remarks on ordered~SC MIMO-OFDM
detectors\citet{paulraj03}, which essentially permute the columns
of~$\mattf\ofsn$ to perform SC~detection of the transmitted data
symbols according to a given sorting criterion (such as, e.g., V-BLAST
sorting\citet{wolniansky98}) to obtain better detection performance
than in the unsorted case. The permutation of the columns of~$\mattf\ofsn$
can be represented by means of right-multiplication of~$\mattf\ofsn$
by an $\ntx\times\ntx$ permutation matrix~$\matperm\ofsn$. The
matrices subjected to QR~decomposition are then given by $\mattf\ofsn\matperm\ofsn,n\in\setdat$.
If $\matperm\ofsn$ is constant across all OFDM tones, i.e., $\matperm\ofsn=\matperm_{0},n\in\setdat$,
we have $\pol{\mattf(s)\matperm_{0}}0L$ and Algorithms~I--III can
be applied to~$\mattf\ofsn\matperm_{0}$. A MIMO-OFDM ordered SC~detector
using Algorithm~II to compute QR~factors of~$\mattf(s)\matperm_{0}$,
along with a strategy for choosing~$\mathbf{P}_{0}$, was presented
in\citet{wuebben06-03}. If $\matperm\ofsn$ varies across~$n$,
the matrices $\mattf\ofsn\matperm\ofsn,n\in\setdat$, in general,
can no longer be seen as samples of a polynomial matrix of maximum
degree~$L\ll\ndat$, so that the interpolation-based QR~decomposition
algorithms presented above can not be applied.

\section{Complexity Analysis}

\label{sec:complexity}

We are next interested in assessing under which circumstances the
interpolation-based Algorithms II and~III offer computational complexity
savings over the brute-force approach in Algorithm~I. To this end,
we propose a simple computational complexity metric, representative
of VLSI circuit complexity as quantified by the product of chip area
and processing delay\citet{kaeslin08}. We note that other important
aspects of VLSI design, including, e.g., wordwidth requirements, memory
access strategies, and datapath architecture, are not accounted for
in our analysis. Nevertheless, the proposed metric is indicative of
the complexity of Algorithms~I--III and allows to quantify the impact
of the system parameters on the potential savings of interpolation-based
QR~decomposition over brute-force per-tone QR~decomposition.

In the remainder of the paper, unless explicitly specified otherwise,
the term \emph{complexity} refers to computational complexity according
to the metric defined in Section~\ref{sub:complmetric} below. We
derive the complexity of individual \emph{computational tasks} (i.e.,
interpolation, QR~decomposition, mapping~$\map$, inverse mapping~$\map^{-1}$,
and reduction step) in Section \ref{sub:pertonecosts}. Then, we proceed
to computing the total complexity of Algorithms~I--III in Section~\ref{sub:overallalgcosts}.
Finally, in Section~\ref{sub:complcomparison} we compare the complexity
results obtained in Section~\ref{sub:overallalgcosts} and we derive
conditions on the system parameters under which Algorithms II and~III
exhibit lower complexity than Algorithm~I.

\subsection{Complexity Metric}

\label{sub:complmetric}

In the VLSI implementation of a given algorithm, a wide range of trade-offs
between silicon area~$A$ and processing delay~$\tau$ can, in general,
be realized\citet{kaeslin08}. Parallel processing reduces~$\tau$
at the expense of a larger~$A$, whereas resource sharing reduces~$A$
at the expense of a larger~$\tau$. However, the corresponding circuit
transformations typically do not affect the area-delay product~$A\tau$~significantly.
For this reason, the area-delay product is considered a relevant indicator
of algorithm complexity\citet{kaeslin08}. In the definition of the
specific complexity metric that will be used subsequently, we only
take into account the arithmetic operations with a significant impact
on~$A\tau$. More specifically, we divide the operations underlying
the algorithms under consideration into three classes, namely i)~multiplications,
ii)~divisions and square roots, and iii)~additions and subtractions.
Class~iii) operations will not be counted as they typically have
a significantly lower VLSI circuit complexity than Class~i) and Class~ii)
operations.

In all algorithms presented in this paper, the number of Class~i)
operations is significantly larger than the number of Class~ii) operations.%
\footnote{We assume that division of an $M$-dimensional vector~$\mathbf{a}$
by a scalar~$\alpha$, such as the divisions in (\ref{eq:q_k}),
(\ref{eq:q_kdemap}), or~(\ref{eq:r_kdemap}), is implemented by
first computing the single division $\beta\triangleq1/\alpha$ and
then multiplying the $M$ entries of~$\mathbf{a}$ by $\beta$, at
the cost of one Class~ii) operation and $M$ Class~i) operations,
respectively.%
} By assuming a VLSI architecture where the Class~ii) operations are
performed by low-area high-delay arithmetical units operating in parallel
to the multipliers performing the Class~i) operations, it follows
that the Class~i) operations dominate the overall complexity and
the Class~ii) operations can be neglected.

Within Class~i), we distinguish between \emph{full multiplications}
(i.e., multiplications of two variable operands) and \emph{constant
multiplications} (i.e., multiplications of a variable operand by a
constant operand%
\footnote{In the context of the interpolation-based algorithms considered in
this paper, all operands that depend on~$\mattf(s)$ are assumed
variable. The coefficients of interpolation filters, e.g., are treated
as constant operands. For a detailed discussion on the difference
between full multiplications and constant multiplications, we refer
to Section~\ref{sub:interpdedicatedmultipliers}.%
}). We define the cost of a full multiplication as the unit of computational
complexity. We do not distinguish between real-valued full multiplications
and complex-valued full multiplications, as we assume that both are
performed by multipliers designed to process two variable complex-valued
operands. The fact, discussed in detail in Section~\ref{sub:interpdedicatedmultipliers},
that a constant multiplication can be implemented in VLSI at significantly
smaller cost than a full multiplication, will be accounted for through
a weighting factor smaller than one.

\subsection{Per-Tone Complexity of Individual Computational Tasks}

\label{sub:pertonecosts}In order to simplify the notation, in the
remainder of this section we drop the dependence of all quantities
on~$s_{n}$. We furthermore introduce the auxiliary variable\[
\nentries_{k}\triangleq\nrx k+\ntx k-\frac{(k-1)k}{2},\qquad\allone k{\ntx}\]
which specifies the maximum total number of nonzero entries in $\subcol{\mathbf{Q}}1k$
and~$\subrow{\mathbf{R}}1k$, and hence also in $\subcol{\tilde{\mathbf{Q}}}1k$
and~$\subrow{\tilde{\mathbf{R}}}1k$, in accordance with the fact
that $\mathbf{R}$ and~$\tilde{\mathbf{R}}$ are upper triangular.

\paragraph*{Interpolation}

We quantify the complexity of interpolating an LP to one target point
through an equivalent of~$\cip$ full multiplications. The dependence
of interpolation complexity on the underlying VLSI implementation
and on the number of base points is assumed to be incorporated into~$\cip$.
Specific strategies for efficient interpolation along with the corresponding
values of~$\cip$ are presented in Section~\ref{sec:efficientinterpolation}.
Since interpolation of an LP~matrix is performed entrywise, the complexity
of interpolating~$\subcol{\mattf}k{\ntx}(s)$ to one target point
is given by\[
\ciphk k=\nrx\bigpar{\ntx-k+1}\cip,\qquad\allone k{\ntx}.\]
Similarly, interpolation of $\tilde{\mathbf{Q}}(s)$ and~$\tilde{\mathbf{R}}(s)$
to one target point has complexity \[
\cipqr=\nentries_{\ntx}\cip\]
and the complexity of interpolating $\tilde{\mathbf{q}}_{k}(s)$ and~$\tilde{\mathbf{r}}_{k}^{T}(s)$
to one target point is given by\[
\cipqrk k=\bigpar{\nrx+\ntx-k+1}\cip,\qquad\allone k{\ntx}.\]

\paragraph*{QR~decomposition}

In order to keep our discussion independent of the QR~decomposition
method, we denote the cost of performing QR~decomposition on an $\nrx\times k$
matrix by~$\cqrk k$ ($\allone k{\ntx}$). Specific expressions for~$\cqrk k$
will only be required in the numerical complexity analysis in Section~\ref{sec:numresults}.

\paragraph*{Mapping $\map$}

We denote the overall cost of mapping $\qrpartnotilde k{\ntx}{}$
to $\qrparttilde k{\ntx}{}$ ($\allone k{\ntx}$) by~$\cmapk k$.
In the case $k=1$, application of the mapping~$\map$ requires computation
of $[\mathbf{R}]_{1,1}$, $[\mathbf{R}]_{1,1}^{2}$, $[\mathbf{R}]_{1,1}^{2}[\mathbf{R}]_{2,2}$,
$[\mathbf{R}]_{1,1}^{2}[\mathbf{R}]_{2,2}^{2},\ldots,$$\prod_{i=1}^{\ntx}[\mathbf{R}]_{i,i}^{2}$,
at the cost of $2\ntx-1$ full multiplications. This step yields both
the scaling factors $\Delta_{k'-1}[\mathbf{R}]_{k',k'}$, $\allone{k'}{\ntx}$,
and the diagonal entries of~$\tilde{\mathbf{R}}$. From~(\ref{eq:q_tilde_k_A})
we can deduce that the first column of~$\tilde{\mathbf{Q}}$ is equal
to the first column of~$\mattf$ and is hence obtained at zero complexity.
The remaining entries of~$\tilde{\mathbf{Q}}$ and the entries of~$\tilde{\mathbf{R}}$
above the main diagonal are obtained by scaling the corresponding
entries of $\mathbf{Q}$ and~$\mathbf{R}$ according to~(\ref{eq:q_k_map})
and~(\ref{eq:r_k_map}), respectively, which requires $\nentries_{\ntx}-\nrx-\ntx$
full multiplications. Hence, we obtain\[
\cmapk 1=\nentries_{\ntx}-\nrx+\ntx-1.\]
Next, we consider the case $k>1$, which only occurs in Step~\ref{enu:alg3demap}
of Algorithm~III, where $\Delta_{k-1}=[\tilde{\mathbf{R}}]_{k-1,k-1}$
is already available from the previous iteration which involves interpolation
of~$\tilde{\mathbf{r}}_{k-1}^{T}(s)$. The application of the mapping~$\map$
first requires computation of $\Delta_{k-1}[\mathbf{R}]_{k,k}$, $\Delta_{k-1}[\mathbf{R}]_{k,k}^{2}$,
$\Delta_{k-1}[\mathbf{R}]_{k,k}^{2}[\mathbf{R}]_{k+1,k+1},\ldots,$
$\Delta_{k-1}\prod_{i=k}^{\ntx}[\mathbf{R}]_{i,i}^{2}$, at the cost
of $2(\ntx-k+1)$ full multiplications. Then, the entries of~$\subcol{\mathbf{Q}}k{\ntx}$
and the entries of~$\subrow{\mathbf{R}}k{\ntx}$ above the main diagonal
of~$\mathbf{R}$ are scaled according to~(\ref{eq:q_k_map}) and~(\ref{eq:r_k_map}),
which requires $\nentries_{\ntx}-\nentries_{k-1}-(\ntx-k+1)$ full
multiplications. In summary, we obtain\[
\cmapk k=\nentries_{\ntx}-\nentries_{k-1}+\ntx-k+1,\qquad\alltwo k{\ntx}.\]

\paragraph*{Inverse mapping $\map^{-1}$}

We denote the overall cost of mapping $\qrparttilde 1k{}$ to $\qrpartnotilde 1k{}$
($\allone k{\ntx}$) by~$\cdemapk k$. Since $\Delta_{0}=1$ and
$[\tilde{\mathbf{R}}]_{1,1}=[\mathbf{R}]_{1,1}^{2}$, by first computing~$([\tilde{\mathbf{R}}]_{1,1})^{1/2}$
and then its inverse, we can obtain both~$[\mathbf{R}]_{1,1}$ and
the scaling factor $(\Delta_{0}[\mathbf{R}]_{1,1})^{-1}=1/[\mathbf{R}]_{1,1}$
at the cost of one square root operation and one division. For $\alltwo{k'}k$,
the scaling factors $(\Delta_{k'-1}[\mathbf{R}]_{k',k'})^{-1}$ can
be obtained according to~(\ref{eq:Delta_scalingfactorforinvmap})
by computing $([\tilde{\mathbf{R}}]_{k'-1,k'-1}[\tilde{\mathbf{R}}]_{k',k'})^{-1/2}$,
at the cost of $k-1$ full multiplications, $k-1$ square root operations,
and $k-1$ divisions. The entries of~$\subcol{\mathbf{Q}}1k$ and
the remaining entries of~$\subrow{\mathbf{R}}1k$ on and above the
main diagonal of~$\mathbf{R}$ are obtained by scaling the corresponding
entries of $\subcol{\tilde{\mathbf{Q}}}1k$ and~$\subrow{\tilde{\mathbf{R}}}1k$
according to~(\ref{eq:q_kdemap}) and~(\ref{eq:r_kdemap}), respectively,
at the cost of $\nentries_{k}-1$ full multiplications. Since we neglect
the impact of square root operations and divisions on complexity,
we obtain\[
\cdemapk k=\nentries_{k}+k-2,\qquad k=1,2,\ldots,\ntx.\]

\paragraph*{Reduction step}

Since matrix subtraction has negligible complexity, for a given $k\in\{1,2,\ldots,\ntx\}$,
the complexity associated with the computation of $\subcol{\mattf}k{\ntx}-\subcol{\mathbf{Q}}1{k-1}\subcolrow{\mathbf{R}}k{\ntx}1{k-1}$,
denoted by~$\credk k$, is given by the complexity associated with
the multiplication of the $\nrx\times(k-1)$ matrix~$\subcol{\mathbf{Q}}1{k-1}$
by the $(k-1)\times\bigpar{\ntx-k+1}$ matrix~$\subcolrow{\mathbf{R}}k{\ntx}1{k-1}$.
Hence, we obtain\[
\credk k=\nrx(k-1)\bigpar{\ntx-k+1}.\]

\subsection{Total Complexity of Algorithms I--III}

\label{sub:overallalgcosts}

\begin{table*}
\caption{Total complexity associated with the individual computational tasks\label{tbl:alg13taskcosts}}

\begin{centering}
\vspace{2mm}
\par\end{centering}

\begin{centering}
\small{\begin{tabular}{ccccc}
\vspace{1mm}Computational task & $\!\!$Symbol$\,^{\textrm{a}}$$\!\!$ & $\!\!$Algorithm~I$\!\!$ & Algorithm~II & Algorithm~III\tabularnewline
\hline 
\vspace{-2mm} &  &  &  & \tabularnewline
Interpolation of $\mattf(s)$ & $\ciphalg A$ & $\ndat\ciphk 1$ & $\nbase_{\ntx}\ciphk 1$ & ${\displaystyle \nbase_{1}\ciphk 1+2L\sum_{k=2}^{\ntx}\ciphk k}$\tabularnewline
$\!\!$Interpolation of $\tilde{\mathbf{Q}}(s)$ and~$\tilde{\mathbf{R}}(s)$$\!\!$ & $\!\!\cipqralg A\!\!$ & $0$ & $\!\!\bigpar{\ndat-\nbase_{\ntx}}\cipqr\!\!$ & ${\displaystyle \sum_{k=1}^{\ntx}\bigpar{\ndat-\nbase_{k}}\cipqrk k}$\tabularnewline
QR~decomposition & $\cqralg A$ & $\ndat\cqrk{\ntx}$ & $\nbase_{\ntx}\cqrk{\ntx}$ & $\!\!{\displaystyle \nbase_{1}\cqrk{\ntx}+2L\sum_{k=2}^{\ntx}\cqrk{(\ntx-k+1)}}\!\!$\tabularnewline
Mapping $\map$ & $\cmapalg A$ & $0$ & $\nbase_{\ntx}\cmapk 1$ & ${\displaystyle \nbase_{1}\cmapk 1+2L\sum_{k=2}^{\ntx}\cmapk k}$\tabularnewline
Inverse mapping $\map^{-1}$ & $\cdemapalg A$ & $0$ & $\!\!\bigpar{\ndat-\nbase_{\ntx}}\cdemapk{\ntx}\!\!$ & ${\displaystyle 2L\sum_{k=2}^{\ntx}\cdemapk{k-1}+\bigpar{\ndat-\nbase_{\ntx}}\cdemapk{\ntx}}$\tabularnewline
\vspace{1mm}Reduction & $\credalg A$ & $0$ & $0$ & ${\displaystyle 2L\sum_{k=2}^{\ntx}\credk k}$\tabularnewline
\hline
\end{tabular}}
\par\end{centering}

\begin{centering}
~
\par\end{centering}

\centering{}\small{$\!\,^{\textrm{a}}\,$The index $\textrm{A}$
is a placeholder for the algorithm number (I, II, or III).}
\end{table*}

The contribution of a given computational task to the overall complexity
of a given algorithm is obtained by multiplying the corresponding
per-tone complexity, computed in the previous section, by the number
of relevant tones. For simplicity of exposition, in the ensuing analysis
we restrict ourselves to the case where $\nbase_{k}=2kL+1$ ($\allone k{\ntx}$)
and $\setidxbase_{1}\subseteq\setidxbase_{2}\subseteq\ldots\subseteq\setidxbase_{\ntx}\subset\setdat$,
for which we obtain $|\setidxbase_{k}\backslash\setidxbase_{k-1}|=2L$
and $|\setdat\backslash\setidxbase_{k}|=\ndat-2kL-1$ ($\allone k{\ntx}$).
With the total complexity of the individual tasks summarized in Table~\ref{tbl:alg13taskcosts},
the complexity associated with Algorithms~I--III is trivially obtained
as \begin{align}
\calg I & =\ciphalg I+\cqralg I\label{eq:calgI}\\
\calg{II} & =\ciphalg{II}+\cipqralg{II}+\cqralg{II}+\cmapalg{II}+\cdemapalg{II}\label{eq:calgII}\\
\calg{III} & =\ciphalg{III}+\cipqralg{III}+\cqralg{III}+\cmapalg{III}+\cdemapalg{III}+\credalg{III}.\label{eq:calgIII}\end{align}

\subsection{Complexity Comparisons}

\label{sub:complcomparison}

In the following, we identify conditions on the system parameters
and on the interpolation cost~$\cip$ that guarantee that Algorithms~II
and~III exhibit smaller complexity than Algorithm~I. We start by
comparing Algorithms~I and~II and note that\begin{align}
\calg I-\calg{II} & =\left(\ndat-\nbase_{\ntx}\right)\biggl(\cqrk{\ntx}-\cdemapk{\ntx}-\frac{\ntx\left(\ntx+1\right)}{2}\cip\biggr)-\nbase_{\ntx}\cmapk 1.\label{eq:cIminuscII}\end{align}
Hence, if $\cip$ satisfies\begin{equation}
\cip<\cipmax{II}\triangleq\frac{2\left(\cqrk{\ntx}-\cdemapk{\ntx}\right)}{\ntx\left(\ntx+1\right)}\label{eq:cipmax2}\end{equation}
then there exists a $\ndat_{\mathrm{min}}$ such that $\calg{II}<\calg I$
for $\ndat\geq\ndat_{\mathrm{min}}$, i.e., Algorithm~II exhibits
a lower complexity than Algorithm~I for a sufficiently high number
of data-carrying tones~$\ndat$. Moreover, for $\cip<\cipmax{II}$,
increasing~$\nbase_{\ntx}$ reduces $\calg I-\calg{II}$. If the
inequality~(\ref{eq:cipmax2}) is met, (\ref{eq:cIminuscII}) implies,
since $\nbase_{\ntx}=2\ntx L+1$, that for increasing~$L$ and with
all other parameters fixed, Algorithm~II exhibits smaller savings.
For larger~$\cqrk{\ntx}\!$, again with all other parameters fixed,
Algorithm~II exhibits larger savings.

In order to compare Algorithms~II and~III, we start from~(\ref{eq:calgII})
and~(\ref{eq:calgIII}) and rewrite $\calg{II}-\calg{III}$ as \begin{equation}
\calg{II}-\calg{III}=\Delta\cqrtot+\Delta\cmaptot+\Delta\ciptot-\credalg{III}\label{eq:dcalg}\end{equation}
where we have introduced \begin{align*}
\Delta\cqrtot & \triangleq\cqralg{II}-\cqralg{III}\\
\Delta\cmaptot & \triangleq\cmapalg{II}+\cdemapalg{II}-\cmapalg{III}-\cdemapalg{III}\\
\Delta\ciptot & \triangleq\ciphalg{II}+\cipqralg{II}-\ciphalg{III}-\cipqralg{III}.\end{align*}
From the results in Table~\ref{tbl:alg13taskcosts} we get\begin{equation}
\Delta\cqrtot=2L\sum_{k=2}^{\ntx}\left(\cqrk{\ntx}-\cqrk{(\ntx-k+1)}\right)\label{eq:dcqr}\end{equation}
which is positive since, obviously, $\cqrk{\ntx}>\cqrk{(\ntx-k+1)}$
($\alltwo k{\ntx}$). Furthermore, again employing the results in
Table~\ref{tbl:alg13taskcosts}, straightforward calculations yield\begin{align}
\Delta\ciptot & =-2L\sum_{k=2}^{\ntx}k(k-1)\cip\nonumber \\
 & =-\frac{2}{3}L\ntx\bigpar{\ntxx-1}\cip\label{eq:dcipHQR}\end{align}
and\begin{align}
\Delta\cmaptot & =\bigpar{\nbase_{1}-\nbase_{\ntx}}\bigpar{\nrx-1}\nonumber \\
 & =-2L\bigpar{\nrx-1}\bigpar{\ntx-1}.\label{eq:dcmapdemap}\end{align}
We observe that (\ref{eq:dcalg})--(\ref{eq:dcmapdemap}), along with
the expression for~$\credalg{III}$ in Table~\ref{tbl:alg13taskcosts},
imply that $\calg{II}-\calg{III}$ does not depend on~$\ndat$ and
is proportional to~$L$. Moreover, it follows from~(\ref{eq:dcalg})
and~(\ref{eq:dcipHQR}) that $\calg{III}<\calg{II}$ is equivalent
to $\cip<\cipmax{III}$ with\begin{equation}
\cipmax{III}\triangleq\frac{\Delta\cqrtot+\Delta\cmaptot-\credalg{III}}{\frac{2}{3}L\ntx\bigpar{\ntxx-1}}.\label{eq:cipmax3}\end{equation}
We note that the RHS of~(\ref{eq:cipmax3}) depends solely on $\ntx$
and~$\nrx$, since $\Delta\cqrtot$, $\Delta\cmaptot$, and $\credalg{III}$
are proportional to~$L$. Hence, if $\Delta\cqrtot+\Delta\cmaptot-\credalg{III}>0$
and for $\cip$ sufficiently small, Algorithm~III has lower complexity
than Algorithm~II.

\section{The MMSE Case}

\label{sec:mmsecase}

In this section, we modify the QR~decomposition algorithms described
in Section~\ref{sec:algorithms} to obtain corresponding algorithms
that compute the MMSE-QR~decomposition, as defined in Section~\ref{sub:receivers},
of the channel matrices $\mattf\ofsn,n\in\setdat$. In Section~\ref{sub:regularizedqr},
we discuss the general concept of regularized QR~decomposition, of
which MMSE-QR decomposition is a special case. In Section~\ref{sub:mmsealgs},
we use the results of Section~\ref{sub:regularizedqr} to formulate
and analyze MMSE-QR~decomposition algorithms for MIMO-OFDM.

\subsection{Regularized QR~Decomposition}

\label{sub:regularizedqr}

In the following, we consider, as done in Section~\ref{sub:QRDecomposition},
a generic matrix $\mathbf{A}\in\mathbb{C}^{P\times M}\!$, with $P\geq M$.
\begin{defn}
The \emph{regularized QR~decomposition} of~\textbf{$\mathbf{A}$}
with the real-valued \emph{regularization parameter}~$\alpha>0$,
is the unique factorization $\mathbf{A}=\mathbf{QR}$, where the \emph{regularized
QR~factors} $\mathbf{Q}\in\mathbb{C}^{P\times M}$ and~$\mathbf{R}\in\mathbb{C}^{M\times M}$
are obtained as follows: $\stkmata=\stkmatq\mathbf{R}$ is the unique
QR~decomposition of the full-rank $(P+M)\times M$ augmented matrix
$\stkmata\triangleq\matcouple{\mathbf{A}^{T}}{\alpha\mathbf{I}_{M}}^{T}\!$,
and $\mathbf{Q}\triangleq\subrow{\stkmatq}1P\!$.
\end{defn}
In the following, we consider GS-based and UT-based algorithms for
computing the regularized QR~decomposition of~$\mathbf{A}$ through
the QR~decomposition of the augmented matrix~$\stkmata$. We will
see that both classes of algorithms exhibit higher complexity than
the corresponding algorithms for QR~decomposition of~$\mathbf{A}$
described in Section~\ref{sub:QRDecomposition}.

GS-based QR~decomposition of~$\stkmata$ produces $\mathbf{Q}$,
$\mathbf{R}$, and, as a by-product, the $M\times M$ matrix~$\subrow{\stkmatq}{P+1}{P+M}\!$.
Since GS-based QR~decomposition according to~(\ref{eq:y_k})--(\ref{eq:r_k})
operates on entire columns of the matrix to be decomposed, the computation
of~$\subrow{\stkmatq}{P+1}{P+M}$ can not be avoided. Thus, GS-based
regularized QR~decomposition of~$\mathbf{A}$ has the same complexity
as GS-based QR~decomposition of~$\stkmata$, which in turn has a
higher complexity than GS-based QR~decomposition of~$\mathbf{A}$.

Representing the UT-based QR~decomposition of~$\stkmata$ in the
standard form~(\ref{eq:utbasedqr}) yields\begin{align}
\mathbf{\Theta}_{U}\cdots\mathbf{\Theta}_{2}\mathbf{\Theta}_{1}\underbrace{\left[\begin{array}{ccc}
\mathbf{A} & \mathbf{I}_{P} & \mathbf{0}\\
\alpha\mathbf{I}_{M} & \mathbf{0} & \mathbf{I}_{M}\end{array}\right]}_{=\matcouple{\stkmata}{\mathbf{I}_{P+M}}} & =\left[\begin{array}{cc}
\mathbf{R} & \stkmatq^{H}\\
\mathbf{0} & (\stkmatq^{\perp})^{H}\end{array}\right]\label{eq:utbasedqrstkA}\end{align}
with the $(P+M)\times(P+M)$ unitary matrices $\mathbf{\Theta}_{u}$,
$\allone uU$, and where $\stkmatq^{\perp}$ is a $(P+M)\times P$
matrix satisfying $(\stkmatq^{\perp})^{H}\stkmatq^{\perp}=\mathbf{I}_{P}$
and $\stkmatq^{H}\stkmatq^{\perp}=\mathbf{0}$. By rewriting the RHS
of~(\ref{eq:utbasedqrstkA}) as \begin{align}
\left[\begin{array}{cc}
\mathbf{R} & \stkmatq^{H}\\
\mathbf{0} & (\stkmatq^{\perp})^{H}\end{array}\right] & =\left[\begin{array}{ccc}
\mathbf{R} & \mathbf{Q}^{H} & (\subrow{\stkmatq}{P+1}{P+M})^{H}\\
\mathbf{0} & (\subrow{(\stkmatq^{\perp})}1P)^{H} & (\subrow{(\stkmatq^{\perp})}{P+1}{P+M})^{H}\end{array}\right]\label{eq:utbasedqrstkArhs}\end{align}
we observe that UT-based regularized QR~decomposition of~$\mathbf{A}$
according to~(\ref{eq:utbasedqrstkA}), besides computing $\mathbf{R}$
and~$\mathbf{Q}^{H}\!$, yields the matrices $(\stkmatq^{\perp})^{H}$
and~$(\subrow{\stkmatq}{P+1}{P+M})^{H}$ as by-products. As observed
previously in\citet{burg05}, the corresponding complexity overhead
can not be eliminated completely, but it can be reduced by removing
the last~$M$ columns on both sides of~(\ref{eq:utbasedqrstkA}).
Thus, using~(\ref{eq:utbasedqrstkArhs}), we obtain the \emph{efficient
UT-based regularized QR~decomposition} described by the standard
form\begin{align}
\mathbf{\Theta}_{U}\cdots\mathbf{\Theta}_{2}\mathbf{\Theta}_{1}\left[\begin{array}{cc}
\mathbf{A} & \mathbf{I}_{P}\\
\alpha\mathbf{I}_{M} & \mathbf{0}\end{array}\right] & =\left[\begin{array}{cc}
\mathbf{R} & \mathbf{Q}^{H}\\
\mathbf{0} & (\subrow{(\stkmatq^{\perp})}1P)^{H}\end{array}\right]\label{eq:utbasedregqr}\end{align}
which yields only $(\subrow{(\stkmatq^{\perp})}1P)^{H}$ as a by-product\citet{burg05}.
We note that since the $P\times P$ matrix $(\subrow{(\stkmatq^{\perp})}1P)^{H}$
is larger than the $(P-M)\times P$ matrix $(\mathbf{Q}^{\perp})^{H}$
in~(\ref{eq:utbasedqr}), obtained as a by-product of UT-based QR~decomposition
of~$\mathbf{A}$, efficient UT-based regularized QR~decomposition
of~$\mathbf{A}$ exhibits higher complexity than UT-based QR~decomposition
of~$\mathbf{A}$.

Finally, we note that since $\mathbf{Q}=\subrow{\stkmatq}1P\!$, applying
the mapping~$\map$ to the regularized QR~factors $\mathbf{Q}$
and~$\mathbf{R}$ of~$\mathbf{A}$ according to (\ref{eq:Delta_k_map})--(\ref{eq:r_k_map})
is equivalent to applying~$\map$ to the QR~factors $\stkmatq$
and~$\mathbf{R}$ of~$\stkmata$ to obtain $\tilde{\stkmatq}$ and~$\tilde{\mathbf{R}}$
followed by extracting $\tilde{\mathbf{Q}}=\subrow{\tilde{\stkmatq}}1P\!$.
With this insight, it is straightforward to verify that Theorem~\ref{thm:qrinterp},
formulated for QR~decomposition of an LP~matrix~$\mathbf{A}(s)$,
is valid for regularized QR~decomposition of~$\mathbf{A}(s)$ as
well.

\subsection{Application to MIMO-OFDM MMSE-Based Detectors}

\label{sub:mmsealgs}

With the definition of regularized QR~decomposition in the previous
section, we recognize that MMSE-QR~decomposition of~$\mattf\ofsn$,
defined in Section~\ref{sub:receivers}, is a special case of regularized
QR~decomposition of~$\mattf\ofsn$ obtained by setting the regularization
parameter~$\alpha$ to~$\sqrt{\ntx}\stddevnoise$. The modification
of Algorithms I and~II to the MMSE case is straightforward and simply
amounts to replacing, in Step~\ref{enu:alg2qr} of both algorithms,
QR~decomposition by MMSE-QR decomposition. The resulting algorithms
are referred to as Algorithm~I-MMSE and Algorithm~II-MMSE, respectively.

In the following, we compare the complexity of Algorithm~I-MMSE and
Algorithm~II-MMSE. By denoting the complexity associated with computing
the MMSE-QR decomposition of an $\nrx\times\ntx$ matrix by~$\cmmseqr$,
the overall complexity of Algorithms I-MMSE and II-MMSE is given by\begin{equation}
\calg{I-MMSE}=\calg I+\ndat\left(\cmmseqr-\cqrk{\ntx}\right)\label{eq:calgIMMSE}\end{equation}
and\begin{equation}
\calg{II-MMSE}=\calg{II}+\nbase_{\ntx}\bigpar{\cmmseqr-\cqrk{\ntx}}\label{eq:calgIIMMSE}\end{equation}
respectively. Since $\cmmseqr>\cqrk{\ntx}\!$, as explained in Section~\ref{sub:regularizedqr},
(\ref{eq:calgIMMSE}) and~(\ref{eq:calgIIMMSE}) imply that $\calg{I-MMSE}>\calg I$
and $\calg{II-MMSE}>\calg{II}$, respectively. Thus, from~(\ref{eq:calgI}),
(\ref{eq:calgII}), (\ref{eq:calgIMMSE}), and~(\ref{eq:calgIIMMSE}),
we get\begin{align}
\frac{\calg{II-MMSE}}{\calg{II}} & =\frac{(\cmapalg{II}+\cipqralg{II}+\cdemapalg{II})+\nbase_{\ntx}\left(\ciphk 1+\cmmseqr\right)}{(\cmapalg{II}+\cipqralg{II}+\cdemapalg{II})+\nbase_{\ntx}\left(\ciphk 1+\cqrk{\ntx}\right)}\nonumber \\
 & <\frac{\ciphk 1+\cmmseqr}{\ciphk 1+\cqrk{\ntx}}\nonumber \\
 & =\frac{\calg{I-MMSE}}{\calg I}\label{eq:cI-II-MMSE}\end{align}
where the inequality follows from the simple property \[
\alpha>\beta>0,\gamma>0\implies\frac{\gamma+\alpha}{\gamma+\beta}<\frac{\alpha}{\beta}.\]
From~(\ref{eq:cI-II-MMSE}) we can therefore conclude that\[
\frac{\calg{II-MMSE}}{\calg{I-MMSE}}<\frac{\calg{II}}{\calg I}\]
which implies, assuming $\calg{II}<\calg I$, that the relative savings
of Algorithm~II-MMSE over Algorithm~I-MMSE are larger than the relative
savings of Algorithm~II over Algorithm~I.

Finally, we briefly discuss the extension of Algorithm~III to the
MMSE case. As a starting point, we consider the straightforward approach
of applying Algorithm~III to the MMSE-augmented channel matrix~$\stkmattf\ofsn$
in~(\ref{eq:Hmmseaugmented}) to produce $\stkmatq\ofsn$ and~$\mathbf{R}\ofsn$
for all $n\in\setdat$. In the following, we denote by $\tilde{\stkmatq}\ofsn$
and~$\tilde{\mathbf{R}}\ofsn$ the matrices resulting from the application
of the mapping~$\map$ to $\stkqrnotilde{\ofsn}$. We observe that
the straightforward approach under consideration is inefficient, since
we are only interested in obtaining $\mathbf{Q}\ofsn=\subrow{\stkmatq}1{\nrx}\ofsn$
and~$\mathbf{R}\ofsn$ for all $n\in\setdat$. Consequently, we would
like to avoid computing the last~$\ntx$ rows of~$\stkmatq\ofsn$
at as many tones as possible. Now, the reduction step (i.e., Step~\ref{enu:alg3red})
in the $k$th iteration of Algorithm~III requires knowledge of~$\subcol{\stkmatq}1{k-1}\ofsn$
at the tones $n\in\setidxbase_{k}\backslash\setidxbase_{k-1}$ ($\alltwo k{\ntx}$).
Hence, at the tones $n\in\setidxbase_{k}\backslash\setidxbase_{k-1}$
we must compute all $\nrx+\ntx$ rows of~$\subcol{\stkmatq}1{k-1}\ofsn$
anyway. In contrast, at the tones $n\in\setdat\backslash\setidxbase_{\ntx}$
the last~$\ntx$ rows of~$\stkmatq\ofsn$ are not required. Therefore,
at the tones $n\in\setdat\backslash\setidxbase_{\ntx}$ we can restrict
interpolation and inverse mapping to $\tilde{\mathbf{Q}}\ofsn=\subrow{\tilde{\stkmatq}}1{\nrx}\ofsn$
and~$\tilde{\mathbf{R}}\ofsn$. 

In the following, we partition~$\tilde{\stkvecq}_{k}\ofsn$, the~$k$th
column of~$\tilde{\stkmatq}\ofsn$, as\[
\tilde{\stkvecq}_{k}\ofsn=\left[\begin{array}{c}
\tilde{\mathbf{q}}_{k}\ofsn\\
\check{\mathbf{q}}_{k}\ofsn\end{array}\right]\!,\qquad\allone k{\ntx}\]
with the $\nrx\times1$ vector~$\tilde{\mathbf{q}}_{k}\ofsn$ and
the $\ntx\times1$ vector~$\check{\mathbf{q}}_{k}\ofsn$. With this
notation, we can formulate the resulting algorithm as follows:

\algobox{Algorithm~III-MMSE}{
\begin{enumerate}
\item Set $k\leftarrow1$.
\item Interpolate $\subcol{\mattf}k{\ntx}(s)$ from $\setidx{\setpil}$
to $\setidx{\setidxbase_{k}\backslash\setidxbase_{k-1}}$.\label{enu:alg3mmseloopstart}
\item For each $n\in\setidxbase_{k}\backslash\setidxbase_{k-1}$, construct
$\subcol{\stkmattf}k{\ntx}\ofsn$ according to~(\ref{eq:Hmmseaugmented}).
\item If $k=1$, go to Step~\ref{enu:alg3mmseqr}. Otherwise, for each
$n\in\setidxbase_{k}\backslash\setidxbase_{k-1}$, apply $\map^{-1}:\stkqrparttilde 1{k-1}{\ofsn}\mapsto\stkqrpartnotilde 1{k-1}{\ofsn}$.
\item For each $n\in\setidxbase_{k}\backslash\setidxbase_{k-1}$, overwrite
$\subcol{\stkmattf}k{\ntx}\ofsn$ by $\subcol{\stkmattf}k{\ntx}\ofsn-\subcol{\stkmatq}1{k-1}\ofsn\subcolrow{\mathbf{R}}k{\ntx}1{k-1}\ofsn$.\label{enu:alg3mmsered}
\item For each $n\in\setidxbase_{k}\backslash\setidxbase_{k-1}$, perform
QR~decomposition on $\subcol{\stkmattf}k{\ntx}\ofsn$ to obtain $\subcol{\stkmatq}k{\ntx}\ofsn$
and~$\subcolrow{\mathbf{R}}k{\ntx}k{\ntx}\ofsn$, and, if $k>1$,
construct $\subrow{\mathbf{R}}k{\ntx}\ofsn=\matcouple{\mathbf{0}}{\subcolrow{\mathbf{R}}k{\ntx}k{\ntx}\ofsn}.$\label{enu:alg3mmseqr}
\item For each $n\in\setidxbase_{k}\backslash\setidxbase_{k-1}$, apply
$\map:\stkqrpartnotilde k{\ntx}{\ofsn}\mapsto\stkqrparttilde k{\ntx}{\ofsn}$.%
\footnote{Since $\check{\mathbf{q}}_{\ntx}\ofsn$ is not needed, its computation
in the $\ntx$th iteration can be skipped. %
}
\item Interpolate $\tilde{\mathbf{q}}_{k}(s)$ and $\tilde{\mathbf{r}}_{k}^{T}(s)$
from~$\setidx{\setidxbase_{k}}$ to~$\setidx{\setdat\backslash\setidxbase_{k}}$.
\item If $k=\ntx$, proceed to Step~\ref{enu:alg3mmsedemap}. Otherwise,
interpolate~$\check{\mathbf{q}}_{k}(s)$ from~$\setidx{\setidxbase_{k}}$
to~$\setidx{\setidxbase_{\ntx}\backslash\setidxbase_{k}}$.
\item Set $k\leftarrow k+1$ and go back to Step~\ref{enu:alg3mmseloopstart}.
\item For each $n\in\setdat\backslash\setidxbase_{\ntx}$, apply $\map^{-1}:\qrtilde{\ofsn}\mapsto\qrnotilde{\ofsn}$.\label{enu:alg3mmsedemap}
\end{enumerate}
}

A detailed complexity analysis of Algorithm~III-MMSE goes beyond
the scope of this paper. We mention, however, the following important
aspect of the comparison of Algorithm~III-MMSE with Algorithms I-MMSE
and II-MMSE. Step~\ref{enu:alg2qr} of Algorithms~I-MMSE and~II-MMSE
requires MMSE-QR decomposition, which is a special case of regularized
QR~decomposition, whereas Step~\ref{enu:alg3mmseqr} of Algorithm~III-MMSE
requires QR~decomposition of an augmented matrix. As shown in Section~\ref{sub:regularizedqr},
the algorithms for regularized QR~decomposition and for QR~decomposition
of an augmented matrix have the same complexity under a GS-based approach,
but not under a UT-based approach. In the latter case, Algorithms~I-MMSE
and~II-MMSE can perform efficient UT-based regularized QR~decomposition
according to the standard form~(\ref{eq:utbasedregqr}), whereas
Algorithm~III-MMSE must perform UT-based QR~decomposition of an
augmented matrix according to the standard form~(\ref{eq:utbasedqrstkA}),
which results in higher complexity. This aspect does not occur in
the comparison of Algorithm~III with Algorithms I and~II and will
be further examined numerically in Section~\ref{sub:numcomparison}.

\section{Efficient Interpolation}

\label{sec:efficientinterpolation}

Throughout this section, we consider interpolation of a generic~LP
$\pol{a(s)}{\dg_{1}}{\dg_{2}}$ of maximum degree $\dg=\dg_{1}+\dg_{2}$
from $\setbase$ to $\settgt$, where $|\setbase|=\nbase$ and $|\settgt|=\ntgt$.
We note that in the context of interpolation in MIMO-OFDM systems,
relevant for the algorithms presented in this paper, all base points
and all target points correspond to OFDM tones. Therefore, in the
following we assume that $\setbase$ and~$\settgt$ satisfy the condition\begin{equation}
\setbase\cup\settgt\subseteq\{s_{0},s_{1},\ldots,s_{\ncar-1}\}.\label{eq:basetgtpointsaretones}\end{equation}

The complexity analysis in Section~\ref{sec:complexity} showed that
interpolation-based QR~decomposition algorithms yield savings over
the brute-force approach only if~$\cip$ is sufficiently small. Straightforward
interpolation of~$a(s)$, which corresponds to direct evaluation
of~(\ref{eq:lpinterp}), is performed by carrying out the multiplication
of the $\ntgt\times\nbase$ interpolation matrix~$\mattgt\matbase^{\dagger}$
by the $\nbase\times1$ vector~$\mathbf{a}_{\setbase}$. The corresponding
complexity is given by~$\ntgt\nbase$, which results in $\cip=\nbase$
full multiplications per target point. In the context of interpolation-based
QR decomposition, this complexity may be too high to get savings over
the brute-force approach in Algorithms~I or I-MMSE, since exact interpolation
of $\pol{\tilde{\mathbf{q}}_{k}(s)}{kL}{kL}$ and $\pol{\tilde{\mathbf{r}}_{k}^{T}(s)}{kL}{kL}$
requires $B\geq2kL+1$ ($\allone k{\ntx}$), with the worst case being
$\nbase\geq2\ntx L+1$. In this section, we present interpolation
methods characterized by significantly smaller values of~$\cip$.
As demonstrated by the numerical results in Section~\ref{sec:numresults},
this can then lead to significant savings of the interpolation-based
approaches for QR~decomposition over the brute-force approach.

\subsection{Interpolation with Dedicated Multipliers}

\label{sub:interpdedicatedmultipliers}

As already noted, the interpolation matrix $\mattgt\matbase^{\dagger}$
is a function of $\setbase$, $\settgt$, $\dg_{1}$ and~$\dg_{2}$,
but not of the realization of the LP~$a(s)$ to be interpolated.
Hence, as long as $\setbase$, $\settgt$, $\dg_{1}$ and~$\dg_{2}$
do not change, multiple LPs can be interpolated using the same interpolation
matrix $\mattgt\matbase^{\dagger}\!$, which can be computed off-line.
This observation leads to the first strategy for efficient interpolation,
which consists of carrying out the matrix-vector product $(\mattgt\matbase^{\dagger})\mathbf{a}_{\setbase}$
in~(\ref{eq:lpinterp}) through $\ntgt\nbase$ constant multiplications,
where the entries of~$\mattgt\matbase^{\dagger}$ are constant and
the entries of~$\mathbf{a}_{\setbase}$ are variable.

In the context of VLSI implementation, full multiplications and constant
multiplications differ significantly. Whereas a full multiplication
must be performed by a \emph{full multiplier} which processes two
variable operands, in a constant multiplication, the fact that one
of the operands, and more specifically its binary representation,
is known a priori, can be exploited to perform binary logic simplifications
that result in a drastically simpler circuit\citet{kaeslin08}. The
resulting multiplier, called a \emph{dedicated multiplier} in the
following, consumes only a fraction of the silicon area (down to~$1/9$,
as reported in\citet{haene06-07} for complex-valued dedicated multipliers)
required by a full multiplier, and exhibits the same processing delay.
Furthermore, we mention that it is possible to obtain further area
savings, again without affecting the processing delay, by merging~$K$
dedicated multipliers into a single block multiplier that jointly
performs the~$K$ multiplications, according to a technique known
as \emph{partial product sharing}\citet{lefevre01-05}, which essentially
exploits common bit patterns in the binary representations of the~$K$
coefficients to obtain circuit simplifications. For simplicity of
exposition, in the sequel we do not consider partial product sharing.

In the remainder of the paper, $\cmultc$ and~$\cmultr$ denote the
complexity associated with a constant multiplication of a complex-valued
variable operand by a complex-valued and by a real-valued constant
coefficient, respectively. Since~$\mattgt\matbase^{\dagger}$ is
real-valued for $\dg_{1}=\dg_{2}$ and complex-valued otherwise, interpolation
through constant multiplications with dedicated multipliers has a
complexity per target point of\[
\cip=\begin{cases}
\cmultr\nbase, & \quad\dg_{1}=\dg_{2}\\
\cmultc\nbase, & \quad\dg_{1}\neq\dg_{2}.\end{cases}\]
By leaving a cautionary implementation margin from the best-effort
value of~$1/9$ reported in\citet{haene06-07}, we assume that $\cmultc=1/4$
in the remainder of the paper. Since the multiplication of two complex-valued
numbers requires (assuming straightforward implementation) four real-valued
multiplications, whereas multiplying a real-valued number by a complex-valued
number requires only two real-valued multiplications, we henceforth
assume that $\cmultr=\cmultc/2$, which leads to $\cmultr=1/8$.

\subsection{Equidistant Base Points}

\label{sub:equidistantbasepoints}

In the following, we say that the points in a set $\{u_{0},u_{1},\ldots,u_{K-1}\}\subset\setunit$
are \emph{equidistant on $\setunit$} if $u_{k}=u_{0}e^{j2\pi k/K}$
for $\allone k{K-1}$. So far, we discussed interpolation of $\pol{a(s)}{\dg_{1}}{\dg_{2}}$
for generic sets $\setbase$ and~$\settgt$. In the remainder of
Section~\ref{sec:efficientinterpolation} we will, however, focus
on the following special case. Given integers $\nbase,\upratio>1$,
we consider the set of~$\nbase$ base points $\setbase=\{b_{k}=e^{j2\pi k/\nbase}:\allzero k{\nbase-1}\}$
and the set of $\ntgt=(\upratio-1)\nbase$ target points $\settgt=\{t_{(\upratio-1)k+r-1}=b_{k}e^{j2\pi r/(\upratio\nbase)}:\allzero k{\nbase-1},\allone r{\upratio-1}\}$.
We note that both the~$\nbase$ points in~$\setbase$ and the~$\upratio\nbase$
points in $\setbase\cup\settgt=\{e^{j2\pi l/(\upratio\nbase)}:\allzero l{\upratio\nbase}-1\}$
are equidistant on~$\setunit$. Hence, interpolation of~$a(s)$
from~$\setbase$ to~$\settgt$ essentially amounts to an $\upratio$-fold
increase in the sampling rate of~$a(s)$ on~$\setunit$, and will
therefore be termed \emph{upsampling of~$a(s)$ from~$\nbase$ equidistant
base points by a factor of~$\upratio$} in the remainder of the paper.
The corresponding base point matrix~$\matbase$ and target point
matrix~$\mattgt$ are constructed according to~(\ref{eq:matbase})
and~(\ref{eq:mattgt}), respectively. We note that for $\nbase\geq\dg+1$,
$\matbase$ satisfies $\matbase^{H}\matbase=\nbase\mathbf{I}_{\nbase}$
and hence $\matbase^{\dagger}=(1/\nbase)\matbase^{H}\!$.

We recall that the number of OFDM~tones~$\ncar$ is typically a
power of two. Therefore, in order to have~$\upratio\nbase$ equidistant
points on~$\setunit$ while satisfying the condition~(\ref{eq:basetgtpointsaretones}),
in the following we constrain both~$\nbase$ and~$\upratio$ to
be powers of two. Finally, in order to satisfy the condition $\nbase\geq\dg+1$
mandated by the requirement of exact interpolation, we set $\nbase=2^{\left\lceil \log(\dg+1)\right\rceil }$.

\subsection{Interpolation by Fast Fourier Transform}

\label{sub:interpifftfft}

In the context of upsampling from~$\nbase$ equidistant base points
by a factor of~$\upratio$, it is straightforward to verify that
the $\nbase\times(\dg+1)$ matrix~$\matbase$ is given by\begin{equation}
\matbase=\matcouple{\subcol{\bigpar{\matdft{\nbase}}}{\nbase-\dg_{1}+1}{\nbase}}{\subcol{\bigpar{\matdft{\nbase}}}1{\dg_{2}+1}}\label{eq:matBfft}\end{equation}
and that the $(\upratio-1)\nbase\times(\dg+1)$ matrix~$\mattgt$
is obtained by removing the rows with indices in $\mathcal{R}\triangleq\{1,\upratio+1,\ldots,(\nbase-1)\upratio+1\}$
from the $\upratio\nbase\times(\dg+1)$ matrix\begin{equation}
\bar{\mattgt}\triangleq\matcouple{\subcol{\bigpar{\matdft{\upratio\nbase}}}{\upratio\nbase-\dg_{1}+1}{\upratio\nbase}}{\subcol{\bigpar{\matdft{\upratio\nbase}}}1{\dg_{2}+1}}.\label{eq:matTbarfft}\end{equation}
As done in Section~\ref{sub:lpandinterp}, we consider the vectors
$\veccoeff=\hvecd{a_{-\dg_{1}}}{a_{-\dg_{1}+1}}{a_{\dg_{2}}}^{T}\!$,
$\mathbf{a}_{\setbase}=\matbase\veccoeff$, and $\mathbf{a}_{\settgt}=\mattgt\veccoeff$.
By defining the $\nbase$-dimensional vector $\mathbf{a}^{(\nbase)}\triangleq\fftinterpdatavector{a_{0}}{a_{1}}{a_{\dg_{2}}}{a_{-\dg_{1}}}{a_{-\dg_{1}+1}}{a_{-1}}^{T}\!$,
which contains $\nbase-(\dg+1)$ zeros between the entries $a_{\dg_{2}}$
and~$a_{-\dg_{1}}$, and by taking~(\ref{eq:matBfft}) into account,
we can write $\mathbf{a}_{\setbase}=\matbase\veccoeff=\matdft{\nbase}\mathbf{a}^{(\nbase)}\!$,
from which follows that $\mathbf{a}^{(\nbase)}=\matdft{\nbase}^{-1}\mathbf{a}_{\setbase}$.
Next, we insert $(\upratio-1)\nbase$ zeros into~$\mathbf{a}^{(\nbase)}$
after the entry~$a_{\dg_{2}}$ to obtain the $\upratio\nbase$-dimensional
vector $\mathbf{a}^{(\upratio\nbase)}\triangleq\fftinterpdatavector{a_{0}}{a_{1}}{a_{\dg_{2}}}{a_{-\dg_{1}}}{a_{-\dg_{1}+1}}{a_{-1}}^{T}\!$.
Further, we define $\mathbf{a}_{\setbase\cup\settgt}\triangleq\hvecd{a(e^{j0})}{a(e^{j2\pi/\upratio\nbase})}{a(e^{j2\pi(\upratio\nbase-1)/\upratio\nbase})}^{T}=\bar{\mattgt}\mathbf{a}$
to be the vector containing the samples of~$a(s)$ at the points
in $\setbase\cup\settgt$. We note that using~(\ref{eq:matTbarfft})
we can write \begin{equation}
\bar{\mattgt}\mathbf{a}=\matdft{\upratio\nbase}\mathbf{a}^{(\upratio\nbase)}.\label{eq:Tbara_FFT}\end{equation}
Next, we observe that by removing the rows with indices in~$\mathcal{R}$
from both sides of the equality $\mathbf{a}_{\setbase\cup\settgt}=\bar{\mattgt}\mathbf{a}$
we obtain the equality $\mathbf{a}_{\settgt}=\mattgt\veccoeff$. The
latter observation, combined with~(\ref{eq:Tbara_FFT}), implies
that~$\mathbf{a}_{\settgt}$ can be obtained by removing the rows
with indices in~$\mathcal{R}$ from the vector $\matdft{\upratio\nbase}\mathbf{a}^{(\upratio\nbase)}\!$.
Finally, we note that since $\nbase$ and~$\upratio\nbase$ are powers
of two, left-multiplication by $\matdft{\nbase}^{-1}$ and~$\matdft{\upratio\nbase}$
can be computed through a $\nbase$-point radix-2 inverse FFT (IFFT)
and an $\upratio\nbase$-point radix-2 FFT, respectively\citet{brigham74}.
We can therefore conclude that FFT-based interpolation of~$a(s)$
from~$\setbase$ to~$\settgt$ can be carried out as follows:
\begin{enumerate}
\item Compute the $\nbase$-point radix-2 IFFT $\mathbf{a}^{(\nbase)}=\matdft{\nbase}^{-1}\mathbf{a}_{\setbase}$.
\item Construct $\mathbf{a}^{(\upratio\nbase)}$ from~$\mathbf{a}^{(\nbase)}$
by inserting $(\upratio-1)\nbase$ zeros after the entry~$a_{\dg_{2}}$
in~$\mathbf{a}^{(\nbase)}$.
\item Compute the $\upratio\nbase$-point radix-2 FFT $\mathbf{a}_{\setbase\cup\settgt}=\matdft{\upratio\nbase}\mathbf{a}^{(\upratio\nbase)}\!$.
\item Extract $\mathbf{a}_{\settgt}$ from~$\mathbf{a}_{\setbase\cup\settgt}$
by removing the entries of~$\mathbf{a}_{\setbase\cup\settgt}$ with
indices in~$\mathcal{R}$.
\end{enumerate}
Now, we note that if generic radix-2 IFFT and FFT algorithms are used
in Steps 1 and~3, respectively, the approach described above does
not exploit the structure of the problem at hand and is inefficient
in the following three aspects. First, neither the IFFT in Step~1
nor the FFT in Step~3 take into account that $\nbase-(\dg+1)$ entries
of~$\mathbf{a}^{(\nbase)}$ (and also, by construction, of~$\mathbf{a}^{(\upratio\nbase)}$)
are zero. As this inefficiency does not arise in the case $\nbase=\dg+1$
and has only marginal impact on interpolation complexity otherwise,
we will not consider it further. Second, the FFT in Step~3 ignores
the fact that~$\mathbf{a}^{(\upratio\nbase)}$ contains the $(\upratio-1)\nbase$
zeros that were inserted in Step~2. Third, the values of~$a(s)$
at the base points, which are already known prior to interpolation,
are unnecessarily computed by the FFT in Step~3 and then discarded
in Step~4. In the following, we present a modified FFT algorithm,
tailored to the problem at hand, which eliminates the latter two inefficiencies
and leads to a significantly lower interpolation complexity than the
generic FFT-based interpolation method described above.

From now on, in order to simplify the notation, we assume that $\ncar=\upratio\nbase$.
Thus, with $s_{n}=e^{j2\pi n/N}\!$, $\allzero n{\ncar-1}$, the base
points and the target points are given by $b_{k}=s_{\upratio k}$
and $t_{(\upratio-1)k+r-1}=s_{\upratio k+r}$ ($\allzero k{\nbase-1}$,
$\allone r{\upratio-1}$), respectively. The derivation presented
in the following will be illustrated through an example obtained by
setting $\nbase=\upratio=4$ and $\dg_{1}=\dg_{2}+1=2$, but is valid
in general for the case where $\dg_{1}$ and~$\dg_{2}$ satisfy the
inequalities $0\leq\dg_{1}\leq\nbase/2$ and $0\leq\dg_{2}\leq\nbase/2-1$,
respectively. We note that these two inequalities, combined with $\nbase=2^{\left\lceil \log(\dg_{1}+\dg_{2}+1)\right\rceil }$,
are satisfied in the case $\dg_{1}=\dg_{2}$. Hence, the following
derivation covers the case of interpolation of the entries of $\pol{\tilde{\mathbf{Q}}(s)}{\ntx L}{\ntx L}$
and~$\pol{\tilde{\mathbf{R}}(s)}{\ntx L}{\ntx L}$, as required in
Algorithms II, III, II-MMSE and~III-MMSE.

\begin{figure}
\begin{centering}
\begin{tabular}{>{\centering}m{0.48\textwidth}>{\centering}m{0.48\textwidth}}
\centering{}\includegraphics{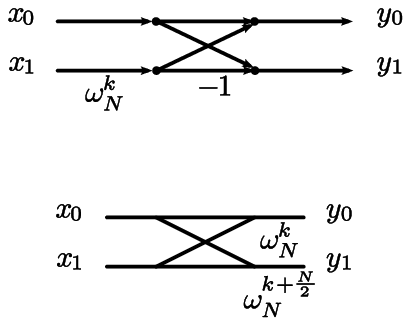}\\
~\\
~\\
\small{(with $\omega_{\ncar}^{k+\ncar/2}=-\omega_{\ncar}^{k}$)} & \centering{}\includegraphics{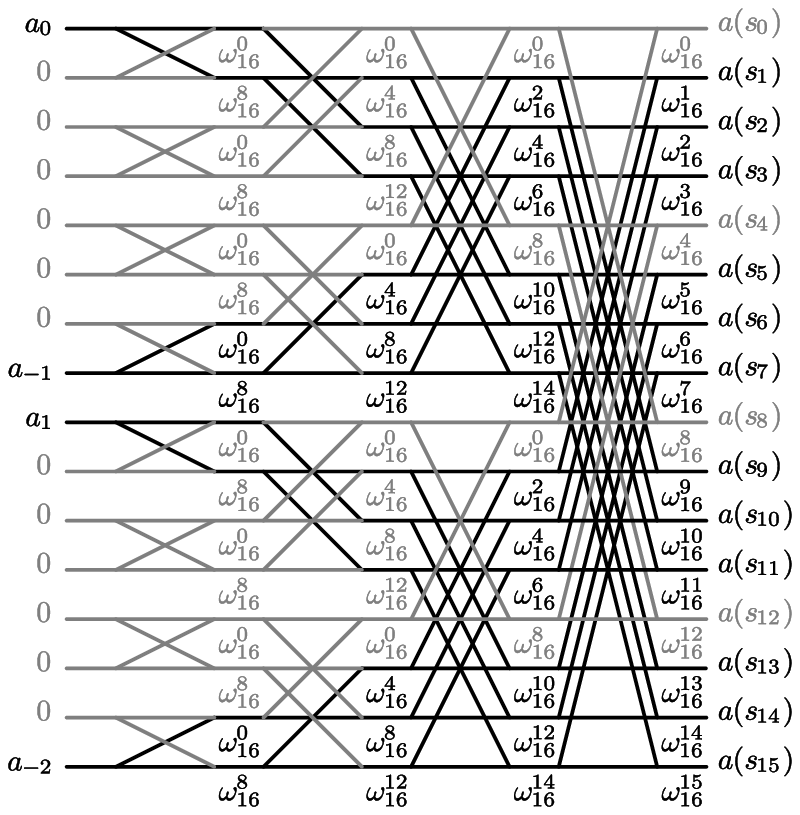}\tabularnewline
\small{(a)} & \small{(b)}\tabularnewline
\end{tabular}
\par\end{centering}

\caption{\label{fig:butterflyfullfft}(a) SFG of a radix-2 butterfly (top)
with twiddle factor $\omega_{\ncar}^{k}$, and alternative, equivalent
representation (bottom) needed for compact illustration in FFT SFGs.
(b) SFG of the full $\ncar$-point radix-2 decimation-in-time FFT,
without the scrambling stage. $\ncar=\upratio\nbase$, $\nbase=\upratio=4$,
$\dg_{1}=\dg_{2}+1=2$. SFG branches depicted in grey will be pruned.}

\end{figure}

The proposed modified FFT is based on a decimation-in-time radix-2
$\ncar$-point FFT, consisting of a scrambling stage followed by $\log\ncar$
computation stages\citet{brigham74}, each containing $\ncar/2$ radix-2
butterflies described by the signal flow graph~(SFG) in Fig.~\ref{fig:butterflyfullfft}a.
The twiddle factors used in the FFT butterflies are powers of $\omega_{\ncar}\triangleq e^{-j2\pi/\ncar}\!$.

\begin{figure}
\begin{centering}
\begin{tabular}{>{\centering}m{0.48\textwidth}>{\centering}m{0.48\textwidth}}
\centering{}\includegraphics{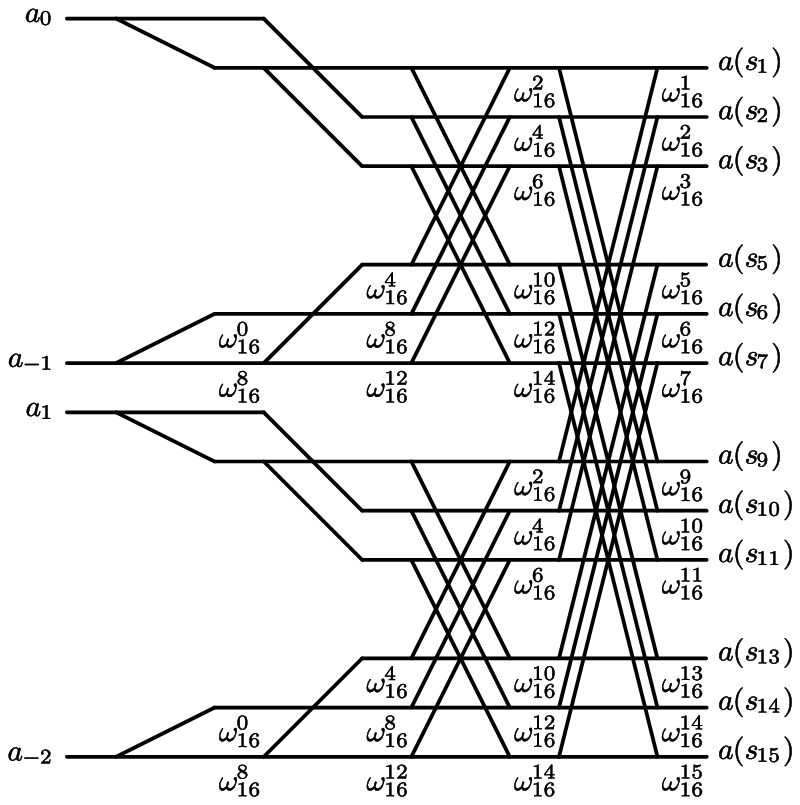} & \centering{}\includegraphics{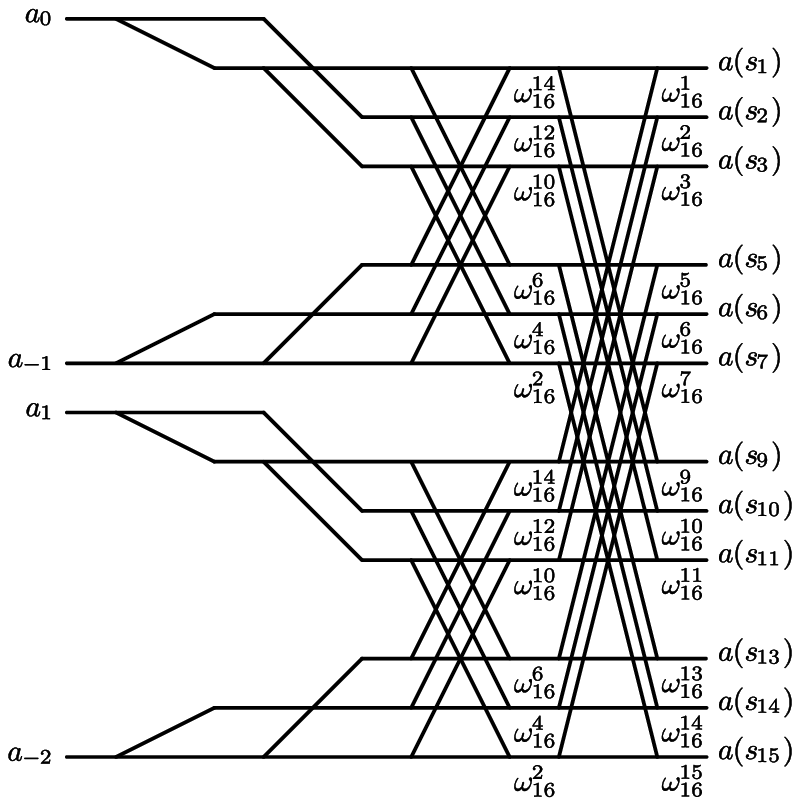}\tabularnewline
\small{(a)} & \small{(b)}\tabularnewline
\end{tabular}
\par\end{centering}

\caption{SFG of the pruned $\ncar$-point FFT, without the scrambling stage,
before (a) and after (b) shifting all multiplications from the first
$\log\upratio$ stages into stage $1+\log\upratio$. $\ncar=\upratio\nbase$,
$\nbase=\upratio=4$, $\dg_{1}=\dg_{2}+1=2$.\label{fig:prunedffts}}

\end{figure}

The SFG of the unmodified $\ncar$-point FFT is shown in Fig.~\ref{fig:butterflyfullfft}b.
We observe that the scrambling stage at the beginning of the FFT (not
depicted in Fig.~\ref{fig:butterflyfullfft}b) causes the nonzero
entries $a_{-\dg_{1}},a_{-\dg_{1}+1},\ldots,a_{\dg_{2}}$ of~$\mathbf{a}^{(\upratio\nbase)}$
to be scattered rather than to appear in blocks as is the case in~$\mathbf{a}^{(\upratio\nbase)}\!$.
The main idea of the proposed approach is to prune all SFG branches
that involve multiplications and additions with operands equal to
zero, as done in\citet{skinner76},%
\footnote{The SFG pruning approach proposed in\citet{skinner76} applies to
the case $\dg_{1}=0$ only.%
} and all SFG branches that lead to the computation of the already
known values of~$a(s)$ at the base points. The SFG of the resulting
pruned FFT is shown in Fig.~\ref{fig:prunedffts}a.

Further complexity reductions can be obtained as follows. We observe
that in the pruned FFT, the SFG branches departing from $a_{0},a_{1},\ldots,a_{\dg_{2}}$
contain no arithmetic operations in the first $\log\upratio$ computation
stages. In contrast, the SFG branches departing from $a_{-\dg_{1}},a_{-\dg_{1}+1},\ldots,a_{-1}$
contain multiplications by twiddle factors in each of the first $\log\upratio$
computation stages. These multiplications can however be shifted into
computation stage $1+\log\upratio$ through basic SFG transformations.
The result is the modified FFT illustrated in Fig.~\ref{fig:prunedffts}b,
for which the first $\log\upratio$ computation stages do not contain
any arithmetic operations and therefore have zero complexity, whereas
the last $\log\nbase$ computation stages contain $(\upratio-1)\nbase/2$
butterflies each. Thus, since each radix-2 butterfly entails one full
multiplication,%
\footnote{We assume that the FFT processor does not use any dedicated multipliers.%
} the total complexity of FFT-based interpolation of~$\mathbf{a}(s)$
from~$\setbase$ to~$\settgt$ is determined by the $(\nbase/2)\log\nbase$
full multiplications required by the $\nbase$-point radix-2 IFFT
$\mathbf{a}^{(\nbase)}=\matdft{\nbase}^{-1}\mathbf{a}_{\setbase}$
and the $(\upratio-1)(\nbase/2)\log\nbase$ full multiplications required
in the last $\log\nbase$ computation stages of the proposed modified
 $\upratio\nbase$-point FFT, which computes~$\mathbf{a}_{\settgt}$
from~$\mathbf{a}^{(\upratio\nbase)}\!$. The corresponding interpolation
complexity per target point is therefore given by\begin{equation}
\cipfft\triangleq\frac{\bigl(\frac{\nbase}{2}\log\nbase\bigr)+\bigl((\upratio-1)\frac{\nbase}{2}\log\nbase\bigr)}{(\upratio-1)\nbase}=\frac{1}{2}\frac{\upratio}{\upratio-1}\log\nbase.\label{eq:cipfft}\end{equation}
We mention that a modified $\upratio\nbase$-point FFT can be derived,
analogously to above, also in the case $\dg_{1}=0$ (for which $\dg=\dg_{2}$
and $\nbase=2^{\left\lceil \log(\dg_{2}+1)\right\rceil }$), relevant
for interpolation of~$\pol{\mattf(s)}0L$ in Algorithms I--III and
I-MMSE through III-MMSE. The corresponding interpolation complexity
per target point is again given by~(\ref{eq:cipfft}). 

Finally, we note that in MIMO-OFDM transceivers the FFT processor
that performs $\ncar$-point IFFT/FFT for OFDM modulation/demodulation
can be reused with slight modifications to carry out the $\nbase$-point
IFFT and the proposed modified $\upratio\nbase$-point FFT that are
needed for interpolation. Such a resource sharing approach reduces
the silicon area associated with interpolation and hence further reduces~$\cipfft$.
The resulting savings will, for the sake of generality of exposition,
not be taken into account in the following.

\subsection{Interpolation by FIR Filtering}

\label{sub:interpfirfiltering}

We consider upsampling of~$a(s)$ from~$\nbase$ equidistant base
points by a factor of~$\upratio$, as defined in Section~\ref{sub:equidistantbasepoints}.
The derivations in this section are valid for arbitrary integers $\nbase,\upratio>1$,
and hence not specific to the case where $\nbase$ and~$\upratio$
are powers of two.
\begin{prop}
\label{pro:TBdaggerproperties}In the context of upsampling from~$\nbase$
equidistant base points by a factor of~$\upratio$, the $\nbase(\upratio-1)\times\nbase$
interpolation matrix~$\mattgt\matbase^{\dagger}$ satisfies the following
properties:
\begin{enumerate}
\item \label{enu:TBdaggerrowcirculant}There exists an $(\upratio-1)\times\nbase$
matrix~$\mathbf{F}_{0}$ such that~$\mattgt\matbase^{\dagger}$
can be written as\begin{equation}
\mattgt\matbase^{\dagger}=\left[\begin{array}{c}
\mathbf{F}_{0}\matcirc{\nbase}{}\\
\mathbf{F}_{0}\matcirc{\nbase}2\\
\vdots\\
\mathbf{F}_{0}\matcirc{\nbase}{\nbase}\end{array}\right]\label{eq:TBdaggerasrowcirculant}\end{equation}
with the $\nbase\times\nbase$ circulant matrix\begin{align*}
\matcirc{\nbase}{} & \triangleq\left[\begin{array}{cc}
\mathbf{0} & \mathbf{I}_{\nbase-1}\\
1 & \mathbf{0}\end{array}\right]\!.\end{align*}

\item \label{enu:TBdaggerconjugatesimmetry}The matrix~$\mathbf{F}_{0}$,
as implicitly defined in~(\ref{eq:TBdaggerasrowcirculant}), satisfies\[
\bigbrk{\mathbf{F}_{0}}_{r,k+1}=\bigbrk{\mathbf{F}_{0}}_{\upratio-r,\nbase-k}^{*},\qquad\allone r{\upratio-1},\ \allzero k{\nbase-1}.\]

\end{enumerate}
\end{prop}
\begin{proof}
Since $\mathbf{B}^{\dagger}=(1/\nbase)\mathbf{B}^{H}\!$, the entries
of~$\mattgt\matbase^{\dagger}$ are given by \begin{eqnarray}
\bigbrk{\mattgt\matbase^{\dagger}}_{k(\upratio-1)+r,k'+1} & = & \frac{1}{\nbase}\sum_{v=-\dg_{1}}^{\dg_{2}}e^{-j2\pi v\frac{\upratio(k-k')+r}{\upratio\nbase}}\label{eq:lpinterpentries}\end{eqnarray}
for $\allzero{k,k'}{\nbase-1}$ and $\allone r{\upratio-1}$. The
two properties are now established as follows:
\begin{enumerate}
\item The RHS of~(\ref{eq:lpinterpentries}) remains unchanged upon replacing
$k$ and~$k'$ by $(k+1)\modulo\nbase$ and $(k'+1)\modulo\nbase$,
respectively. Hence, for a given $r\in\{1,2,\ldots,\upratio-1\}$,
the $\nbase\times\nbase$ matrix obtained by stacking the rows indexed
by $r,(\upratio-1)+r,\ldots,(\nbase-1)(\upratio-1)+r$ (in this order)
of~$\mattgt\matbase^{\dagger}$ is circulant. By taking~$\mathbf{F}_{0}$
to consist of the last $\upratio-1$ rows of~$\mattgt\matbase^{\dagger}\!$,
and using $\matcirc{\nbase}{\nbase}=\mathbf{I}_{\nbase}$, along with
the fact that for $b\in\mathbb{Z}$, the multiplication $\mathbf{F}_{0}\matcirc{\nbase}b$
corresponds to circularly shifting the columns of~$\mathbf{F}_{0}$
to the right by $b\modulo\nbase$ positions, we obtain~(\ref{eq:TBdaggerasrowcirculant}).
\item The entries of~$\mathbf{F}_{0}$ are obtained by setting $k=\nbase-1$
in~(\ref{eq:lpinterpentries}) and are given by \begin{align*}
[\mathbf{F}_{0}]_{r,k'+1} & =\frac{1}{\nbase}\sum_{v=-\dg_{1}}^{\dg_{2}}e^{-j2\pi v\frac{r-\upratio(k'+1)}{\upratio\nbase}},\qquad\allone r{\upratio-1},\mbox{ }\allzero{k'}{\nbase-1}.\end{align*}
Hence, for $\allone r{\upratio-1}$ and $\allzero{k'}{\nbase-1}$,
we obtain\begin{align*}
[\mathbf{F}_{0}]_{\upratio-r,\nbase-k'}^{*} & =\frac{1}{\nbase}\sum_{v=-\dg_{1}}^{\dg_{2}}e^{j2\pi v\frac{\upratio-r-\upratio(\nbase-k')}{\upratio\nbase}}\\
 & =\frac{1}{\nbase}\sum_{v=-\dg_{1}}^{\dg_{2}}e^{-j2\pi v\frac{r-\upratio(k'+1)}{\upratio\nbase}}\\
 & =[\mathbf{F}_{0}]_{r,k'+1}.\end{align*}

\end{enumerate}
\end{proof}
We note that Property~\ref{enu:TBdaggerrowcirculant} in Proposition~\ref{pro:TBdaggerproperties}
implies that the matrix-vector multiplication $(\mattgt\matbase^{\dagger})\mathbf{a}_{\setbase}$
in~(\ref{eq:lpinterp}) can be carried out through the application
of $\upratio-1$ FIR filters. Specifically, for $\allone r{\upratio-1}$,
the entries $r,r+R,\dots,r+(\nbase-1)\upratio$ of~$\mathbf{a}_{\settgt}$
can be obtained by computing the circular convolution of~$\mathbf{a}_{\setbase}$
with the impulse response of length~$\nbase$ contained in the $r$th
row of~$\mathbf{F}_{0}$. In the remainder of the paper, we will
say that the $\upratio-1$ FIR filters are \emph{defined} by~$\mathbf{F}_{0}$.
By allocating~$\nbase$ dedicated multipliers per FIR filter (one
per impulse response tap), we would need a total of $(\upratio-1)\nbase$
dedicated multipliers. We will next see that the complex-conjugate
symmetry in the rows of~$\mathbf{F}_{0}$, formulated as Property~\ref{enu:TBdaggerconjugatesimmetry}
in Proposition~\ref{pro:TBdaggerproperties}, allows to reduce the
number of dedicated multipliers and the interpolation complexity by
a factor of two.

In the following, we assume that the multiplications of a variable
complex-valued operand by a constant $\gamma\in\mathbb{C}$ and by
its complex conjugate~$\gamma^{*}$ can be carried out using the
same dedicated multiplier, and that the resulting complexity is comparable
to the complexity of multiplication by~$\gamma$ alone. This is justified
as the multiplication by~$\gamma^{*}\!$, compared to the multiplication
by~$\gamma$, involves the same four underlying real-valued multiplications
and only requires two additional sign flips, which have significantly
smaller complexity than the real-valued multiplications. Thus, we
can perform multiplication by the coefficients~$\bigbrk{\mathbf{F}_{0}}_{r,k+1}$
and $\bigbrk{\mathbf{F}_{0}}_{\upratio-r,\nbase-k}=\bigbrk{\mathbf{F}_{0}}_{r,k+1}^{*}$
through a single dedicated multiplier ($\allone r{\upratio/2}$, $\allzero k{\nbase/2-1}$).
This resource sharing approach leads to\begin{equation}
\cip=\begin{cases}
\frac{\cmultr}{2}\nbase, & \quad\dg_{1}=\dg_{2}\\
\frac{\cmultc}{2}\nbase, & \quad\dg_{1}\neq\dg_{2}.\end{cases}\label{eq:cipfirfilteringB}\end{equation}

So far, we assumed that~$a(s)$ is interpolated from the~$\nbase=2^{\left\lceil \log(\dg+1)\right\rceil }$
base points in~$\setbase$, resulting in~$\cip$ according to~(\ref{eq:cipfirfilteringB}).
We will next show that the interpolation complexity can be further
reduced by using a smaller number of base points $\nbase'<\nbase$.
Interpolation will be exact as long as the condition $\nbase'\geq\dg+1$
is satisfied.

As done above, we assume knowledge of the~$\nbase$ samples $a(s),s\in\setbase$.
In the following, however, we require that for a given target point~$t_{r}$,
the sample~$a(t_{r})$ is obtained by interpolation from only~$\nbase'$
base points, picked from the $\nbase$ elements of~$\setbase$ as
a function of~$t_{r}$. For simplicity of exposition, we assume that~$\nbase'$
is even, and for every $t_{r}\in\settgt$ we choose the~$\nbase'$
elements of~$\setbase$ that are located closest to~$t_{r}$ on~$\setunit$.
We will next show that the resulting interpolation of~$a(s)$ from~$\setbase$
to~$\settgt$ can be performed through FIR~filtering.

In the following, we define~$\nbase$ disjoint subsets~$\settgt_{k}$
of~$\settgt$ (satisfying $\settgt_{0}\cup\settgt_{1}\cup\ldots\cup\settgt_{\nbase-1}=\settgt$)
and consider the corresponding subsets $\setbase_{k}$ of~$\setbase$,
defined such that for all points in~$\settgt_{k}$, the~$\nbase'$
closest base points are given by the elements of~$\setbase_{k}$
($\allzero k{\nbase-1}$). We next show that the interpolation matrix
corresponding to interpolation of~$a(s)$ from~$\setbase_{k}$ to~$\settgt_{k}$
is independent of~$k$. To this end, we first consider the set of
target points $\settgt_{0}\triangleq\{t_{(\nbase-1)(\upratio-1)+r-1}\!:\allone r{\upratio-1}\}$,
containing the $\upratio-1$ target points located on~$\setunit$
between the base points $b_{\nbase-1}$ and~$b_{0}$. The subset
of~$\setbase$ containing the~$B'$ points that are closest to every
point in~$\settgt_{0}$ is given by $\setbase_{0}\triangleq\{b_{0},b_{1},\ldots,b_{\nbase'/2}$,
$b_{\nbase-\nbase'/2},b_{\nbase-\nbase'/2+1},\ldots,b_{\nbase-1}\}$.
Interpolation of~$a(s)$ from~$\setbase_{0}$ to~$\settgt_{0}$
involves the base point matrix~$\matbase_{0}$, the target point
matrix~$\mattgt_{0}$, and the interpolation matrix~$\mattgt_{0}\matbase_{0}^{\dagger}$,
constructed as described in Section~\ref{sub:lpandinterp}. Next,
for $\allone k{\nbase-1}$, we denote by $\setbase_{k}$ and~$\settgt_{k}$
the sets obtained by multiplying all elements of $\setbase_{0}$ and~$\settgt_{0}$,
respectively, by~$e^{j2\pi k/\nbase}\!$. We note that~$\settgt_{k}$
contains the $\upratio-1$ target points located on~$\setunit$ between
the base points $b_{k-1}$ and~$b_{k}$, and that~$\setbase_{k}$
is the subset of~$\setbase$ containing the~$\nbase'$ points that
are closest to every point in~$\settgt_{k}$. With the unitary matrix
$\mathbf{S}_{k}\triangleq\diag((e^{j2\pi k/B})^{\dg_{1}},(e^{j2\pi k/B})^{\dg_{1}-1},\ldots,(e^{j2\pi k/B})^{-\dg_{2}})$,
interpolation of~$a(s)$ from $\setbase_{k}$ to~$\settgt_{k}$
involves the base point matrix $\matbase_{k}=\matbase_{0}\mathbf{S}_{k}$,
with pseudoinverse $\matbase_{k}^{\dagger}=\mathbf{S}_{k}^{-1}\matbase_{0}^{\dagger}$,
the target point matrix $\mattgt_{k}=\mattgt_{0}\mathbf{S}_{k}$,
and the interpolation matrix $\mattgt_{k}\matbase_{k}^{\dagger}=\mattgt_{0}\mathbf{S}_{k}\mathbf{S}_{k}^{-1}\matbase_{0}^{\dagger}=\mattgt_{0}\matbase_{0}^{\dagger}$
($\allone k{\nbase-1}$). Hence, the interpolation matrix is independent
of~$k$ and is the same as in the interpolation of~$a(s)$ from~$\setbase_{0}$
to~$\settgt_{0}$.

Now, interpolation of~$a(s)$ from~$\setbase$ to~$\settgt$, with
the constraint that the sample of~$a(s)$ at every target point is
computed only from the samples of~$a(s)$ at the~$\nbase'$ closest
base points, amounts to performing interpolation of~$a(s)$ from~$\setbase_{k}$
to~$\settgt_{k}$ for all $\allzero k{\nbase-1}$, and can be written
in a single equation as $\mathbf{a}_{\settgt}=\mathbf{F}\mathbf{a}_{\setbase}$.
Here, the $(\upratio-1)\nbase\times\nbase$ interpolation matrix~$\mathbf{F}$
is equal to the RHS of~(\ref{eq:TBdaggerasrowcirculant}), with the
$(\upratio-1)\times\nbase$ matrix\begin{align}
\mathbf{F}_{0} & =\matthreesome{\subcol{(\mattgt_{0}\matbase_{0}^{\dagger})}1{\nbase'/2}}{\mathbf{0}}{\subcol{(\mattgt_{0}\matbase_{0}^{\dagger})}{\nbase-\nbase'/2+1}{\nbase}}\label{eq:F0firfiltering}\end{align}
which contains an all-zero submatrix of dimension $(\upratio-1)\times(\nbase-\nbase')$.
Hence, $\mathbf{F}$ satisfies Property~\ref{enu:TBdaggerrowcirculant}
of Proposition~\ref{pro:TBdaggerproperties}, with~$\mathbf{F}_{0}$
given by~(\ref{eq:F0firfiltering}). In addition, we state without
proof that~$\mathbf{F}_{0}$ in~(\ref{eq:F0firfiltering}) satisfies
Property~\ref{enu:TBdaggerconjugatesimmetry} of Proposition~\ref{pro:TBdaggerproperties}.
We can therefore conclude that interpolation from the closest~$\nbase'$
base points maintains the structural properties of interpolation from
all~$\nbase$ base points and, as above, can be performed by FIR~filtering
using $\upratio-1$ filters with dedicated multipliers that exploit
the conjugate symmetry in the rows of~$\mathbf{F}_{0}$. Since the
rows of~$\mathbf{F}_{0}$ in~(\ref{eq:F0firfiltering}) contain
$\nbase-\nbase'$ zeros, the $\upratio-1$ impulse responses now have
length~$\nbase'$, and we obtain\begin{equation}
\cip=\begin{cases}
\frac{\cmultr}{2}\nbase', & \quad\dg_{1}=\dg_{2}\\
\frac{\cmultc}{2}\nbase', & \quad\dg_{1}\neq\dg_{2}.\end{cases}\label{eq:cipfirfilteringBprime}\end{equation}

\subsection{Inexact Interpolation}

\label{sub:inexactinterp}

The interpolation complexity~(\ref{eq:cipfirfilteringBprime}) of
the approach described in Section~\ref{sub:interpfirfiltering} can
be further reduced by choosing~$\nbase'$ to be smaller than $\dg+1$.
This comes, however, at the cost of a systematic interpolation error
and consequently leads to a trade-off between interpolation complexity
and interpolation accuracy. In the context of MIMO-OFDM detectors,
it is demonstrated in Section~\ref{sub:numperfdegradation} that
the performance degradation resulting from this systematic interpolation
error is often negligible. In the following, we propose an ad-hoc
method for inexact interpolation. The basic idea consists of introducing
an interpolation error metric and formulating a corresponding optimization
problem, which yields the matrix~$\mathbf{F}_{0}$ that defines the
FIR~filters for inexact interpolation.

For simplicity of exposition, we restrict our discussion to inexact
interpolation of $\pol{\tilde{\mathbf{Q}}(s)}{\ntx L}{\ntx L}$ and
$\pol{\tilde{\mathbf{R}}(s)}{\ntx L}{\ntx L}$ with $\dg_{1}=\dg_{2}=\ntx L$,
as required in Step~\ref{enu:alg2interpQR} of Algorithm~II. For
random-valued MIMO channel taps $\mattap_{0},\mattap_{1},\ldots,\mattap_{L}$,
we propose to quantify the interpolation error according to\begin{equation}
e(\mathbf{F}_{0})\triangleq\expval\left[\sum_{n\in\setdat\backslash\setidxbase_{\ntx}}\|\mathbf{Q}^{H}\ofsn\mattf\ofsn-\mathbf{R}\ofsn\|_{2}^{2}\right]\label{eq:iperrormetric}\end{equation}
where the expectation is taken over $\mattap_{0},\mattap_{1},\ldots,\mattap_{L}$,
and where the dependence of the RHS of~(\ref{eq:iperrormetric})
on~$\mathbf{F}_{0}$ is implicit through the fact that within Algorithm~II,
the computation of~$\mathbf{Q}\ofsn$ and~$\mathbf{R}\ofsn$ at
the tones $n\in\setdat\backslash\setidxbase_{\ntx}$ involves interpolation
through the FIR~filters defined by~$\mathbf{F}_{0}$. We mention
that the metric~$e(\mathbf{F}_{0})$ in~(\ref{eq:iperrormetric})
is relevant for MIMO-OFDM sphere decoding, and that minimization of~$e(\mathbf{F}_{0})$
does not necessarily lead to optimal detection performance. Other
applications involving QR~decomposition of polynomial matrices may
require alternative error metrics.

For upsampling from~$\nbase$ equidistant base points by a factor
of~$\upratio$, under the condition $\dg_{1}=\dg_{2}$, the matrix~$\mathbf{F}_{0}$
in~(\ref{eq:F0firfiltering}) is a function of $\ncar,\upratio,\nbase,\nbase'$,
and~$\dg_{1}$. Now, we have that~$\ncar$ is a fixed system parameter
and $\nbase=2^{\left\lceil \log(2\ntx L+1)\right\rceil }$. Moreover,
$\upratio$ is determined by~$\ncar$, $\nbase$, and~$\setdat$,
since~$\upratio$ is either given by  $\upratio=\ncar/\nbase$ in
the case $|\setdat|=\ncar$ or is a function of~$\nbase$ and~$\setdat$
in the case $|\setdat|<\ncar$. Finally, under a fixed complexity
budget (i.e., a given value for~$\cip$), $\nbase'$ is constrained
by~(\ref{eq:cipfirfilteringBprime}). Now, $\pol{\tilde{\mathbf{Q}}(s),\tilde{\mathbf{R}}(s)}{\ntx L}{\ntx L}$
determines $\dg_{1}=\ntx L$, but we propose, instead, to consider~$\dg_{1}$
as a variable parameter, so that $\mathbf{F}_{0}=\mathbf{F}_{0}(\dg_{1})$.
The interpolation error is then minimized by first determining\[
\dg_{1}'\triangleq\argmin{\dg_{1}\in\{1,2,\ldots,\ntx L\}}e(\mathbf{F}_{0}(\dg_{1}))\]
numerically, and then performing interpolation through the FIR filters
defined by~$\mathbf{F}_{0}(\dg_{1}')$.

\section{Numerical Results}

\label{sec:numresults}

The results presented so far do not depend on a specific QR~decomposition
method. For the numerical complexity comparisons presented in this
section, we will get more specific and assume UT-based QR~decomposition
performed through Givens rotations and coordinate rotation digital
computer~(CORDIC) operations\citet{volder59,walther00}, which is
the method of choice in VLSI implementations\citet{burg05,lightbody03}.
For $\mathbf{A}\in\mathbb{C}^{P\times M}$ with $P\geq M$, it was
shown in\citet{burg05} that the complexity of UT-based QR~decomposition
of~$\mathbf{A}$ according to the standard form~(\ref{eq:utbasedqr}),
as required in Algorithms~I--III, is given by\begin{align*}
\cqrpm & \triangleq\frac{3}{2}(P^{2}M+PM^{2})-M^{3}-\frac{1}{2}(P^{2}-P+M^{2}+M)\end{align*}
and that the complexity of efficient UT-based regularized QR~decomposition
of~$\mathbf{A}$ according to the standard form~(\ref{eq:utbasedregqr}),
as required in Algorithms I-MMSE and~II-MMSE, is given%
\footnote{In\citet{burg05}, the last term on the RHS of~(\ref{eq:cmmseqrgivens})
was erroneously specified as $-(1/2)P$.%
} by \begin{equation}
\cmmseqrpm\triangleq\frac{3}{2}(P^{2}M+PM^{2})-\frac{1}{2}P^{2}+\frac{1}{2}P.\label{eq:cmmseqrgivens}\end{equation}
The results in\citet{burg05} carry over, in a straightforward fashion,
to UT-based QR~decomposition of the augmented matrix $\matcouple{\mathbf{A}^{T}}{\alpha\mathbf{I}_{M}}^{T}$
according to the standard form~(\ref{eq:utbasedqrstkA}), as required
in Algorithm~III-MMSE, to yield\[
\cthreemmseqrpm\triangleq\cmmseqrpm+\frac{3}{2}PM^{2}+\frac{1}{2}PM.\]

\subsection{Efficient Interpolation and Performance Degradation}

\label{sub:numperfdegradation}

We start by quantifying the trade-off between interpolation complexity
and detection performance, described in Section~\ref{sub:inexactinterp}.
Specifically, we evaluate the loss in detection performance as we
gradually reduce~$\nbase'$, and hence also~$\cip$, in the interpolation
of~$\tilde{\mathbf{Q}}(s)$ and~$\tilde{\mathbf{R}}(s)$, as required
by Algorithm~II. The corresponding analysis for the interpolation
of $\tilde{\mathbf{q}}_{k}(s)$ and~$\tilde{\mathbf{r}}_{k}^{T}(s)$,
$\allone k{\ntx}$, as required by Algorithm~III, is more involved
and does not yield any additional insight into the trade-off under
consideration. The numerical results presented in the following demonstrate
that for Algorithm~II to have smaller complexity than Algorithm~I,
setting~$\nbase'$ to a value smaller than $\dg+1$, and hence accepting
a systematic interpolation error, may be necessary. On the other hand,
 we will also see that the resulting performance degradation, in terms
of both coded and uncoded bit error rate (BER), can be negligible
even for values of~$\nbase'$ that are significantly smaller than
$\dg+1$.

In the following, we consider a MIMO-OFDM system with $\ndat=\ncar=512$,
$\nrx=4$, and either $\ntx=2$ or $\ntx=4$, operating over a frequency-selective
channel with $L=15$. The data symbols are drawn from a 16-QAM constellation.
In the coded case, a rate~$1/2$ convolutional code with constraint
length~$7$ and generator polynomials $[133_{o}\,171_{o}]$ is used.
The receiver performs maximum-likelihood detection through hard-output
sphere decoding. Our results are obtained through Monte Carlo simulation,
where averaging is performed over the channel impulse response taps
$\mattap_{0},\mattap_{1},\ldots,\mattap_{L}$ assumed i.i.d. $\mathcal{CN}(0,1/(L+1))$.
This assumption on the channel statistics, along with the average
transmit power being given by $\expval[\vectx_{n}^{H}\vectx_{n}]=1$
and the noise variance~$\varnoise$, implies that the per-antenna
receive signal-to-noise ratio~(SNR) is~$1/\sigma_{w}^{2}$. The
receiver employs either Algorithm~I or Algorithm~II to compute $\mathbf{Q}\ofsn$
and~$\mathbf{R}\ofsn$ at all tones. We assume that in Step~\ref{enu:alg2interpH}
of both algorithms, $\pol{\mattf(s)}0L$ is interpolated exactly from
$\nbase=L+1=16$ equidistant base points by FIR~filtering. Since
$0=\dg_{1}\neq\dg_{2}=L$, the corresponding interpolation complexity
per target point is obtained from~(\ref{eq:cipfirfilteringB}) as
$\cipH\triangleq(L+1)\cmultc/2$. With $\cmultc=1/4$, as assumed
in Section~\ref{sub:interpdedicatedmultipliers}, we get%
\footnote{Performing interpolation of~$\mattf(s)$ by FFT would lead to $\cipH$
according to~(\ref{eq:cipfft}), which with $\nbase=16$ and $\upratio=\ncar/\nbase=32$
results in $\cipH=64/31\approx2.06$. Hence, in this case interpolation
of~$\mattf(s)$ by FIR filtering and by FFT have comparable complexity.%
} $\cipH=2$. In Step~\ref{enu:alg2interpQR} of Algorithm~II, we
interpolate $\pol{\tilde{\mathbf{Q}}(s)}{\ntx L}{\ntx L}$ and $\pol{\tilde{\mathbf{R}}(s)}{\ntx L}{\ntx L}$,
with maximum degree $\dg=2\ntx L$, through FIR filtering from $\nbase'\leq\nbase=2^{\left\lceil \log(\dg+1)\right\rceil }$
base points. With $\dg_{1}=\dg_{2}=\ntx L$, the corresponding interpolation
complexity per target point is obtained from~(\ref{eq:cipfirfilteringBprime})
as $\cipqr\triangleq\cmultr\nbase'/2$ with $\cmultr=1/8$, as assumed
in Section~\ref{sub:interpdedicatedmultipliers}. We ensure that
systematic interpolation errors are the sole source of detection performance
degradation by performing all computations in double-precision floating-point
arithmetic. Under inexact interpolation, for every value of $\nbase'<\dg+1$
we determine the value of~$\dg_{1}'$ that minimizes the interpolation
error~$e(\mathbf{F}_{0})$ in~(\ref{eq:iperrormetric}) according
to the procedure described in Section~\ref{sub:inexactinterp}.

\begin{table}
\caption{Simulation parameters\label{tbl:simparams}}

\begin{centering}
\vspace{2mm}
\par\end{centering}

\begin{centering}
\small{\begin{tabular}{cccccc}
$\ntx$ & $\nbase'$ & $\dg_{1}'$ & $\cipqr$ & $\calg{II}/\calg I$ & Interpolation method\vspace{0.5ex}\tabularnewline
\hline
\hline 
$2$ & $64$ & $30$ & $3.43$ & $0.74$ & FFT, exact\tabularnewline
$2$ & $64$ & $30$ & $4$ & $0.82$ & FIR filtering, exact\tabularnewline
$2$ & $32$ & $27$ & $2$ & $0.55$ & FIR filtering, inexact\tabularnewline
$2$ & $16$ & $25$ & $1$ & $0.41$ & FIR filtering, inexact\tabularnewline
$2$ & $12$ & $23$ & $0.75$ & $0.37$ & FIR filtering, inexact\tabularnewline
$2$ & $8$ & $21$ & $0.5$ & $0.34$ & FIR filtering, inexact\tabularnewline
\hline
$4$ & $128$ & $60$ & $4.67$ & $1.08$ & FFT, exact\tabularnewline
$4$ & $128$ & $60$ & $8$ & $1.54$ & FIR filtering, exact\tabularnewline
$4$ & $32$ & $50$ & $2$ & $0.71$ & FIR filtering, inexact\tabularnewline
$4$ & $24$ & $48$ & $1.5$ & $0.64$ & FIR filtering, inexact\tabularnewline
$4$ & $16$ & $42$ & $1$ & $0.57$ & FIR filtering, inexact\tabularnewline
$4$ & $8$ & $31$ & $0.5$ & $0.50$ & FIR filtering, inexact\tabularnewline
\hline
\end{tabular}}
\par\end{centering}

\centering{}\small{~\\Common to all simulations are the parameters
$\ndat=\ncar=512$, $L=15$, $\nrx=4$, and $\cipH=2$.}
\end{table}

Table~\ref{tbl:simparams} summarizes the simulation parameters,
along with the corresponding values of the interpolation complexity
per target point~$\cipqr$ and the resulting algorithm complexity
ratio $\calg{II}/\calg I$, which quantifies the savings of Algorithm~II
over Algorithm~I. The values of $\calg{II}/\calg I$ for the case
where $\tilde{\mathbf{Q}}(s)$ and~$\tilde{\mathbf{R}}(s)$ are interpolated
exactly by FFT are provided for reference. We note that for $\ntx=4$,
exact interpolation, both FFT-based and through FIR filtering, results
in $\calg{II}>\calg I$. Hence, in this case inexact interpolation
is necessary to obtain complexity savings of Algorithm~II over Algorithm~I.
In contrast, for $\ntx=2$, Algorithm~II exhibits lower complexity
than Algorithm~I even in the case of exact interpolation.

\begin{figure}
\begin{centering}
\begin{tabular}{>{\centering}m{0.48\textwidth}>{\centering}m{0.48\textwidth}}
\centering{}\includegraphics[width=0.48\textwidth]{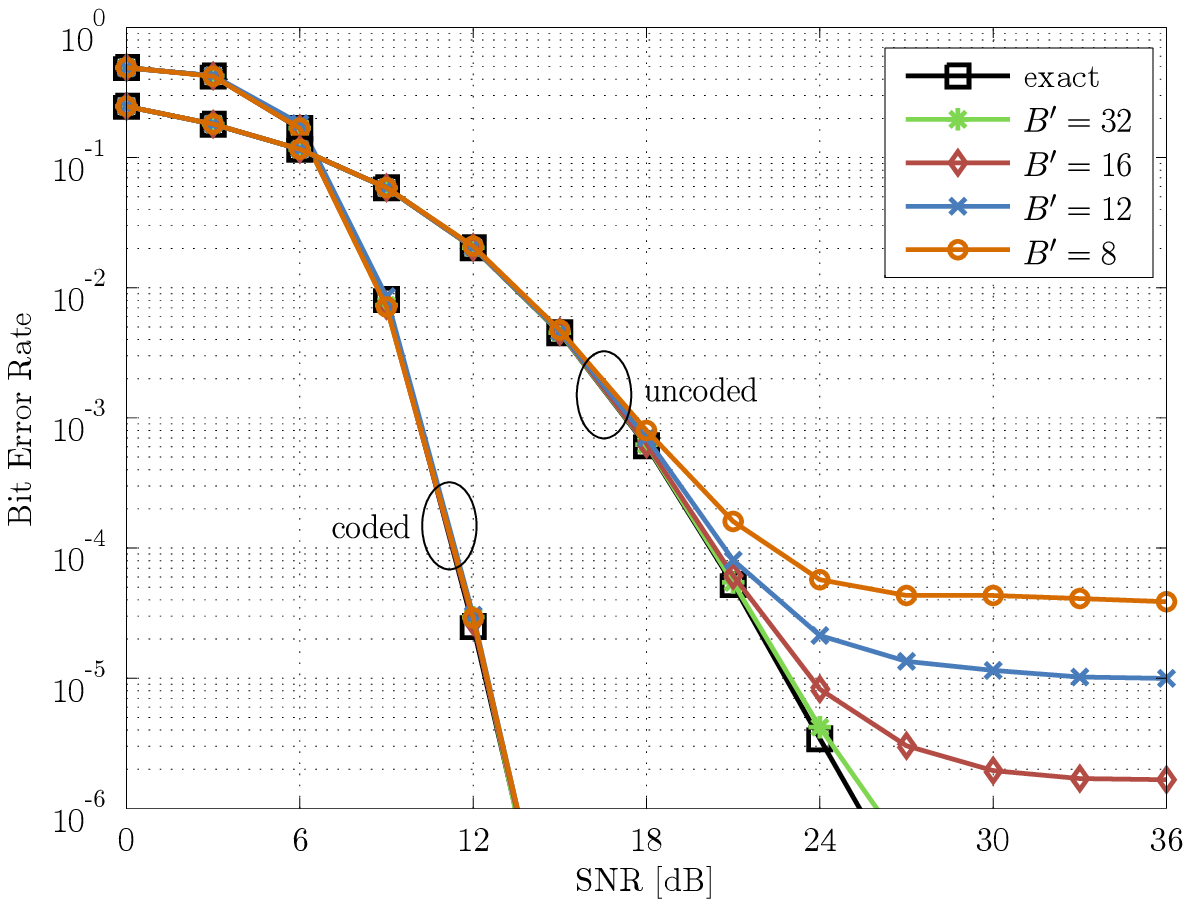} & \centering{}\includegraphics[width=0.48\textwidth]{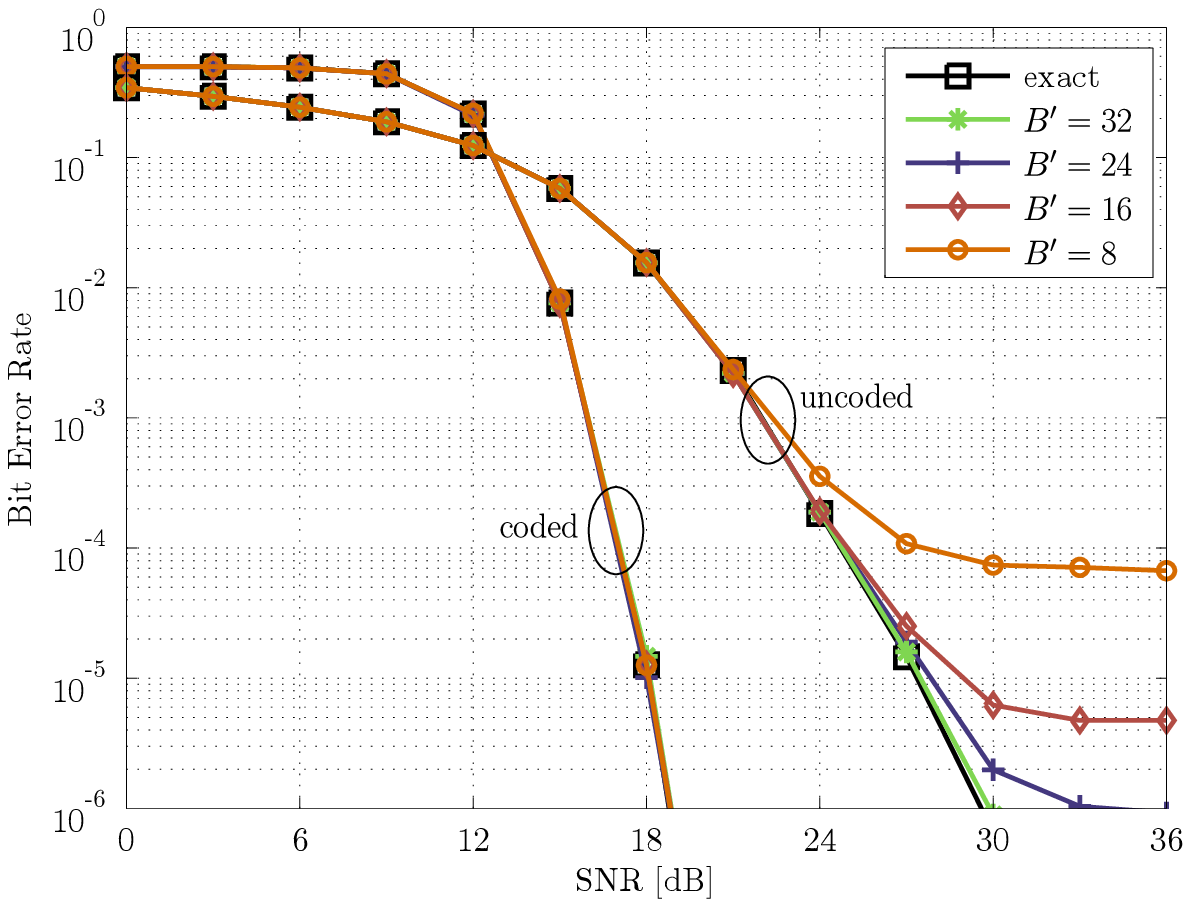}\tabularnewline
\small{(a)} & \small{(b)}\tabularnewline
\end{tabular}
\par\end{centering}

\caption{\label{fig:suboptipBERplots}Bit error rates as a function of SNR
for different interpolation filter lengths, with and without channel
coding, for (a) $\ntx=2$ and (b) $\ntx=4$. The results corresponding
to exact QR~decomposition are provided for reference.}

\end{figure}

Figs.~\ref{fig:suboptipBERplots}a and~\ref{fig:suboptipBERplots}b
show the resulting BER performance for $\ntx=2$ and $\ntx=4$, respectively,
both for the coded and the uncoded case. For uncoded transmission
and inexact interpolation, we observe an error floor at high SNR which
rises with decreasing~$\nbase'$. For $\ntx=2$ and uncoded transmission,
we can see in Fig.~\ref{fig:suboptipBERplots}a and Table~\ref{tbl:simparams},
respectively, that an interpolation filter length of $\nbase'=8$
results in negligible performance loss for SNR values of up to 18~dB,
and yields complexity savings of Algorithm~II over Algorithm~I of~66\%.
Choosing $\nbase'=16$ yields close-to-optimum performance for SNR
values of up to 24~dB and complexity savings of~59\%. For $\ntx=4$
and uncoded transmission, Fig.~\ref{fig:suboptipBERplots}b and Table~\ref{tbl:simparams}
show that the interpolation filter length can be shortened from $\nbase'=128$
to $\nbase'=8$, leading to complexity savings of Algorithm~II over
Algorithm~I of 50\%, at virtually no performance loss in the SNR
range of up to 21~dB. Setting $\nbase'=32$ results in a performance
loss, compared to exact interpolation, of less than 1~dB at $\mbox{BER}=10^{-6}$
and in complexity savings of~29\%. In the coded case, both for $\ntx=2$
and $\ntx=4$, we can see in Figs.~\ref{fig:suboptipBERplots}a and
\ref{fig:suboptipBERplots}b that the BER curves for Algorithm~II,
for all values of~$\nbase'$ under consideration, essentially overlap
with the corresponding curves for Algorithm~I for BERs down to~$10^{-6}\!$.
This observation suggests that for a given target BER and a given
tolerated performance loss of Algorithm~II over Algorithm~I, the
use of channel coding allows to employ significantly shorter interpolation
filters (corresponding to a smaller~$\cipqr$ and hence to a lower
$\calg{II}$, which in turn implies higher savings of Algorithm~II
over Algorithm~I) than in the uncoded case. We conclude that in the
practically relevant case of coded transmission, complexity savings
of Algorithm~II over Algorithm~I can be obtained at negligible detection
performance loss.

\subsection{Algorithm Complexity Comparisons}

\label{sub:numcomparison}

The discussion in Section~\ref{sec:efficientinterpolation} and the
numerical results in Section~\ref{sub:numperfdegradation} demonstrated
that for the case of upsampling from equidistant base points, small
values of~$\cip$ can be achieved and inexact interpolation does
not necessarily induce a significant detection performance loss. Therefore,
in the following we assume that for all $\allone k{\ntx}$, the set~$\mathcal{I}_{k}$
is such that~$\setidx{\setidxbase_{k}}$ contains $\nbase_{k}=|\setidxbase_{k}|=2^{\left\lceil \log_{2}(2kL+1)\right\rceil }$
base points that are equidistant on~$\setunit$, and assume that
$\cip=2$. The latter assumption is in line with the values of~$\cipH$
and~$\cipqr$ found in Section~\ref{sub:numperfdegradation}.

\begin{figure}
\begin{centering}
\begin{tabular}{>{\centering}m{0.48\textwidth}>{\centering}m{0.48\textwidth}}
\centering{}\includegraphics[width=0.48\textwidth]{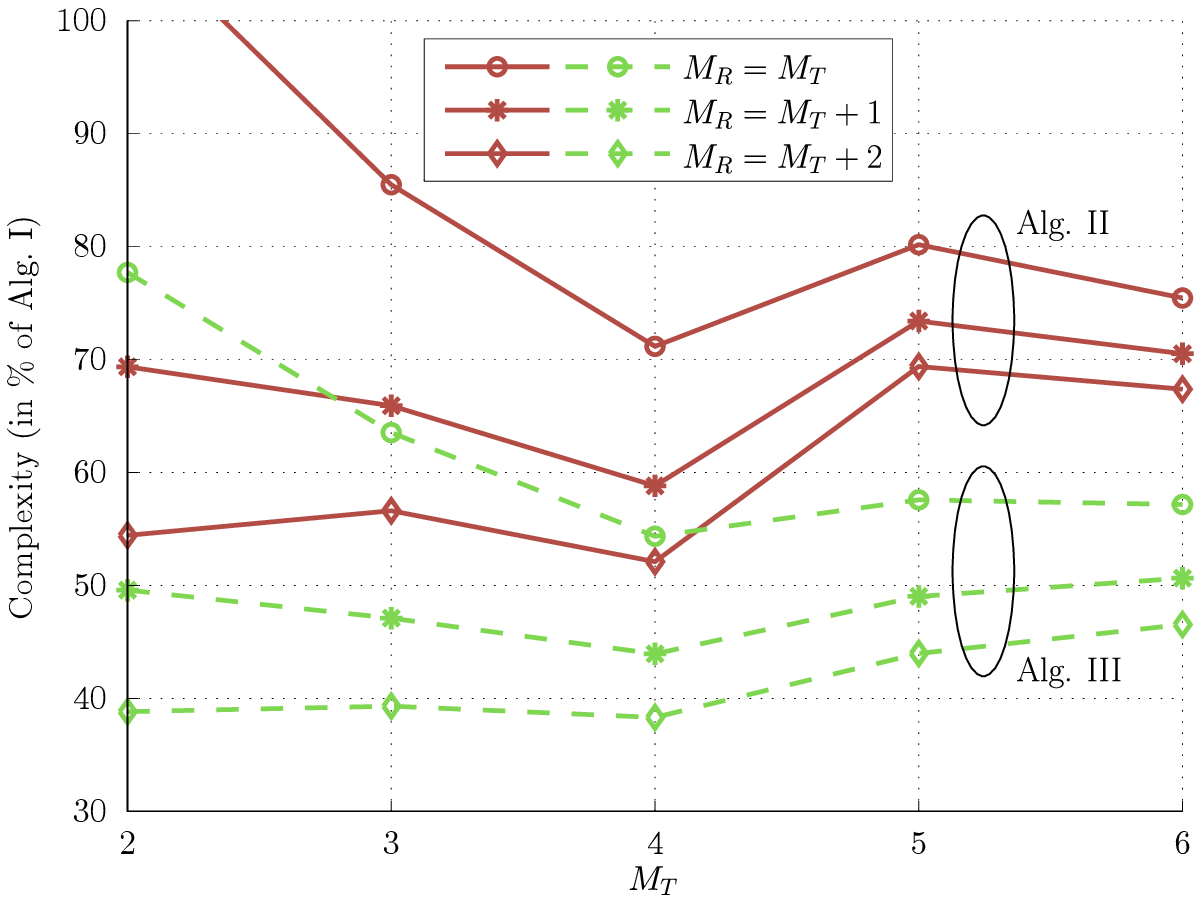} & \centering{}\includegraphics[width=0.48\textwidth]{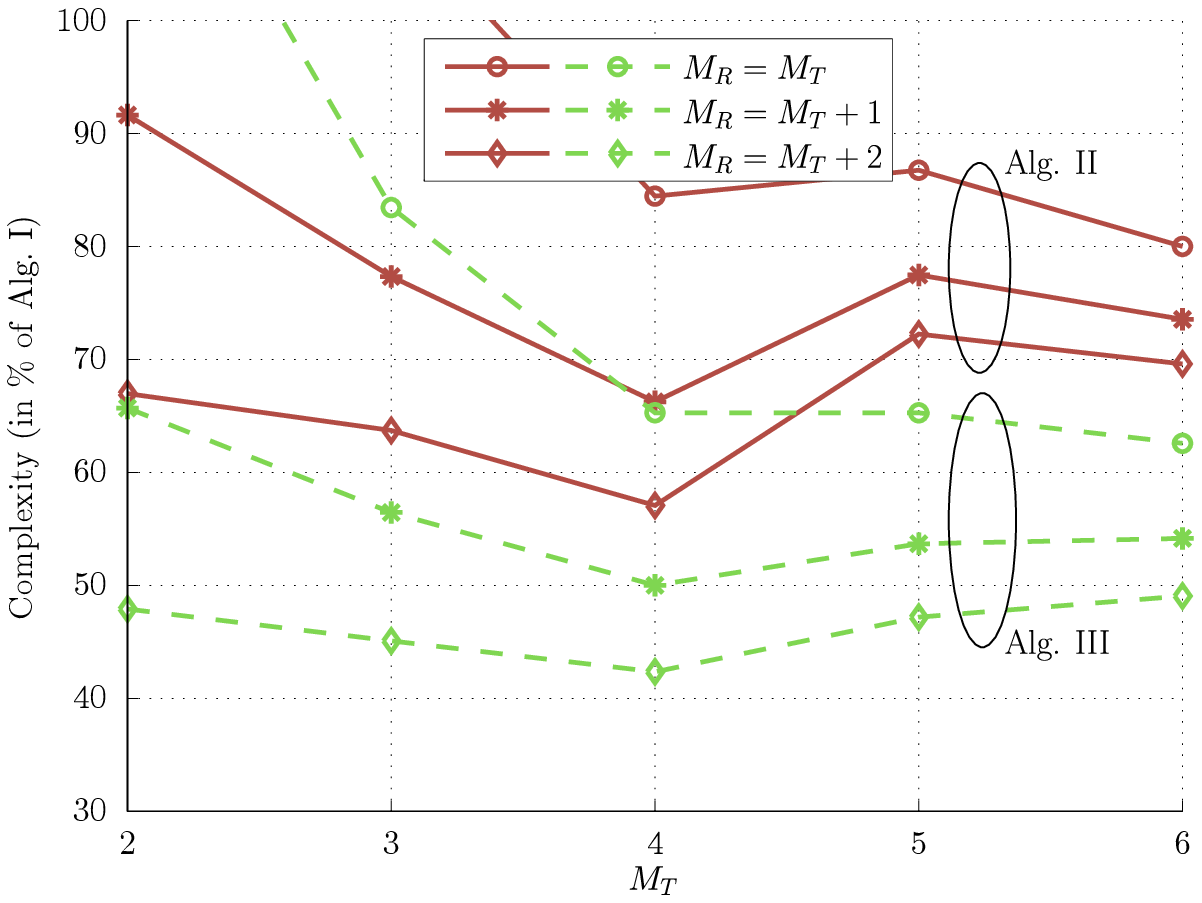}\tabularnewline
\small{(a)} & \small{(b)}\tabularnewline
\end{tabular}
\par\end{centering}

\caption{Complexity of Algorithms~II and~III as percentage of complexity
of Algorithm~I for $\ndat=500$, and $L=15$, (a) including and (b)
excluding the complexity of interpolation of~$\mattf(s)$.\label{fig:mtmrdiff}}

\end{figure}

For $\ndat=500$, $L=15$, and different values of~$\ntx$ and~$\nrx$,
Fig.~\ref{fig:mtmrdiff}a shows the complexity of Algorithms~II
and~III as percentage of the complexity of Algorithm~I. We observe
savings of Algorithms~II and~III over Algorithm~I as high as 48\%
and 62\%, respectively. Furthermore, we can see that Algorithm~III
exhibits a lower complexity than Algorithm~II in all considered configurations.
We note that the latter behavior is a consequence of the small value
of~$\cip$ and of Algorithm~III, with respect to Algorithm~II,
trading a lower QR~decomposition cost against a higher interpolation
cost. Moreover, we observe that the savings of Algorithms II and~III
over Algorithm~I are more pronounced for larger $\nrx-\ntx$. For
the special case $\setpil=\setdat$, where interpolation of~$\mattf(s)$
is not necessary and Algorithm~I simplifies to the computation of~$\ndat$
QR~decompositions, Fig.~\ref{fig:mtmrdiff}b shows that the relative
savings of Algorithms II and~III over Algorithm~I are somewhat reduced,
but still significant. We can therefore conclude that interpolation-based
QR~decomposition, provided that the complexity of interpolation is
sufficiently small, yields fundamental complexity savings.

\begin{figure}
\begin{centering}
\begin{tabular}{>{\centering}m{0.48\textwidth}>{\centering}m{0.48\textwidth}}
\centering{}\includegraphics[width=0.48\textwidth]{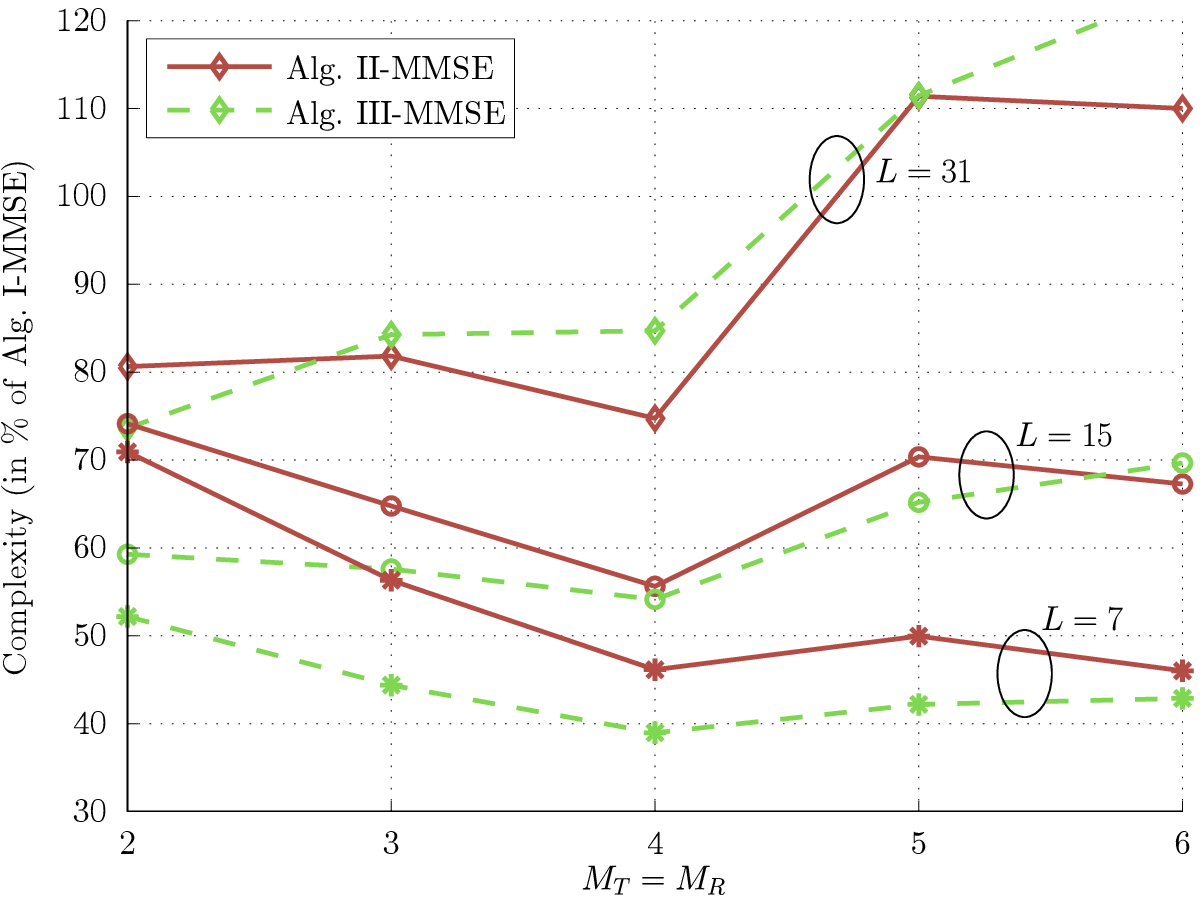} & \centering{}\includegraphics[width=0.48\textwidth]{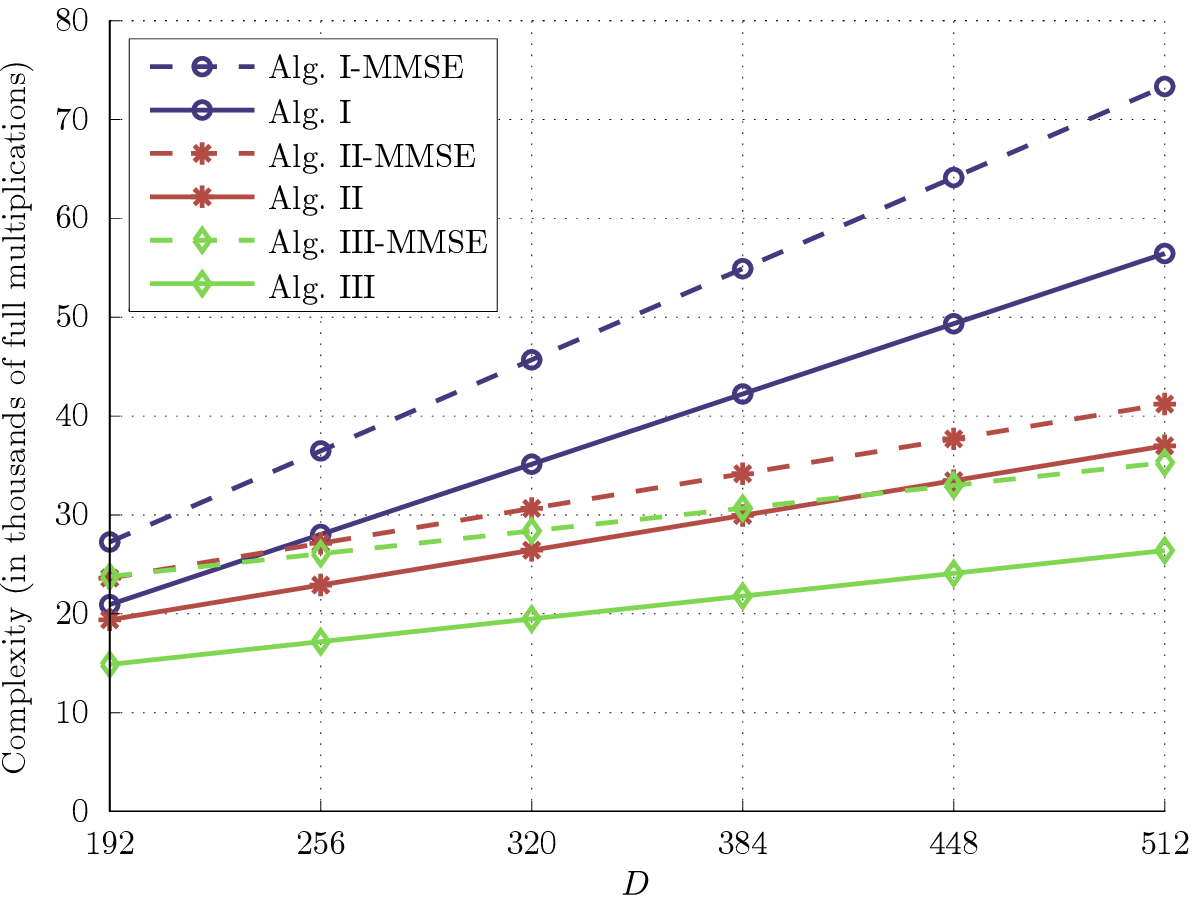}\tabularnewline
\small{(a)} & \small{(b)}\tabularnewline
\end{tabular}
\par\end{centering}

\caption{(a) Complexity of Algorithms~II-MMSE and~III-MMSE as percentage
of complexity of Algorithm~I-MMSE for $\ndat=500$ and $L=15$. (b)
Absolute complexity of Algorithms I--III and I-MMSE through III-MMSE,
for $\ntx=3$, $\nrx=4$, and $L=15$.\label{fig:mmseandabsolute}}

\end{figure}

For $\ndat=500$, $\ntx=\nrx$, and different values of~$L$, Fig.~\ref{fig:mmseandabsolute}a
shows the complexity of Algorithms~II-MMSE and III-MMSE as percentage
of the complexity of Algorithm~I-MMSE. The fact (which also carries
over to the savings of Algorithms II and~III over Algorithm I) that
the savings of Algorithms II-MMSE and III-MMSE over Algorithm I-MMSE
are more pronounced for smaller values of~$L$ is a consequence of~$\nbase_{k}$
being an increasing function of~$L$. In Fig.~\ref{fig:mmseandabsolute}a,
we can see that despite the low interpolation complexity implied by
$\cip=2$, Algorithm~III-MMSE may exhibit a higher complexity than
Algorithm~II-MMSE. This is a consequence of the fact that for some
values of~$\ntx$, $\nrx$, and~$L$, the overall complexity of
the UT-based QR~decompositions with standard form~(\ref{eq:utbasedqrstkA})
required in Algorithm III-MMSE is larger than the overall complexity
of the efficient UT-based regularized MMSE-QR decompositions with
standard form~(\ref{eq:utbasedregqr}) required in Algorithm~II-MMSE.

Finally, Fig.~\ref{fig:mmseandabsolute}b shows the absolute  complexity
of Algorithms I--III and I-MMSE through III-MMSE as a function of~$\ndat$,
for $\ntx=3$, $\nrx=4$, and $L=15.$ We observe that the complexity
savings of Algorithms II and~III over Algorithm I and the savings
of Algorithms II-MMSE and III-MMSE over Algorithm I-MMSE grow linearly
in~$\ndat$. This behavior was predicted for Algorithms I and~II
by the analysis in Section~\ref{sub:complcomparison}, where we showed
that $\calg I-\calg{II}$ is an affine function of~$\ndat$ and is
positive for small~$\cip$ and large~$\ndat$.

\section{Conclusions and Outlook}

\label{sec:conclusions}

On the basis of a new result on the QR~decomposition of LP~matrices,
we formulated interpolation-based algorithms for computationally efficient
QR~decomposition of polynomial matrices that are oversampled on the
unit circle. These algorithms are of practical relevance as they allow
for an (often drastic) reduction of the receiver complexity in MIMO-OFDM
systems. Using a complexity metric relevant for VLSI implementations,
we demonstrated significant and fundamental complexity savings of
the proposed new class of algorithms over brute-force per-tone QR~decomposition.
The savings are more pronounced for larger numbers of data-carrying
tones and smaller channel orders. We furthermore provided strategies
for low-complexity interpolation exploiting the specific structure
of the problem at hand.

The fact that the maximum degree of the LP~matrices $\tilde{\mathbf{Q}}(s)$
and~$\tilde{\mathbf{R}}(s)$ is $2\ntx L$, although the polynomial
MIMO transfer function matrix~$\mattf(s)$ has maximum degree~$L$,
gives rise to the following open questions:
\begin{itemize}
\item Is the mapping $\map$ optimal in the sense of delivering LP matrices
with the lowest maximum degree?
\item Would interpolation-based algorithms for QR~decomposition that explicitly
make use of the unitarity of~$\mathbf{Q}(s)$ allow to further reduce
the number of base points required and hence lead to further complexity
savings?
\end{itemize}
Additional challenges include the extension of the ideas presented
in this paper to sparse channel impulse responses, for which only
few of the impulse response tap matrices are nonzero.

\section*{Acknowledgments}

The authors would like to thank Andreas Burg and Simon Haene for many
inspiring and helpful discussions, Jan Hansen and Moritz Borgmann
for their contributions in early stages of this work, and Gerhard
Doblinger for bringing\citet{skinner76} to their attention.

\bibliographystyle{elsart-num-sort}
\bibliography{/home/dcescato/ctgsvn/ctgtoolchest/texmf.ctg/bibtex/bib/IEEEtran/IEEEabrv,/home/dcescato/ctgsvn/ctgtoolchest/texmf.ctg/bibtex/bib/ctg/confs-jrnls,/home/dcescato/ctgsvn/ctgtoolchest/texmf.ctg/bibtex/bib/ctg/publishers,/home/dcescato/ctgsvn/ctgtoolchest/texmf.ctg/bibtex/bib/ctg/newctgrefs}

\end{document}